\RequirePackage{fix-cm}

\documentclass[fleqn,usenatbib]{mnras}

\usepackage[T1]{fontenc}
\usepackage{lmodern}
\usepackage{pdflscape}
\usepackage{subcaption}
\usepackage{xfrac}

\DeclareRobustCommand{\VAN}[3]{#2}
\let\VANthebibliography\thebibliography
\def\thebibliography{\DeclareRobustCommand{\VAN}[3]{##3}\VANthebibliography}

\usepackage{graphicx}	
\usepackage{amsmath}	
\usepackage{amssymb}	

\title[Data-driven analysis of absorption indices]{A novel data-driven approach to extract stellar population properties from galaxy spectra using absorption indices}

 \author[Sharbaf et al.]{Zahra Sharbaf$^{1, 2,3,4}$\thanks{zsharbaf@ice.csic.es}, Ignacio Ferreras$^{5,2,3}$\thanks{Corresponding author: i.ferreras@ucl.ac.uk}, Anna R. Gallazzi$^4$, Stefano Zibetti$^{4,6}$, \and Daniele Mattolini$^{4,7}$, Laura Scholz-D\'\i az$^4$\\
$^1$ Institute of Space Sciences (ICE, CSIC), Campus UAB, Carrer de Can Magrans, s/n, 08193, Cerdanyola del Valles, Spain\\
$^2$ Instituto de Astrof\'\i sica de Canarias, C/ V\'\i a L\'actea s/n, La Laguna, E-38200 La Laguna, Tenerife, Spain\\
$^3$ Departamento de Astrof\'\i sica, Universidad de La Laguna, E-38205 La Laguna, Tenerife, Spain\\
$^4$ INAF-Osservatorio Astrofisico di Arcetri, Largo Enrico Fermi 5, I-50125 Firenze, Italy\\
$^5$ Department of Physics and Astronomy, University College London, Gower Street, London WC1E 6BT, UK\\
$^6$ Dipartimento di Fisica e Astronomia, Universit\`a degli Studi di Firenze, Via G. Sansone 1, I-50019 Sesto Fiorentino, Italy\\
$^7$ Dipartimento di Fisica, Universit\`a di Trento, Via Sommarive 14, I-38123 Povo (TN), Italy}

\date{Accepted 2026 August 11. Received 2026 August 11; in original form 2026 June 10}

\pubyear{2026}

\begin{document}
\label{firstpage}
\pagerange{\pageref{firstpage}--\pageref{lastpage}}
\maketitle

\begin{abstract}
In an era of highly complex machine learning methods that often are informative but not straightforward to interpret, Principal Component Analysis (PCA)
offers a simple, easily interpretable approach. With no fitting parameters,
it extracts the most salient statistical trends in data without the
need for training sets. In this paper, we explore a large range of composite stellar population models defined for
detailed analyses of galaxy spectra from surveys. Six of the most
prominent spectral indices are targeted to visualize a PCA-based
latent space created by the model data. The age-metallicity degeneracy
is broken in the 3-dimensional space spanned by the first three
eigenvectors, but we emphasize that non-trivial combinations of all
six absorption indices are needed for this.  Moreover, the last
eigenvector suggests an intriguing tug of war between two Balmer
indices: H$\gamma_A$ and $H\delta_A$, that can help discern the
presence of recent bursting behaviour, as it exploits the different
behaviour of the two indices over timescales $\sim$0.5-1\,Gyr.
Comparisons can be made between SDSS and LEGA-C galaxy spectra based on the
latent space created by the models. This method, based on pure data,
produces excellent results in agreement with standard SPS model fitting
techniques, allowing for the study of stellar populations in a variety
of surveys or observational/synthetic databases on solid ground.
\end{abstract}



\section{Introduction}
\label{Sec:Intro}

Depending on the star formation activity, galaxies can generally be classified into two main categories: star-forming and quiescent \citep[e.g.,][]{Williams2009, Brammer2011, Pozzetti2010, Patel2013}. They can be further characterised by their morphology, colour, gas content, and kinematics \citep[e.g.,][]{Strateva2001, Kauffmann2003, bell2012,Sachdeva2020, Taylor2022}. Galaxies have complex evolutionary journeys shaped by both internal processes (such as star formation, gas infall, gas exhaustion, and feedback from active galactic nuclei) and external factors (such as mergers, interactions, and environmental influences). The specific route -- whether a rapid, dramatic transformation through mergers and AGN feedback or a slow, gradual decline due to gas starvation and environmental effects, or even a combination of all the above -- is largely determined by these processes. \citep[e.g.,][]{peng2010,Somerville2015, Henriques2015, Wang2025}. Studies of stellar populations provide crucial insight into the evolutionary paths of galaxies, as their physical properties are a consequence of the star-formation history and metal enrichment. Galaxy spectra, in particular, encode essential information about stellar ages, chemical composition, and kinematics, providing key observational evidence to understand galaxy evolution.

One established method to study the stellar population content of galaxies involves measuring the strengths of specific absorption features in galaxy spectra—most notably through the Lick index system \citep{Worthey1994}. By analysing Lick index measurements, it is possible to constrain the mean stellar ages and metallicities of galaxies \citep[to name a few, see, e.g.,][]{Jorgensen1999, Trager2000, Kuntschner2000, Kauffmann2003, Thomas2003, Gallazzi:05, Gallazzi2014, Rogers:10, McDermi2015, wu2018}. Advancements in the development of stellar population synthesis (SPS) models \citep[e.g.,][]{Bruzual2003, Vazdekis2016, Choi2016, Maraston2005, Maraston2020}, based on moderate resolution stellar spectra -- both empirical and theoretical -- have enabled the generation of detailed synthetic galaxy spectra. Index measurements are compared to grids of SPS model predictions, allowing researchers to characterise key stellar population properties, including metallicity and age. Accurate estimates of these parameters are crucial for reconstructing the chemical enrichment history (CEH) and star formation history (SFH) of galaxies. However, extracting this information remains challenging due to the significant degeneracy among different model parameters, which limits our ability to derive detailed constraints beyond broad estimates such as average age and metallicity. Multivariate analysis methods can be applied to the spectroscopic data, reducing the problem to blind source separation, making use of data-driven diagnostics such as covariance, clustering, or statistical independence, to overcome the entanglement \citep[see, e.g.,][]{Madgwick:03, Kaban2005, Lu2006, Ferreras2006, Rogers2007, Nolan2008, Wild2014}.

Our study aims to develop a data-driven approach for physical comparison of galaxy samples based on their variance in spectral diagnostics. We apply this approach to galaxy samples at different redshifts from SDSS and LEGA-C to study population evolution across redshift. The spectral (co)variance can be combined with stellar population synthesis models to extract information from galaxy spectra \citep{Wild2009, Chen2012, Sharbaf2025}. Our approach via PCA in this work is novel in that the information vectors are defined from the models, and the observed spectra are subsequently projected onto this model-based basis. By applying PCA to the absorption index measurements of the composite stellar population models, we map the data into a new coordinate space shaped by the main sources of variance in these synthetic galaxy spectra. We emphasise that the variance of the input data in this work only relates to sets of the absorption indices in the photospheres of stellar populations. We can trace the data-driven components back to the real physical meaning by exploring the physical properties of the models in the latent space, which helps us decipher the stellar population content of galaxies and their variations. The novel aspect of this study is that PCA is applied only to a subset of absorption index measurements from models, not the entire spectrum. Note that the aim is not dimensionality reduction; it is to rearrange the underlying information in terms of variance in order to decipher its astrophysical meaning. Finally, to investigate the evolution of galaxies across redshift, we take two samples of galaxies from the SDSS and LEGA-C surveys, projecting their absorption indices onto the eigenvectors obtained from the models to explore their evolution.

The structure of the paper is as follows: the composite population models and observational data are presented in \S\ref{Sec:samples}, followed by an overview of the methodology in \S\ref{Sec:methodology}, including the adopted absorption indices (\S\S\ref{Ssec:LS}) and a brief description of Principal Component Analysis (\S\S\ref{Ssec:PCA}). Latent space is presented in \S\ref{Sec:Latent}, with emphasis on assessing the role of recent bursting episodes (\S\S\ref{Ssec:BurstoNo}). In \S\ref{Sec:Real} we confront the model-derived latent space with real galaxies covering a wider redshift range. Finally, we discuss and summarise our results in \S\ref{Sec:Disc}.

\section{Synthetic and observational samples}
\label{Sec:samples}

This work adopts as a theoretical reference a sample of synthetic spectra, the composite stellar population models (CSPs) introduced by \citet{Zibetti2017}. Comparisons with real data are performed for two samples of optical spectra retrieved from the Sloan Digital Sky Survey \citep[SDSS,][]{SDSS}, and the Large Early Galaxy Census Survey \citep[LEGA-C,][]{LEGAC}. The analysis relies on absorption features measured in the same way on model and observed spectra. To ignore the effects of broadening of absorption indices due to resolution and stellar motion, and to compare data and models at the same broadening, all the samples (the synthetic one and the two observational ones) are restricted to the velocity dispersion range of 150-250 km/s. 

\subsection{Composite Stellar Population (CSP) models}
\label{CSPs}

Our approach is hybrid in nature: while we adopt a data-driven statistical method, we rely on a model library in which the sampling of parameter space is defined a priori.
In this study, we use a CSP model library containing 500,000 synthetic spectra, created following the approach presented in \citeauthor{Zibetti2017} (\citeyear{Zibetti2017}; BaStA\footnote{\url{https://www.basta.inaf.it}}), which, in turn, represents a development of the method introduced by \citet{Kauffmann2003} and \citet{Gallazzi:05}. The library’s actual realisation is the same as that employed by \citet{Mattolini:25}, implementing the latest version of the BC03 SSP models \citep[CB19, see appendix A of][for more details]{Sanchez:22}, with updated SFH and chemical enrichment prescriptions (see below and M25 for more details).

The library is constructed to populate the observable parameter space (i.e. the space of absorption indices) as comprehensively and uniformly as possible, while there is no physical relation imposed by the SFH, CEH and dust. We allow for a wide variety of SFHs, adopting a flexible functional form for the secular component and superimposing stochastic bursts spanning a broad range of intensities and ages. Metallicity is treated as independent of the mean stellar age, except for the constraint of monotonic increase with time. The distribution of the models is equalised in the H$\gamma_A$+H$\delta_A$ versus D$_n$(4000) plane to ensure uniform coverage of the allowed parameter space. This provides a library with a maximum physically possible variance in the observed parameter space and provides a base for PCA.

PCA in this model-defined space captures the intrinsic correlations among observables in a purely theoretical framework, excluding by construction correlations driven by galaxy-scale astrophysical processes unrelated to fundamental stellar physics. A succinct breakdown of the model components and their prescriptions follows:

\bigskip

\noindent
{1) Simple Stellar Populations: These are the building blocks of the CSPs. The library is based on the SSPs from the 2019 version of \citet{Bruzual2003}. These SSPs assume a Chabrier Initial Mass Function \citep[IMF,][]{Chabrier2003} with a maximum stellar mass of $100\, M_{\odot}$. The models adopt the MILES spectral library \citep{Sanchez2006}, and stellar evolution follows the PARSEC evolutionary tracks \citep{Bressan2012, Marigo2013, Chen2015}.

\medskip
  
\noindent
{2) Star Formation History:} The SFH for each CSP consists of a continuous parametric component, with additional random bursts of star formation, to produce a more realistic depiction of the actual formation histories.

\medskip

\noindent
-- {Continuous (Secular) Component:} This is described by a delayed Gaussian function \citep{Sandage1986, Gavazzi2002}, that allows for both increasing and decaying phases in the SFR:
    \begin{equation}
    \mathrm{SFR}_\tau(t) \propto \frac{(t - t_{\mathrm{form}})}{\tau} \exp\left[-\frac{(t - t_{\mathrm{form}})^2}{2\tau^2}\right],
    \label{Eq.1}
    \end{equation}
where $t_{\mathrm{form}}$ is the cosmological time when the SFH starts, and $\tau$ represents a formation timescale.

\noindent
-- {Burst Component:} This component adds stochastic variability to the continuous star formation history (SFH). Two-thirds of the models include bursts, ranging from one to six events, distributed uniformly in log space between $10^5$\,yr and the age of the secular component ($t_{\mathrm{form}}$). Each burst is represented as a single-age, single-metallicity population.

\medskip

\noindent
{3) Chemical Enrichment History:} The models enforce the following evolution of the metallicity:

\begin{equation} 
Z_\star(t) = Z_{\star, \mathrm{final}} - (Z_{\star, \mathrm{final}} - Z_{\star,0}) \left[1 - \frac{M(t)}{M_{\mathrm{final}}}\right]^\alpha,
\label{Eq.2}
\end{equation}

\noindent
where $M(t)$ and $M_{\mathrm{final}}$ denote the stellar mass formed by time $t$ and the total stellar mass formed throughout the star formation history (SFH), respectively. The parameter $\alpha \geq 0$ quantifies the rate at which chemical enrichment progresses relative to star formation.

\medskip

\noindent
4) Dust Attenuation: The stellar population spectra are attenuated using the \cite{Charlot2000} two-component model. This model distinguishes between young stars ($\le 10^7$\,yr), attenuated by dust in their birth clouds, as well as the diffuse interstellar medium (ISM), and older populations ($>10^7$\,yr), that are only attenuated by the diffuse ISM. Therefore, in each CSP, the populations are split with respect to this age cut and attenuated accordingly.

\medskip

\noindent
5) Equalisation of the parameter space: To ensure a smooth and uniform distribution of models across the observable parameter space, particularly in diagnostic diagrams such as the H$\gamma_A$$+$H$\delta_A$, versus D$_n$(4000) plane, the final set is resampled. This step addresses imbalances that can arise when generating a library of CSP models, where certain regions become ``overdense'' (i.e. dominated by clustered models) while others are sparsely sampled. These imbalances are especially pronounced in areas corresponding to young or bursty star formation histories, where small variations in SFH or metallicity can cause large shifts in spectral indices.

For a detailed description of these models, see \citet{Zibetti2017} and \citet{Mattolini:25}.

\subsection{Observational data}

The main goal of this paper is to explore the mapping of population parameters of realistic spectral models of composite stellar populations in the latent space produced by a data-driven method based on PCA. We will apply the results to real observed galaxies, by projecting on latent space the measurements of high quality subsets from two state-of-the-art deep surveys of large and representative samples in the local Universe and at intermediate redshift, SDSS and LEGA-C, respectively. A brief description of the data follows.

\subsubsection{SDSS}
\label{Ssec:SDSS}

The low redshift spectroscopic sample is taken from the Sloan Digital Sky Survey \citep[SDSS,][]{SDSS}, in particular the Legacy dataset, that contains single fibre ($3^{\prime\prime}$ diameter aperture) spectroscopy at R$\approx$2,000 resolution over a wavelength range $\lambda$$\in$[3800,9200]\AA\ \citep{Smee:13} from Data Release 7 \citep{Abazajian2009}. The index measurements in this study are based on the catalogue presented by \citet{Mattolini:25}, which includes index measurements from emission line subtracted spectra that have been corrected for aperture effects \citep{Zibetti2026} and combining duplicate observations, resulting in 825263 unique galaxies. This catalogue covers a redshift range of 0.005 to z=0.22, and quality cuts have been applied to define a robust sample for stellar population analysis (see \citet{Mattolini:25} for further details). Accurate determination of stellar population properties requires a minimum signal-to-noise ratio (S/N) per pixel \citep{Gallazzi:05}; an S/N threshold of 10 was imposed, resulting in a final sample of 332,467 spectra. This selection is particularly important for the subsequent variance analysis, as it minimises the influence of observational noise and helps avoid systematic effects introduced by low-SNR data, thereby ensuring that the sample variance reflects intrinsic galaxy properties rather than measurement uncertainties. Finally, to balance a comparable sample with respect to the higher-redshift LEGA-C survey and sufficient effective spectral resolution, we restrict the stellar velocity dispersion to the interval $\sigma\in$[150,250]\,km/s, resulting in a final working sample of 138,659 galaxies.

\subsubsection{LEGA-C}
\label{Ssec:LEGAC}

An intermediate redshift sample is taken from the LEGA-C survey third data release (DR3), with an effective spectral resolution of R$=$3500, across the wavelength range $\lambda$$\in$[6300,8800]\,\AA\ \citep{LEGAC, Straatman2018}, and with integration time per target of  20\,h. A total of 4,081 galaxies were observed, of which 3,029 were primary targets selected from the UltraVISTA catalogue \citep{Muzzin2013} with redshift 0.6$<$z$<$1 and $K_{s} \leq 20.7 - 7.5 \log((1 + z)/1.8)$\,AB. We selected galaxies with no sign of AGN contamination, and a reliable measurement of stellar velocity dispersion, with spectroscopic redshift 0.55$<$z$<$1.1. This represents a sample of 2,864 spectra. The index measurements were calculated by \citet{Gallazzi:2025}, and are available in their catalogue. The stellar absorption features are measured on emission-line-subtracted spectra. Similarly to the SDSS data, the indices are not translated to the Lick/IDS spectral resolution as defined in the original system \citep{Worthey:94}, but are measured on the observed spectra at their native resolution, keeping the original definitions of the central and sidebands \citep[e.g.,][]{Vazdekis:10}. We exclude index measurements with poor quality, following \citet{LEGAC}, and only estimates with less than $1/3$ of the central or side band pixels flagged are considered reliable. In the case of duplicate observations available for a galaxy, the error-weighted mean of the index measurements is computed for those objects, further resulting in 2,588 unique galaxies. From this set, \citet{Gallazzi:2025} extracted a sample of galaxies which requires the measurement of at least one of the Balmer absorption indices (H$\beta$, H$\gamma_A$, H$\delta_A$) and at least one composite metal-sensitive index. In addition to the commonly used [MgFe]$^\prime$ and [Mg$_2$Fe], they also consider alternative definitions such as MgFe50$^\prime$ and MgFe52$^\prime$, which probe similar stellar population properties while extending less into the red part of the spectrum. This allows the inclusion of galaxies for which the standard indices are not covered at higher redshifts. Imposing these criteria, together with a requirement of low spectroscopic uncertainties, yields 553 unique galaxies (the SILVER sample) up to  z$\lesssim$0.77. This significant reduction in sample size resulted from the wavelength coverage of the LEGA-C spectra, which can not cover the  Mg and Fe features at z$\gtrsim$0.8. We restrict the sample to the stellar velocity dispersion to the common interval with our SDSS galaxies, shown above, i.e., 150-250\,km/s, resulting in a final set of 284 LEGA-C galaxies. Finally, our study also requires all six index measurements (see section \ref{Ssec:LS}) simultaneously for each galaxy, producing a final working sample of 34 galaxies in LEGA-C.

\section{Methodology}
\label{Sec:methodology}

A covariance analysis (explained in the subsection \ref{Ssec:PCA}) of galaxy spectra can be done at various levels. One could use the full spectrum as an
information vector \citep[as in][]{variance}, or make use of the lower dimensional description with absorption indices.
Given that most of the information is typically encoded in a relatively small number of spectral windows \citep{Ferreras2023}, we will adopt here a set of absorption indices, as shown below.

\subsection{Stellar absorption indices}
\label{Ssec:LS}

This study is focused on a carefully selected set of six key indices: D$_n$(4000), H$\beta$, H$\gamma_A$, H$\delta_A$, [MgFe]$^\prime$, and [Mg$_2$Fe]. As well-established spectral diagnostics, these indices provide robust constraints on stellar age and metallicity, two of the most fundamental properties of stellar content in galaxies \citep[e.g.,][]{Worthey1994,WO:97,SCT:98,Gallazzi:05, Gallazzi2014}. The D$_n$(4000) index \citep{Balogh:99} measures the prominent break at 4000\,\AA, a strong age (and metallicity) sensitive index, often used in stellar population analyses as it produces a measurable signal even at low S/N. Balmer absorption is highly sensitive to recent star formation activity, and is associated with the presence of warm young stars of spectral types from late B to early F \citep{Kauffmann2003}. These absorption indices can be contaminated by emission from the nebular gaseous phase, particularly H$\beta$, so corrections are necessary in the observational data. We note that a well-known correlation exists between the higher-order Balmer absorption lines and the $[\alpha/\mathrm{Fe}]$ abundance ratio \citep{Thomas2004}, which can lead to slight biases in the derived ages from these indices. However, the CSP models employed here are based on scaled-solar SSPs and do not account for variable elemental abundance ratios. Therefore, any possible impact of $[\alpha/\mathrm{Fe}]$ variations on our variance analysis is expected to be a secondary effect rather than a driving parameter. To address the age-metallicity degeneracy and related challenges, Balmer absorption indices are typically used in combination with metallicity-sensitive indices, such as the composite magnesium and iron indices [MgFe]$^\prime$ and [Mg$_2$Fe], defined in equations~\ref{Eq.3} and \ref{Eq.4}, \citep{Thomas2003, Poggianti1997}.

\begin{equation}
\mathrm{[MgFe]^\prime}= \sqrt{(\mathrm{Mgb} (0.72\mathrm{Fe5270} + 0.28 \mathrm{Fe5335}))}
\label{Eq.3}
\end{equation}

\begin{equation}
\mathrm{[Mg_{2}Fe]}= 0.6 \mathrm{Mg_2} + 0.4  \log(\mathrm{Fe4531} + \mathrm{Fe5015})
\label{Eq.4}
\end{equation}
[MgFe]$^\prime$ increases with overall metallicity and, to a lower degree, with age. The definition of [Mg$_2$Fe] reflects its sensitivity to both magnesium and iron features, making it a reliable indicator of metallicity. Both are constructed to trace total metallicity while minimising sensitivity to non-solar abundance ratios, most notably [Mg/Fe]. The strategic combination of age-sensitive indices (Balmer indices and D$_n$(4000) with metallicity-sensitive ones (Mg–Fe composites) is essential to mitigate the effects of age-metallicity degeneracy \citep{ameg,Gallazzi:05, Gallazzi2014}.

\subsection{Principal Component Analysis (PCA)}
\label{Ssec:PCA}

We encode each galaxy spectrum -- from models or observations -- as a set of six absorption indices, as shown above, i.e. each galaxy is represented by a six-dimensional vector. The ensemble can be described either by the full set of $N_g$ vectors, packed into a $N_g\times 6$ matrix, with $N_g$ being the number of galaxies in the ensemble (equation~\ref{eq:ensemble}). Alternatively, when applying Principal Component Analysis (PCA), we redefine the sample, with respect to a transformed matrix for which the $6\times 6$ covariance matrix is diagonalised. In this new frame, the ``coordinates'' are no longer the absorption indices of each galaxy, but their projections on the eigenvectors of the transformation. PCA can be used to classify, compress, or denoise spectra \citep[see, e.g.][]{folkes1996, Madgwick:03, Rowlands:2018, Nersesian:21}, or alternatively to construct a latent space to decipher the input dataset \citep[see, e.g.][]{Wild2007, Ferreras2006, Rogers2007, Rogers:10, Chen2012, Safarzadeh2016, variance, Sharbaf2025}. In this study, we apply PCA to the $6\times 6$ covariance matrix formed by the ensemble of models defined by the six absorption indices introduced above.
\begin{equation}
    \Phi_k \equiv  \left( I_k^{(1)}, I_k^{(2)}, \dots, I_k^{(6)} \right), \qquad k=1,\cdots,N_g
\label{eq:ensemble}
\end{equation}
Before applying PCA to the input dataset, we standardise the data to ensure that all absorption indices contribute equally and are on a comparable scale. This step is necessary to homogenise the information across different indices, as they may have varying units and dynamic ranges. Standardisation is performed by subtracting the mean and dividing by the standard deviation of each index, computed across all synthetic spectra (see equation~\ref{Eq.6}). This results in each index having a distribution with zero mean and unit variance.
\begin{equation}
\overline{\Phi}_k \equiv
\frac{\Bigl(\Phi_k(I_i) - \langle\Phi_s(I_i)\rangle_s\Bigr)}
{\sqrt{\langle\Phi_s^2(I_i)\rangle_s - \langle\Phi_s(I_i)\rangle_s^2}},
\label{Eq.6}
\end{equation}
where $\langle\cdots\rangle_s$ represents averaging over the ensemble. The diagonalisation of the covariance matrix produces six orthogonal, six-dimensional vectors called the principal components (${\hat{e}_\alpha}$). They are arranged in decreasing order of their variance contribution, quantified by their respective eigenvalues. The projections on to the data for each of the $k$ (synthetic) galaxies are defined as the standard dot product:
\begin{equation}
{\rm PC}\alpha_{k,syn} = \overline{\Phi}_{k, {\rm syn}}\cdot\hat{e}_\alpha = \sum\limits_{i=1}^6\overline{\Phi}_{k, {\rm syn}}(I_{i})\hat{e}_\alpha(I_{i}).
\label{eq:PC1}
\end{equation}

\noindent
The eigenvectors obtained by applying PCA to the standardised CSP absorption index measurements are treated as reference information vectors to explore the intrinsic variance in the observational data. Since these vectors are derived from the standardised absorption indices of the CSPs, the same mean and standard deviation vectors of the CSPs must be used to standardise the absorption-index measurements of both observational datasets, producing the projections of the observed data that represent their location in latent space.
\begin{equation}
{\rm PC}\alpha_{k, {\rm Obs}} = \overline{\Phi}_{k, {\rm Obs}}\cdot \hat{e}_\alpha = \sum\limits_{i=1}^6\overline{\Phi}_{k, {\rm Obs}}(I_{i})e_\alpha(I_{i}).
\label{eq:PC1_obs}
\end{equation}

We note that the covariance matrix is sign-invariant, i.e., a change in the sign of a principal component (and thus its projections) does not affect the matrix. Therefore, it is only the relative signs between each index in a given eigenvector that has to be taken into account. If we simultaneously flip all the signs in an eigenvector, the results are unchanged.

\section{Exploring the latent space of synthetic data}
\label{Sec:Latent}

We apply PCA to a set of composite models selected from the original set  after adopting a few additional constraints. Our sample excludes CSP models with formation times older than 14\,Gyr, a limit imposed by the oldest ages of the SSPs. This threshold reduces the sample size from 500,000 models to 414,514 models. The models have an intrinsic spectral resolution (FWHM)
of 2.5\AA\ \citep{MILES:11}, whereas the absorption indices
of the models are measured on spectra smoothed by a range of
additional stellar velocity dispersions, added in quadrature.
As our study is limited to samples of relatively massive galaxies, we choose index measurements with a velocity dispersion
$\sigma$=200\,km/s. By projecting the absorption indices onto the eigenvectors of the covariance matrix of the models, we visualise the latent space of the models, to be contrasted with
real data. In this section, we focus only on the properties
of the models in this alternative representation of the data.

\subsection{Latent space including all CSP models}
\label{CSPs_all}

\begin{figure} 
    \centering
    \includegraphics[width=\columnwidth]{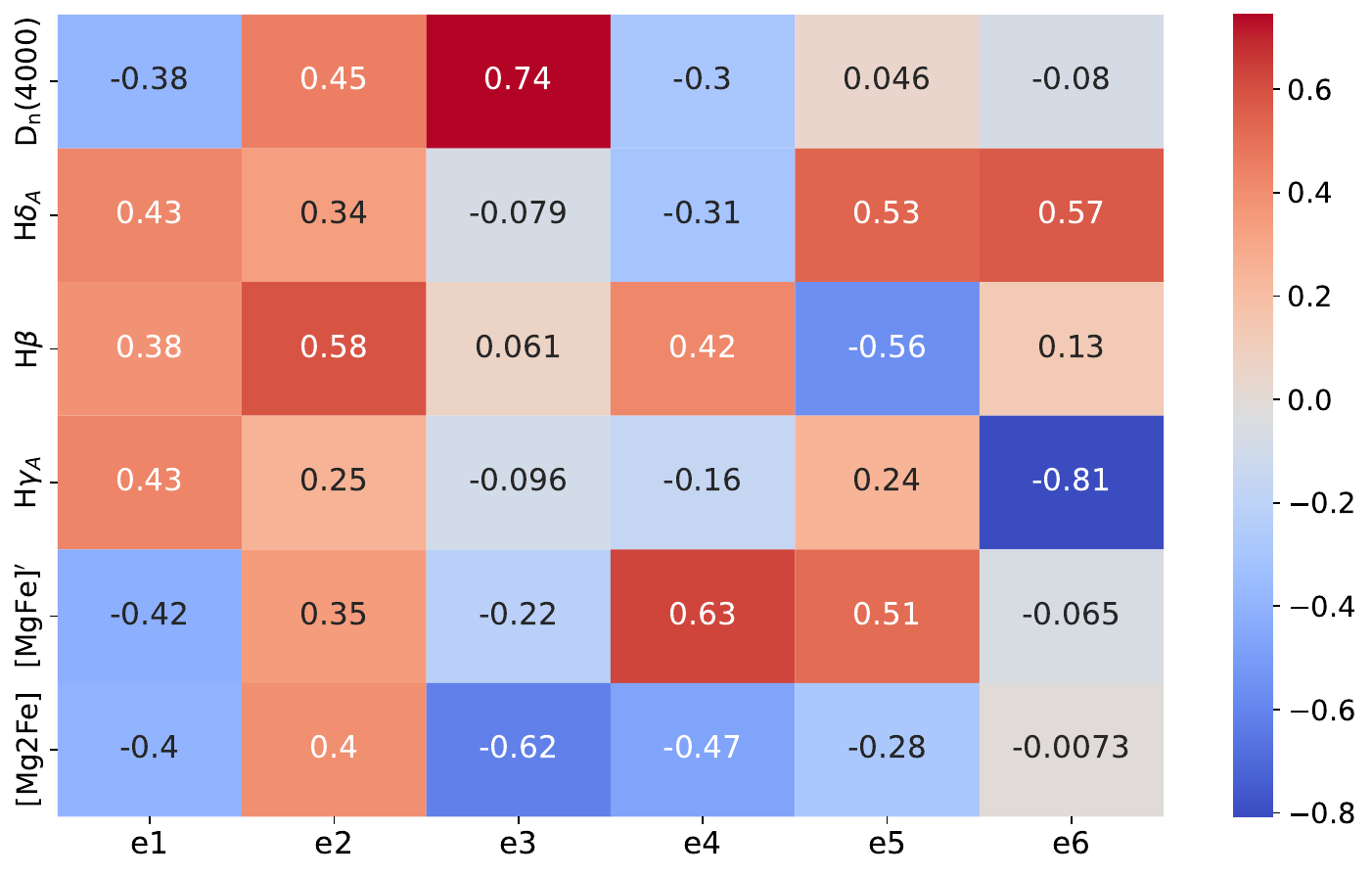}
    \caption{Weights of the six absorption indices of the eigenvectors produced by PCA on the complete set of models.}
    \label{figapp:heatmap_allmodels}
\end{figure}

The fraction of the explained variance for each PC, also known as a scree
plot, from PC1 to PC6 are 83.66\%, 12.34\%, 3.30\%, 0.48\%, 0.19\%, and 0.04\%. The SSP models are constructed from the high-S/N empirical spectra of the MILES stellar library, for which the typical S/N$\gtrsim$100\,\AA$^{-1}$. Therefore, random noise inherited from the input stellar spectra is not expected to dominate the variance captured by the principal components. This interpretation is supported by the systematic correlations found between the higher order PCs, including PC6, and the physical parameters of the models. Nevertheless, because these components explain only a small fraction of the total variance, we interpret them cautiously. 

The first component gives approximately equal weights to all indices. The sign difference for the Balmer indices is just a result of the different variation of the index with respect to age. The second component combines all indices with the same sign, a substantial difference with respect to the first eigenvector. The third component strongly plays D$_n$(4000) against [Mg$_2$Fe], a result that appears to be potentially very interesting, as we will see below, concerning metallicity. Regarding the other, higher-order eigenvectors, PC6 is a very interesting component as it shows strong weights with opposite sign for H$\delta_{A}$ and H$\gamma_{A}$, with very little weight given
to all other indices. We will explore this component in the
discussion section below.

\begin{figure*}
    \centering
    \begin{subfigure}[t]{0.43\textwidth}
        \centering
        \captionsetup{justification=centering}
        \caption*{PC1 vs Spectral Indices}
        \includegraphics[width=\textwidth]{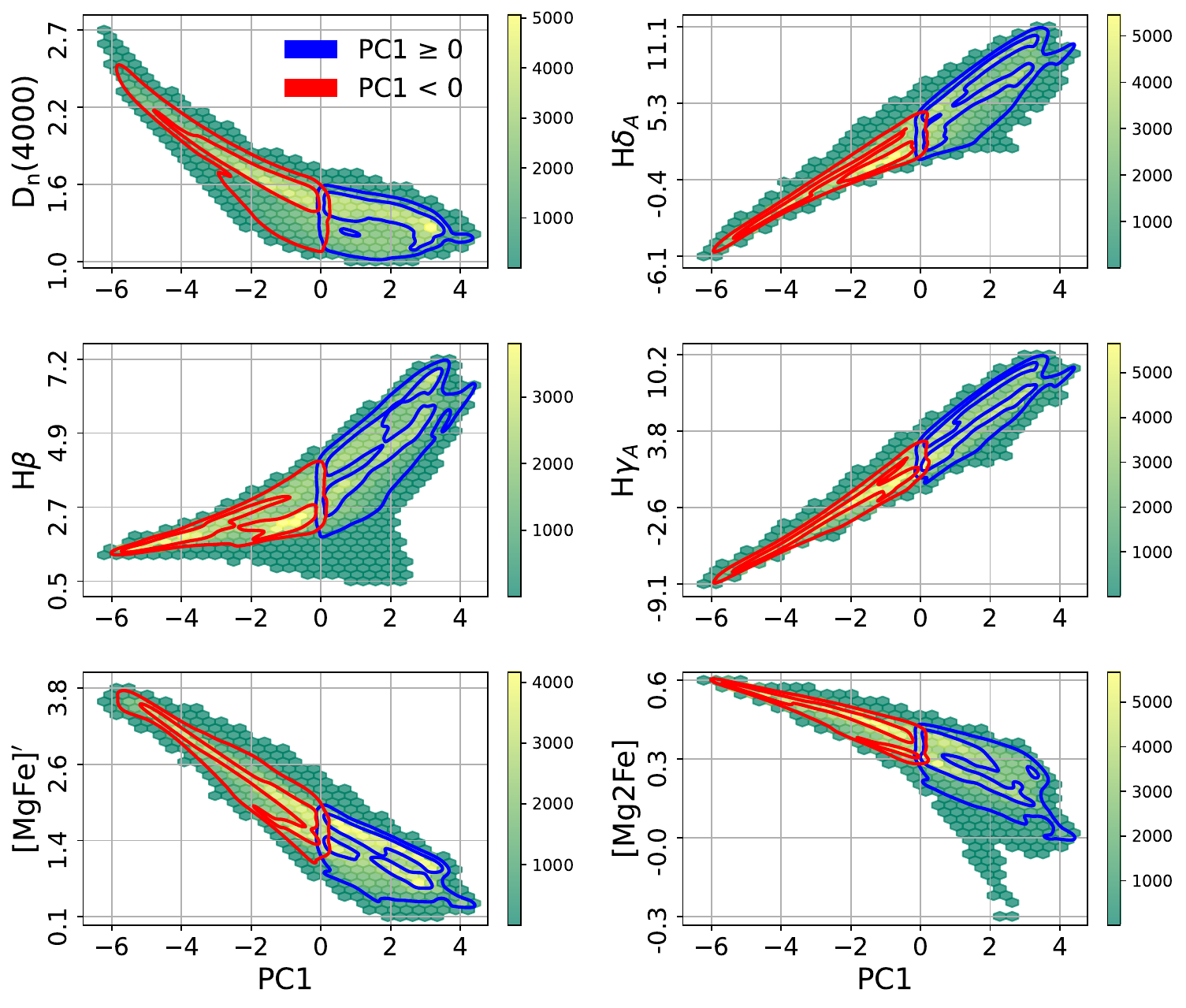}
    \end{subfigure}\hfill
    \begin{subfigure}[t]{0.43\textwidth}
        \centering
        \captionsetup{justification=centering}
        \caption*{PC2 vs Spectral Indices}
        \includegraphics[width=\textwidth]{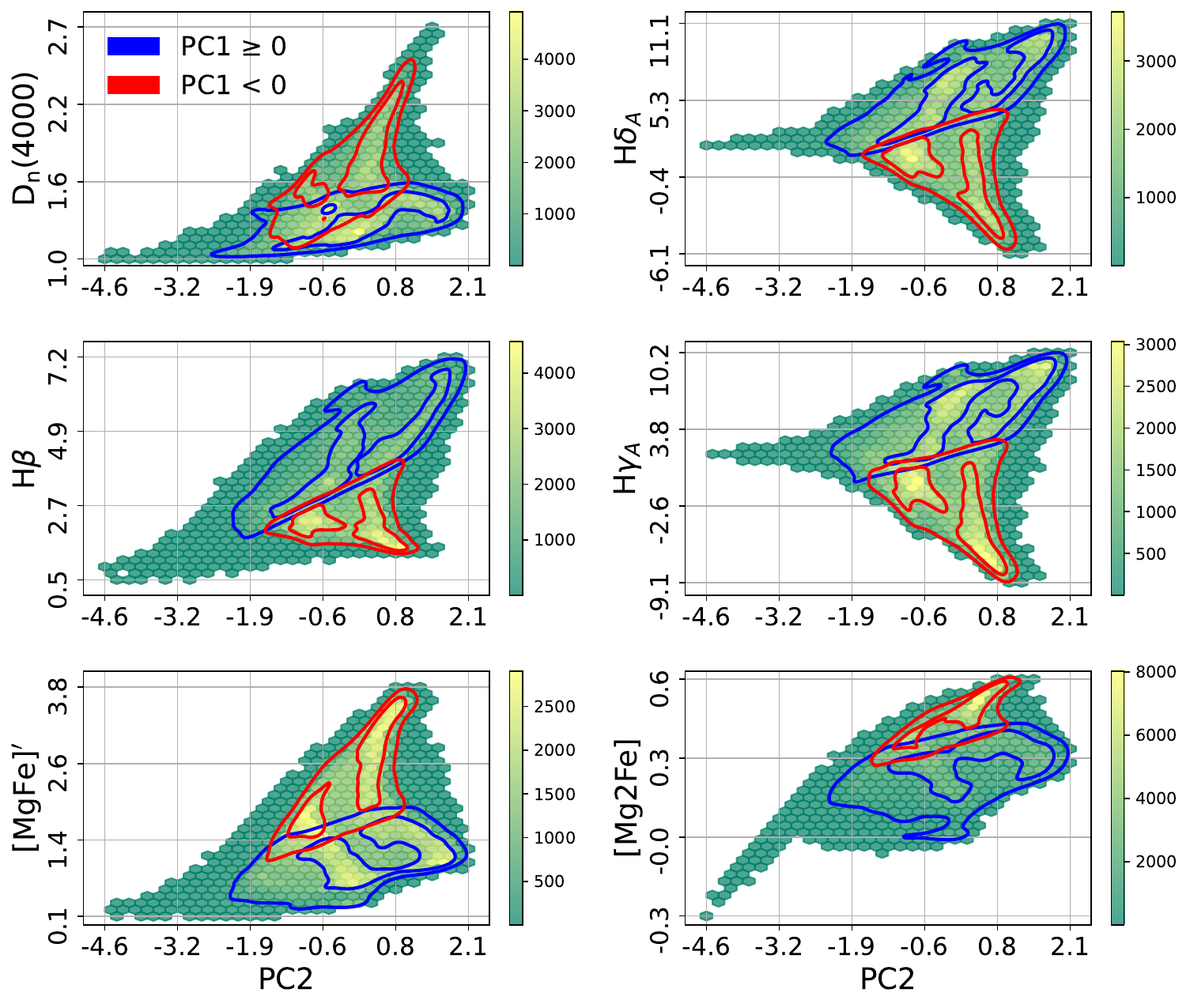}
    \end{subfigure}
    \begin{subfigure}[t]{0.43\textwidth}
        \centering
        \captionsetup{justification=centering}
        \caption*{PC3 vs Spectral Indices}
        \includegraphics[width=\textwidth]{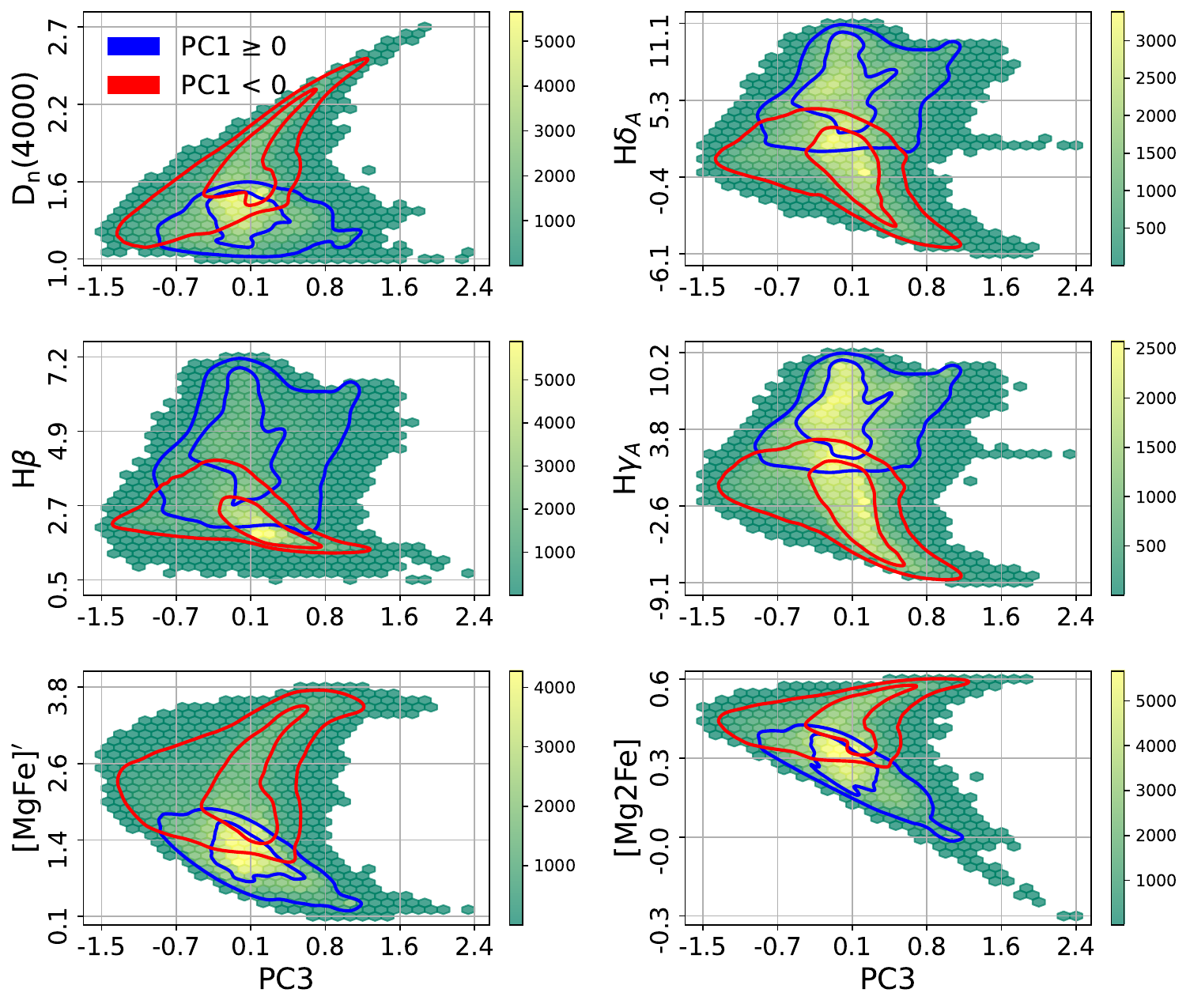}
    \end{subfigure}\hfill
    \begin{subfigure}[t]{0.43\textwidth}
        \centering
        \captionsetup{justification=centering}
        \caption*{PC4 vs Spectral Indices}
        \includegraphics[width=\textwidth]{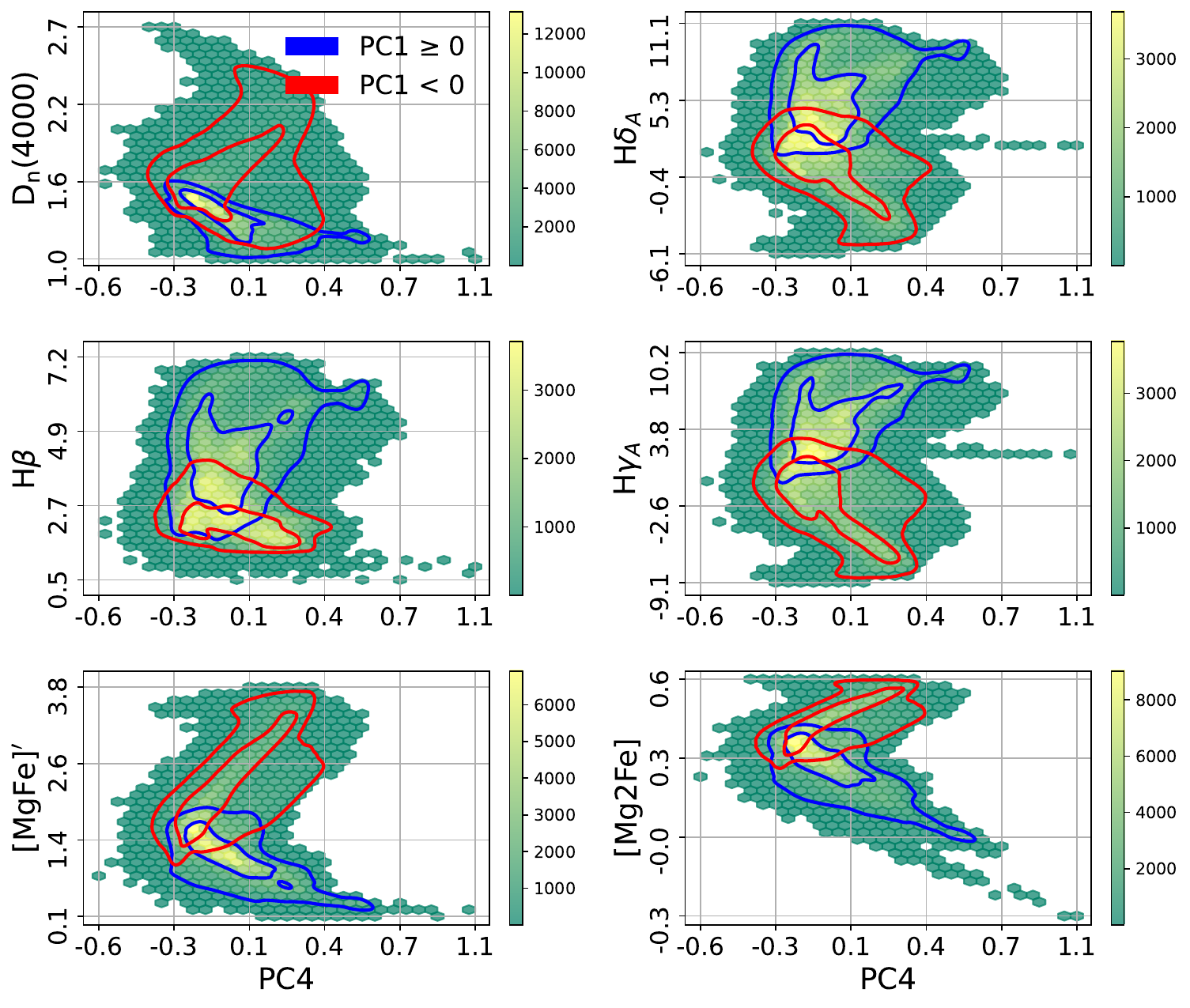}
    \end{subfigure}

    \begin{subfigure}[t]{0.43\textwidth}
        \centering
        \captionsetup{justification=centering}
        \caption*{PC5 vs Spectral Indices}
        \includegraphics[width=\textwidth]{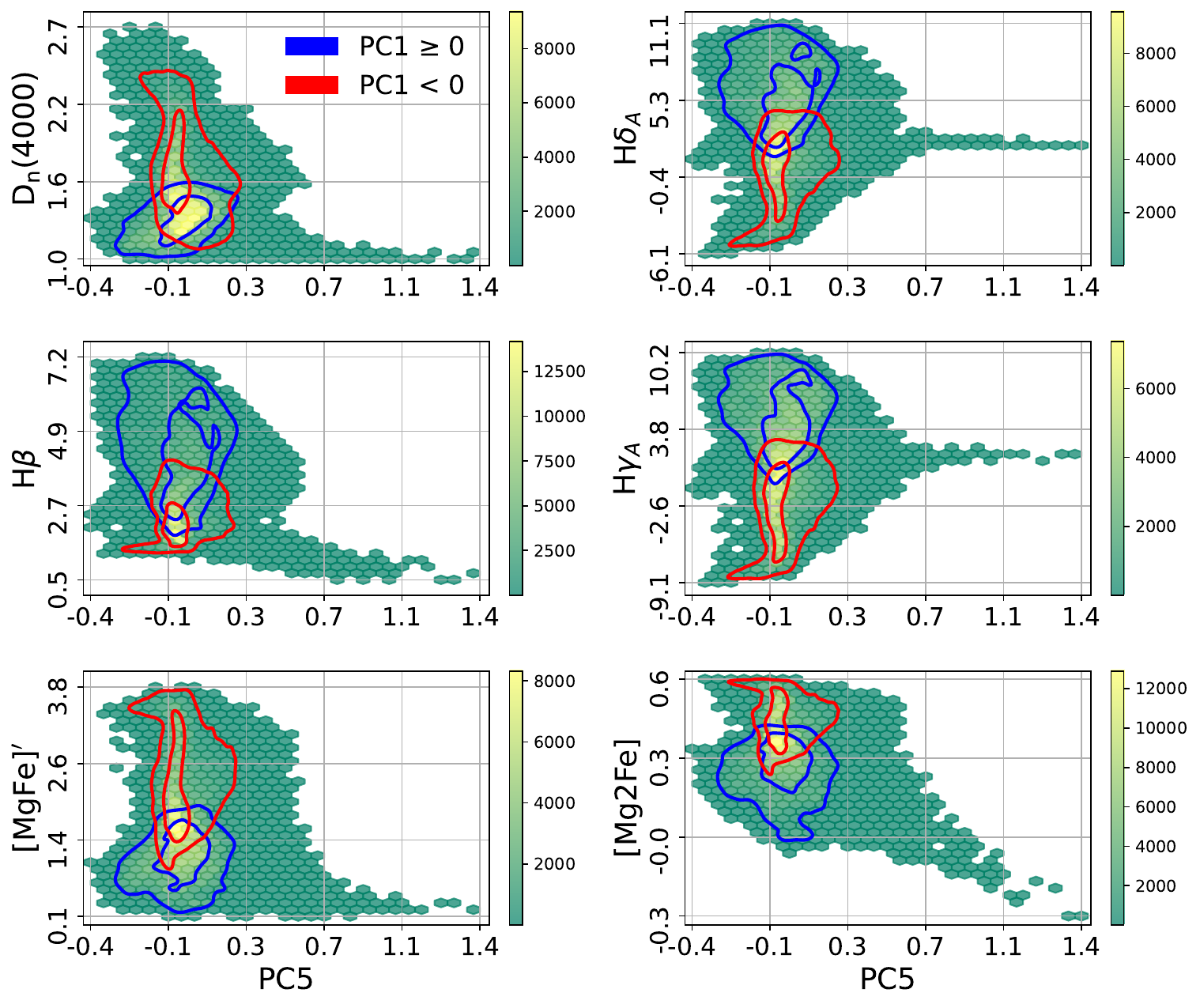}
    \end{subfigure}\hfill
    \begin{subfigure}[t]{0.43\textwidth}
        \centering
        \captionsetup{justification=centering}
        \caption*{PC6 vs Spectral Indices}
        \includegraphics[width=\textwidth]{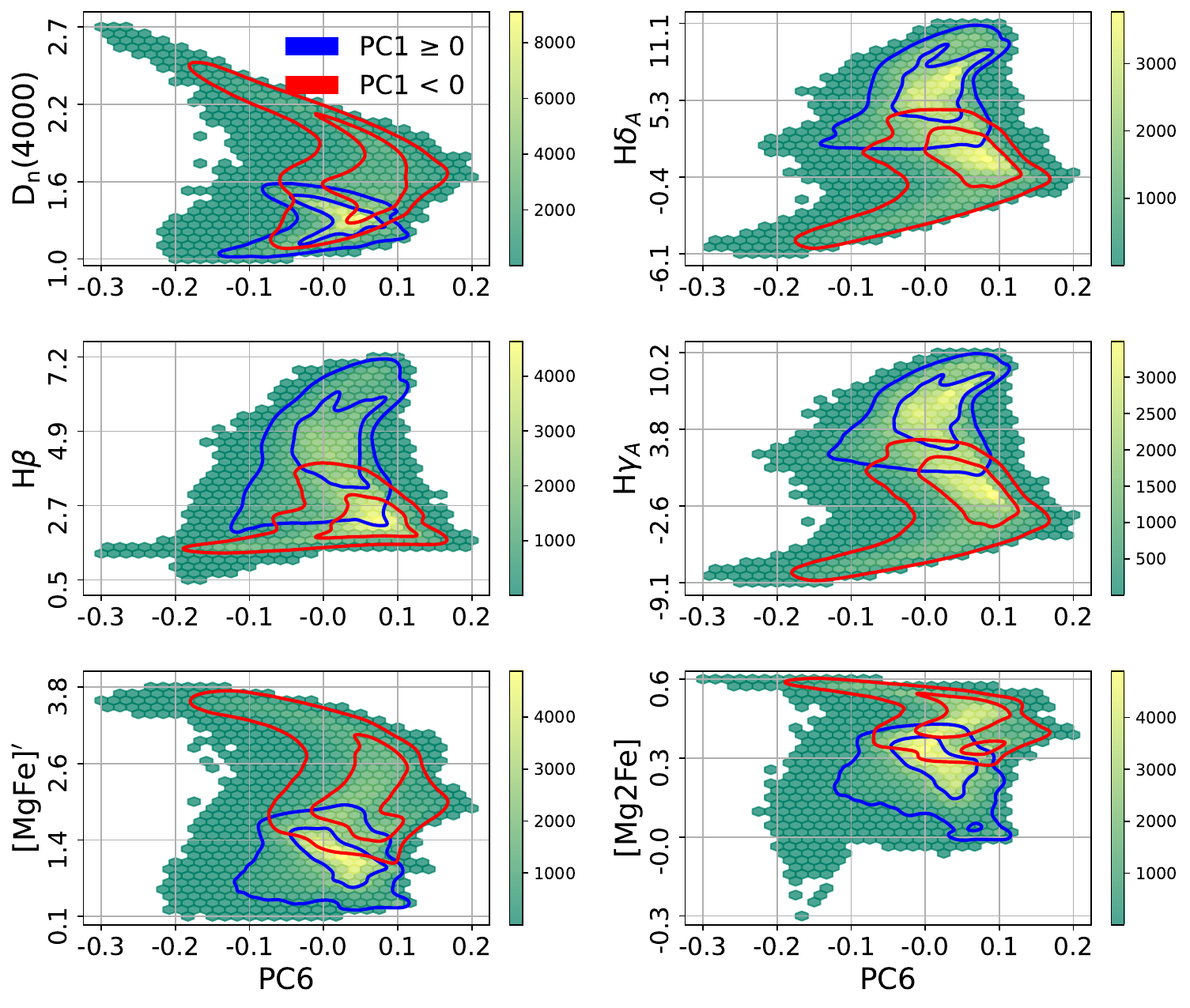}
    \end{subfigure}
    \caption{Correlation of the index measurements with principal components (the projections of the index measurement for all of the CSPs onto the information vectors). The blue (red) contours with three levels illustrate the distribution for models with positive (negative) PC1 values. The color bar indicates the number of models (i.e., the density of points).}
    \label{fig:allmodels}
\end{figure*}

The distribution of absorption indices as a function of principal component projections are shown in Fig.~\ref{fig:allmodels}. The first eigenvector reveals a nearly uniform correlation with all six index measurements. The difference in the relative sign of the Balmer indices versus the other ones is simply a manifestation of the different trends with respect to age. Except for too young populations (younger than a few $10^5$ yr), Balmer indices decrease with age, whereas D$_n$(4000), [MgFe]$^\prime$ and [Mg$_2$Fe] increase with age. This behaviour implies that the projections on to the first component (PC1) encode a global mode of variance, where each index contributes approximately equally. Such a configuration suggests that PC1 does not emphasize specific spectral features, but instead, reflects the overall amplitude -- or dynamical range -- of the combined indices, capturing their coherent variation with respect to stellar age.

To further explore this result, we split the CSP models with respect to the sign of their PC1 projection (shown as coloured contour lines in Fig.~\ref{fig:allmodels}). Models with negative PC1 (red contours) tend to exhibit low-to-intermediate Balmer absorption, while those with positive PC1 values (blue lines) show systematically higher Balmer index values. For the D$_n$(4000) index, negative PC1 values span the full observed range, while positive PC1 data are more tightly associated with the lower end of D$_n$(4000). In contrast, [MgFe]$^\prime$ and [Mg$_2$Fe] display higher to intermediate values in the negative PC1 regime, while positive PC1 models correspond to lower strengths in these metal-sensitive features. Taken together, these trends suggest that negative and positive PC1 values are associated with spectra dominated by cool and hot stars, respectively. In turn, these can be interpreted as old and/or metal-rich as opposed to young and/or metal poor. Thus, PC1 effectively captures a composite axis of stellar population evolution, tracing both age and metallicity effects, through its coherent impact on the index set. The separation based on PC1 sign also aligns clearly with the two branches observed in the PC2 to PC6 maps of Fig.~\ref{fig:allmodels}. When examining the correlations between PC2 to PC6, and the absorption indices, two distinct branches emerge, each corresponding to either the positive or negative PC1 regimes. These branches in latent space reflect a meaningful division in stellar population properties, a result purely driven by the statistical distribution of the sample.

\begin{figure*}
  \includegraphics[width=57mm]{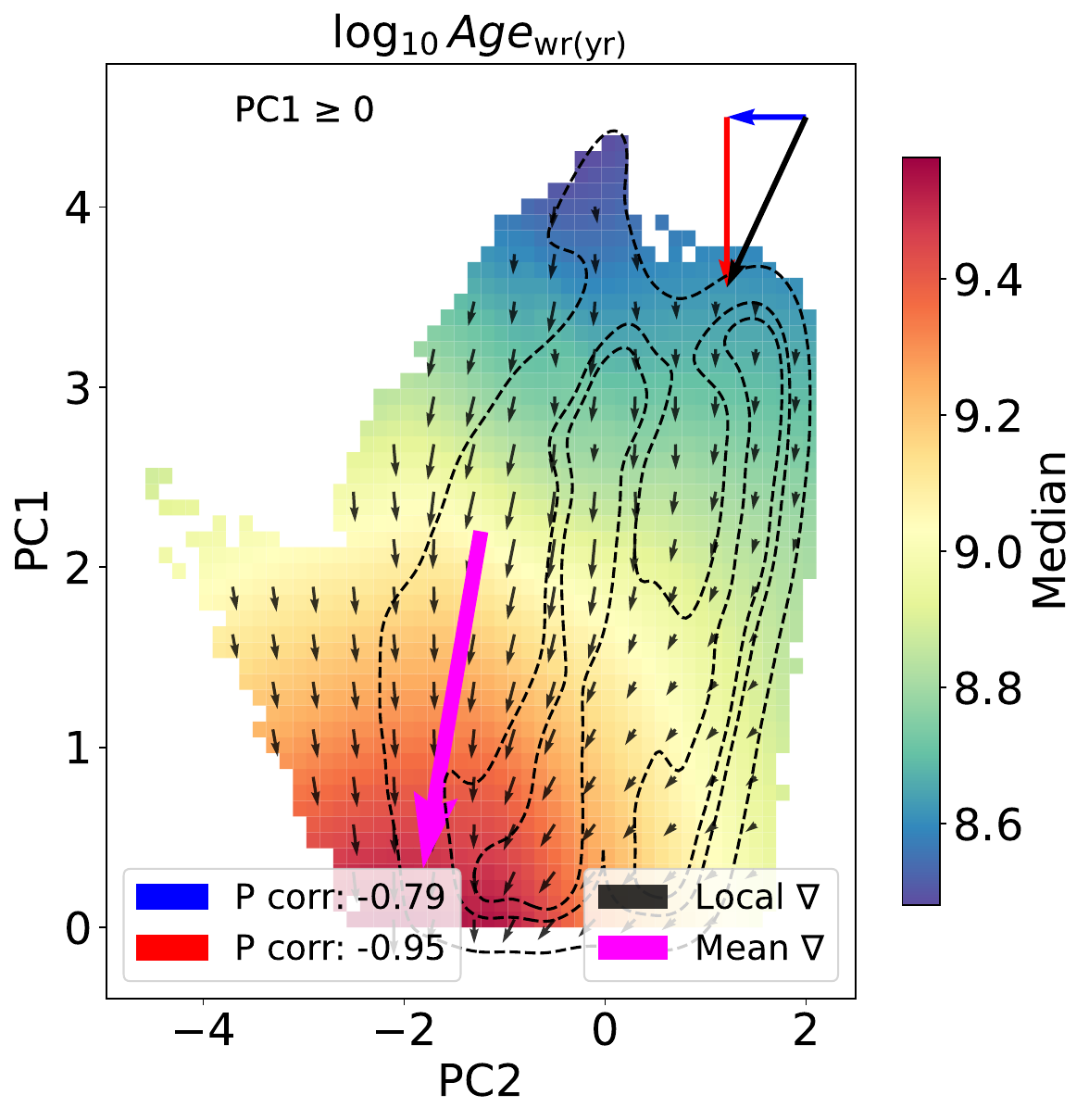}
  \includegraphics[width=57mm]{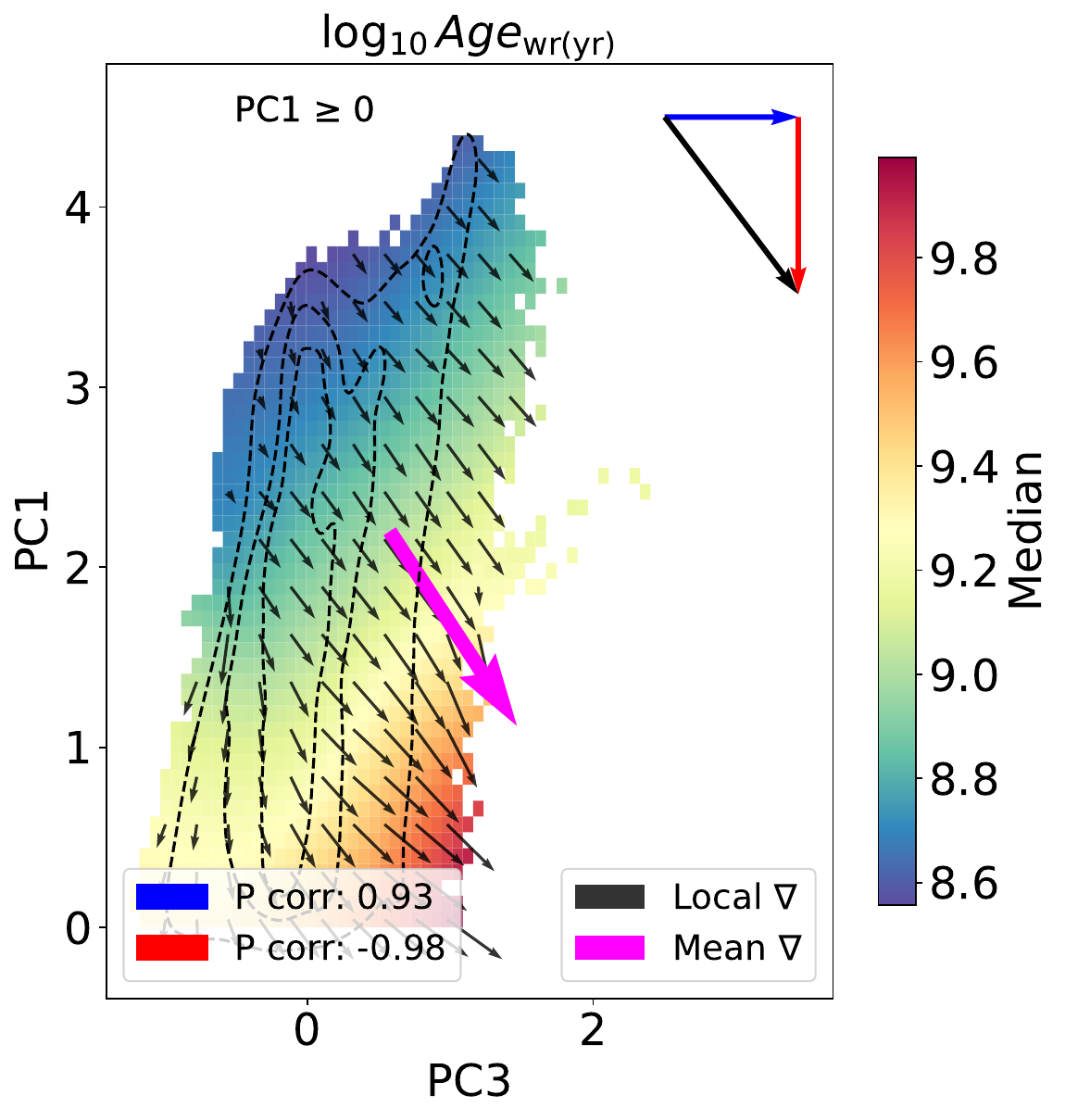}
  \includegraphics[width=57mm]{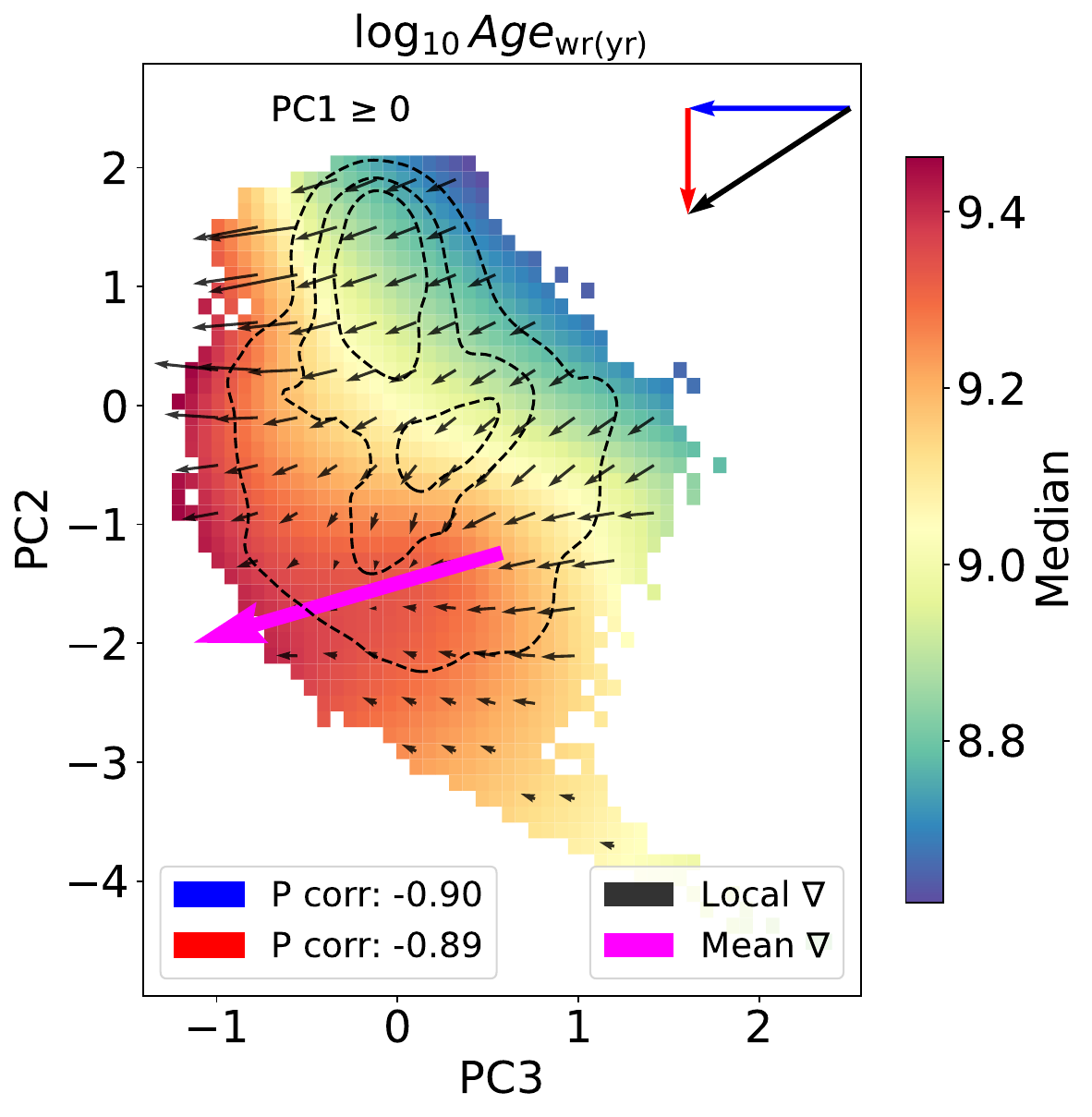}
  \includegraphics[width=57mm]{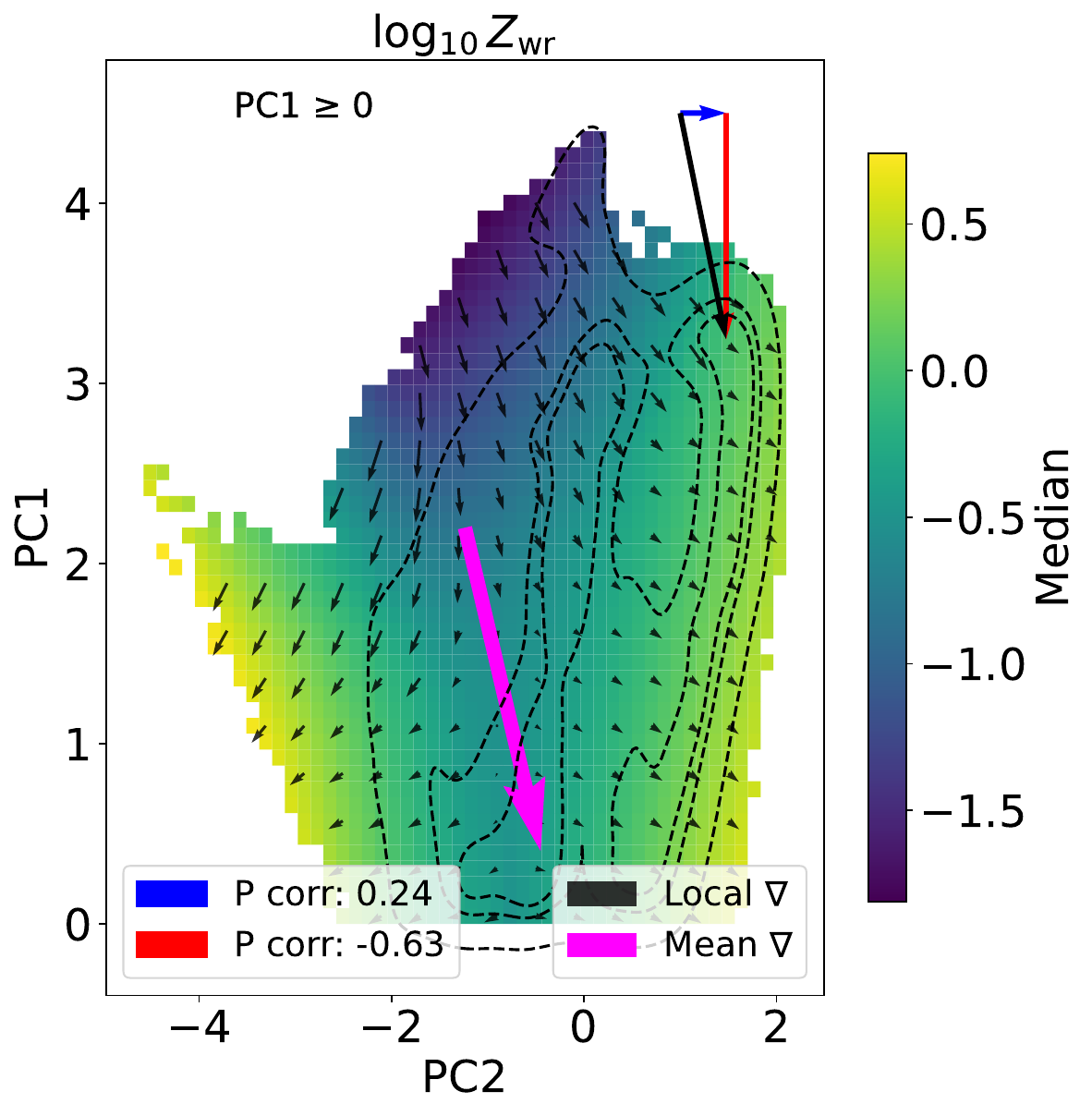}
  \includegraphics[width=57mm]{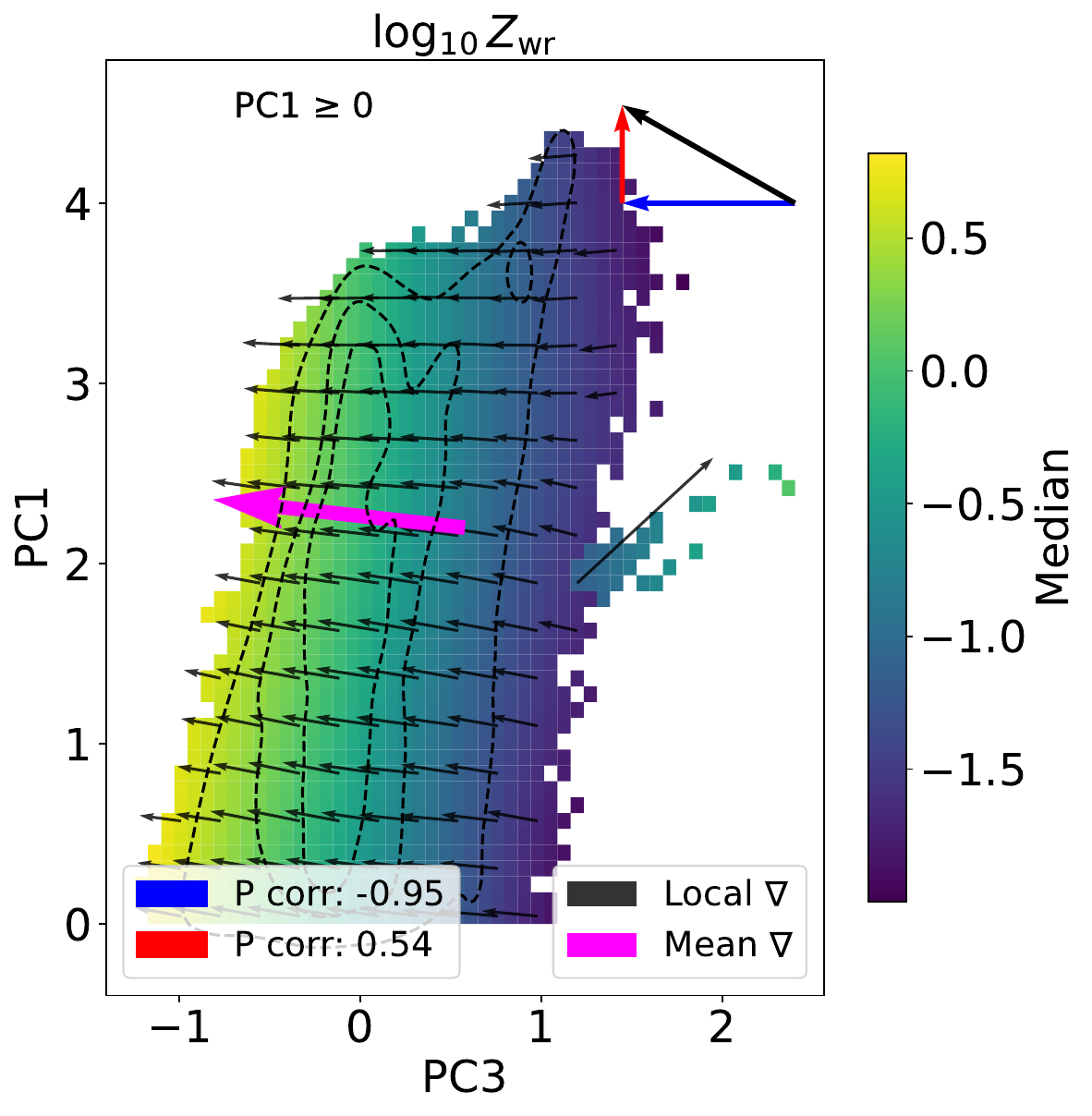}
  \includegraphics[width=57mm]{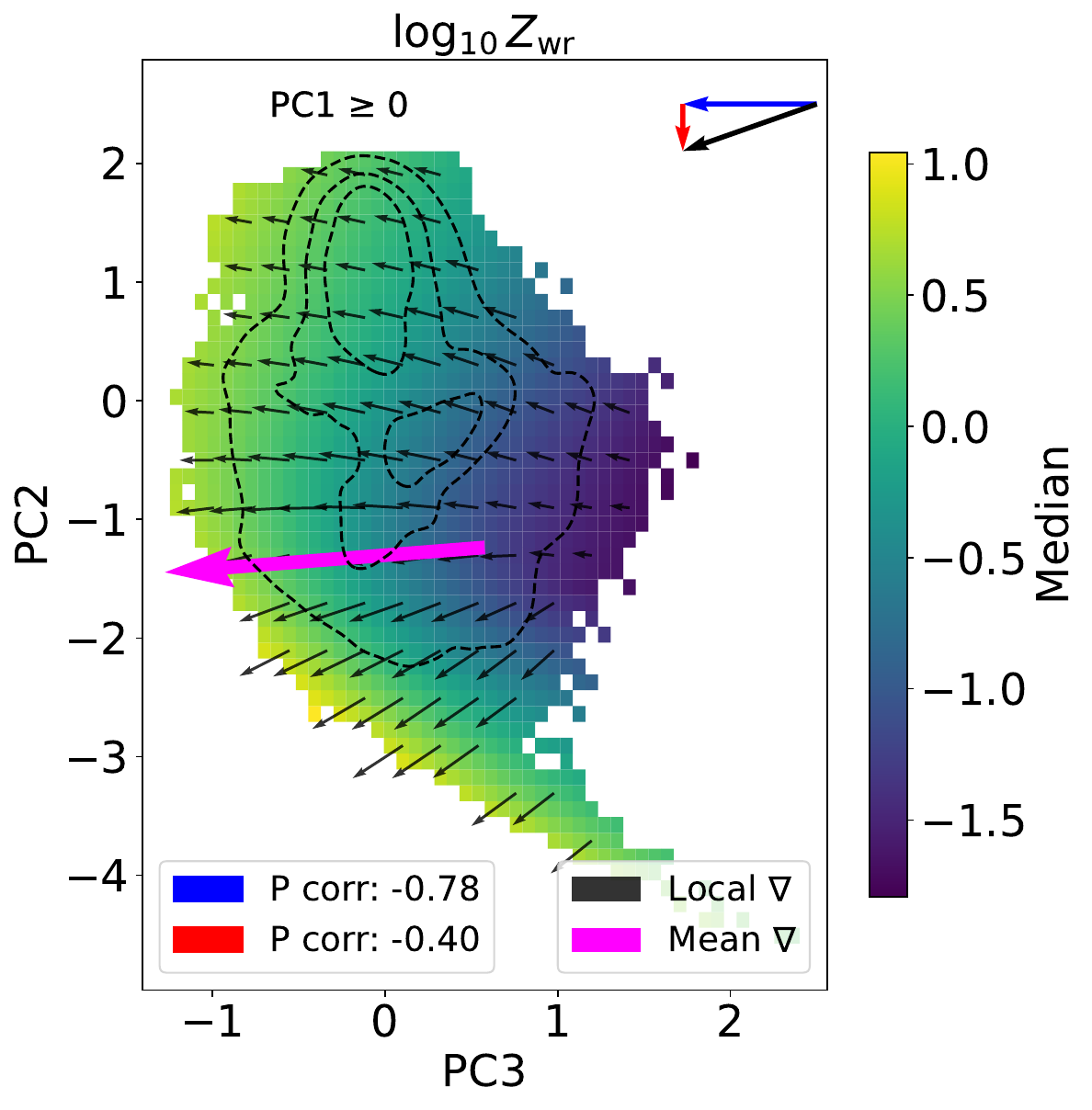}
  \caption{ Maps between the first three principal components for models 
    with positive PC1, which are color-coded based on the median values of the corresponding physical properties, as labelled in the title of each panel. In this case, we show age and metallicity, weighted with respect to SDSS$r$. Each panel includes a pink vector representing the average colour gradient (i.e. the gradient of the physical parameters in 2D latent space). In addition, local gradients within each bin are shown as small black arrows, considering neighbouring bins, mapping the local variations. The partial correlation coefficients are shown as vectors in the top-right corner of each panel (with
the numerical values quoted in the legend). The full version, including models with positive and negative PC1, can be found in the appendix.}
\label{fig:pcmapsALL123}

\end{figure*}

\begin{figure*}
\centering
\includegraphics[width=34mm]{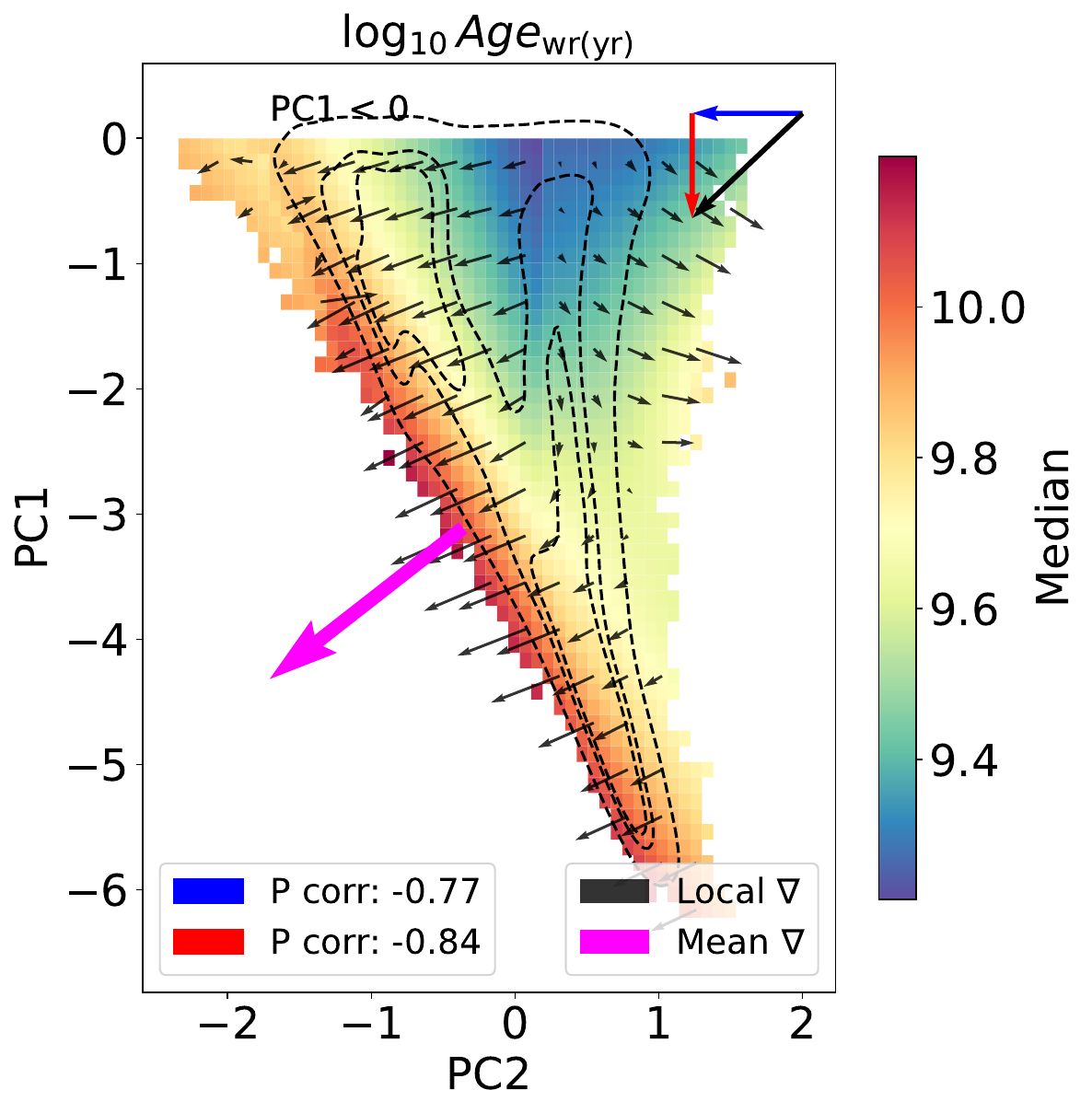}
\includegraphics[width=34mm]{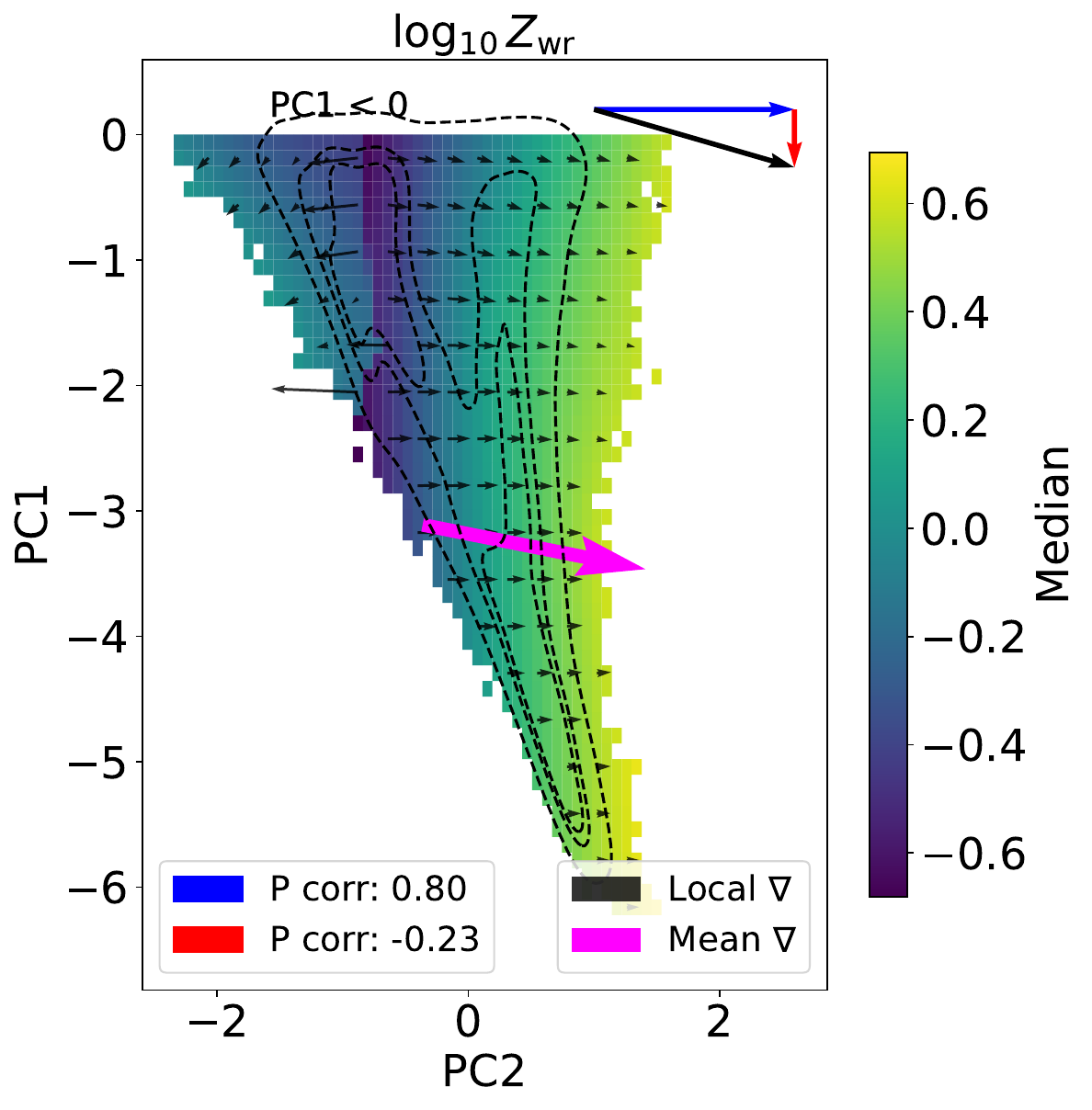}
\includegraphics[width=34mm]{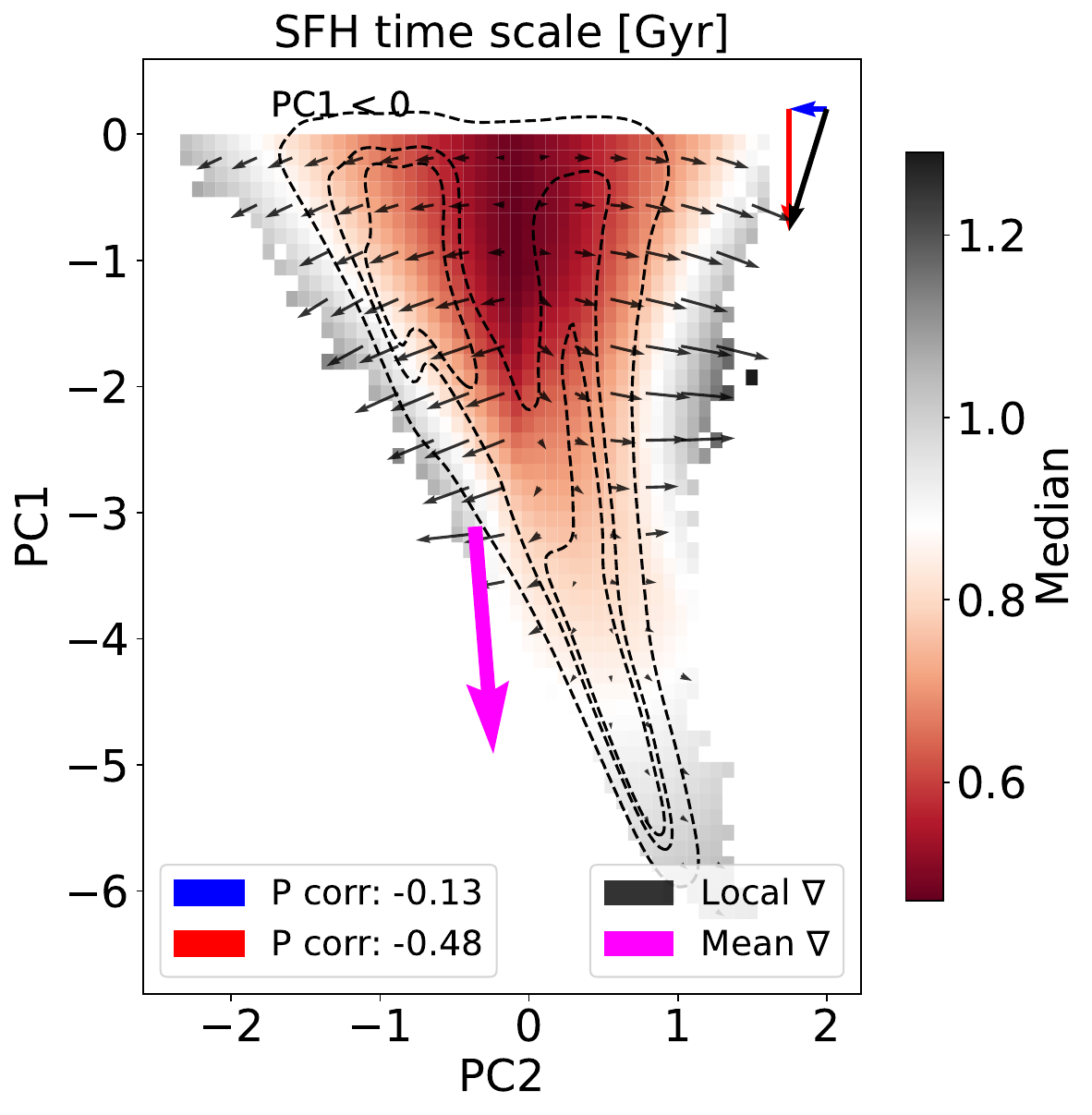}
\includegraphics[width=34mm]{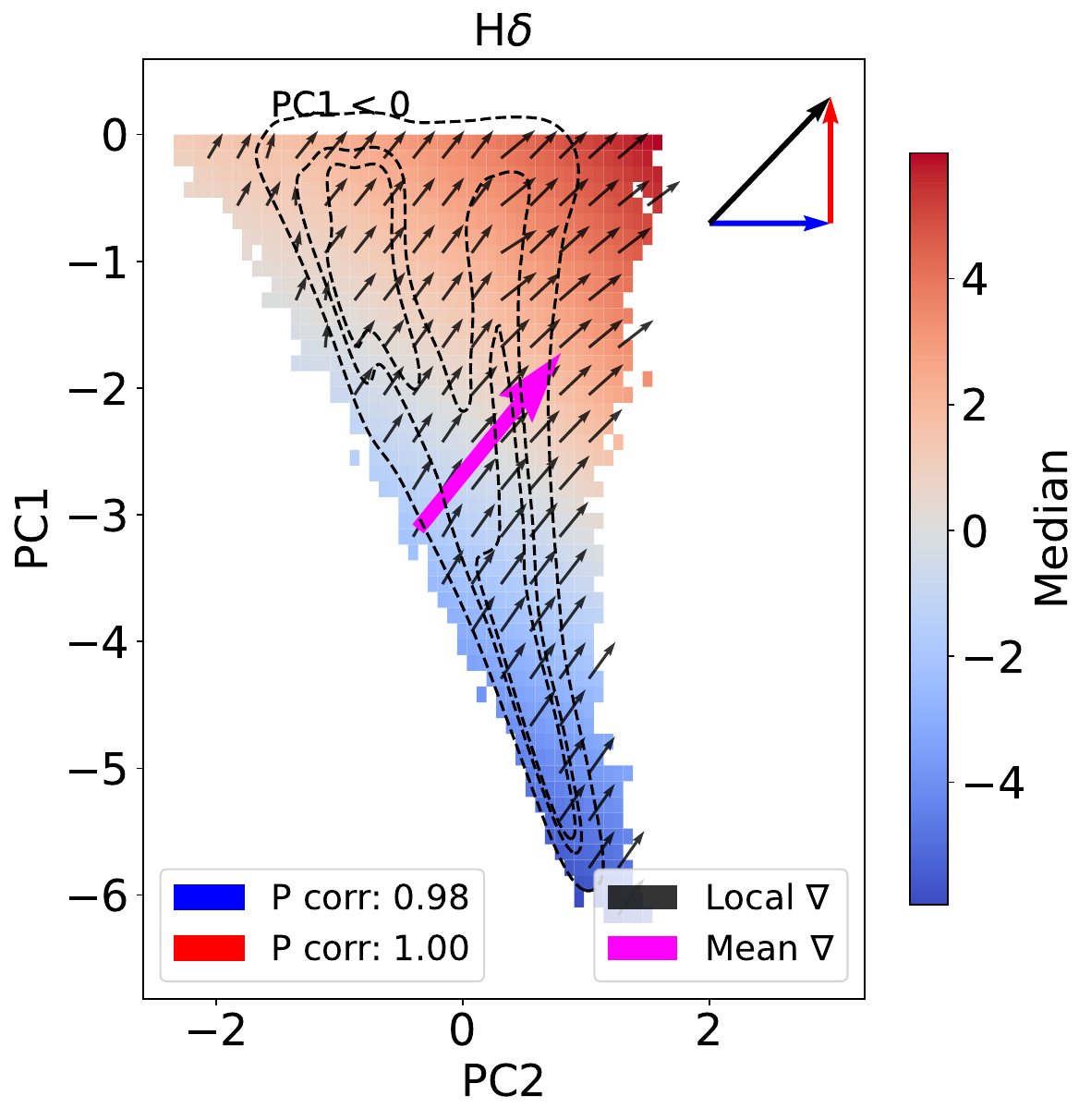}
\includegraphics[width=34mm]{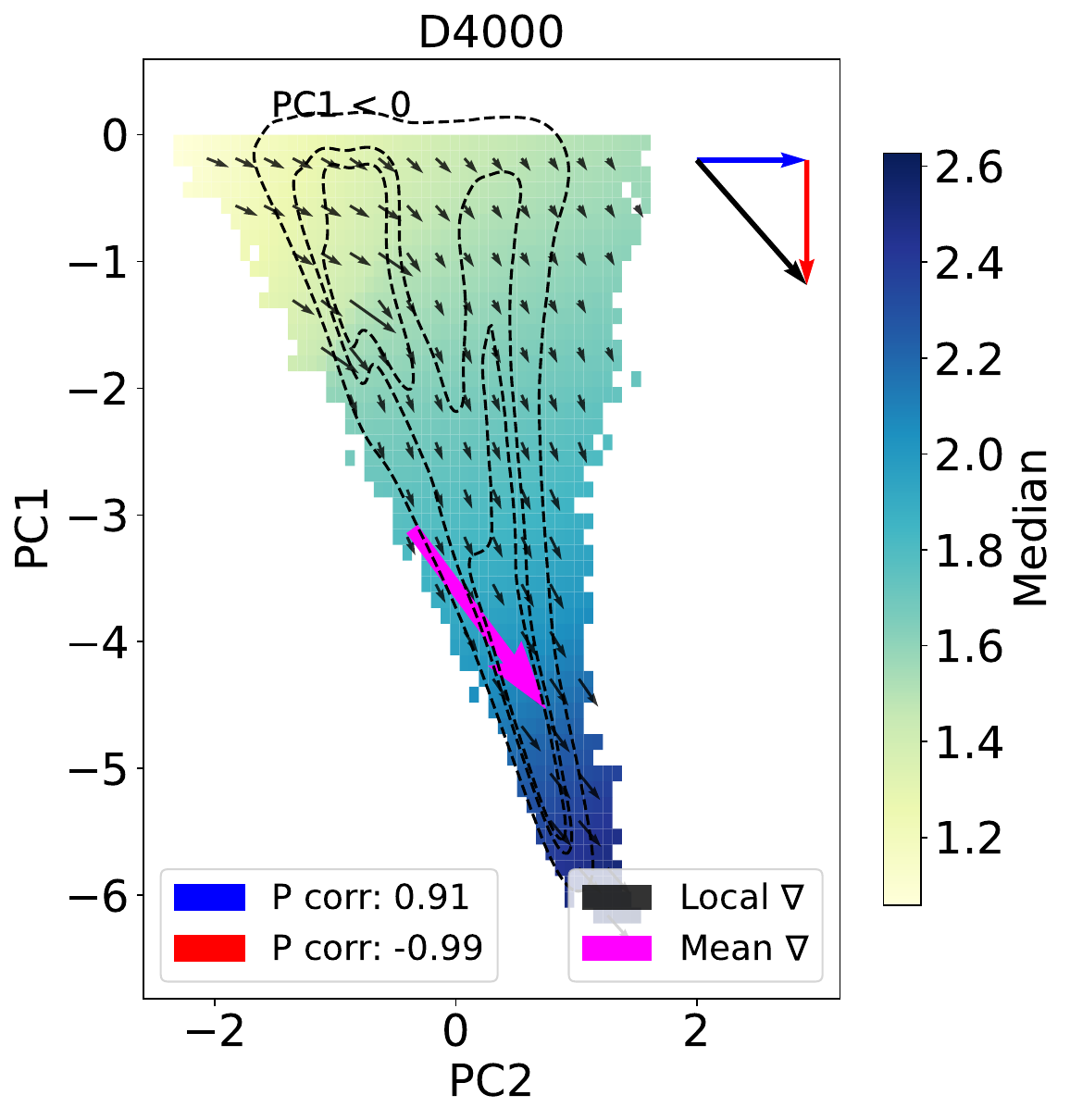}
\includegraphics[width=34mm]{figs/PCMaps/PC12p_all_Age_3.pdf}
\includegraphics[width=34mm]{figs/PCMaps/PC12p_all_Met_3.pdf}
\includegraphics[width=34mm]{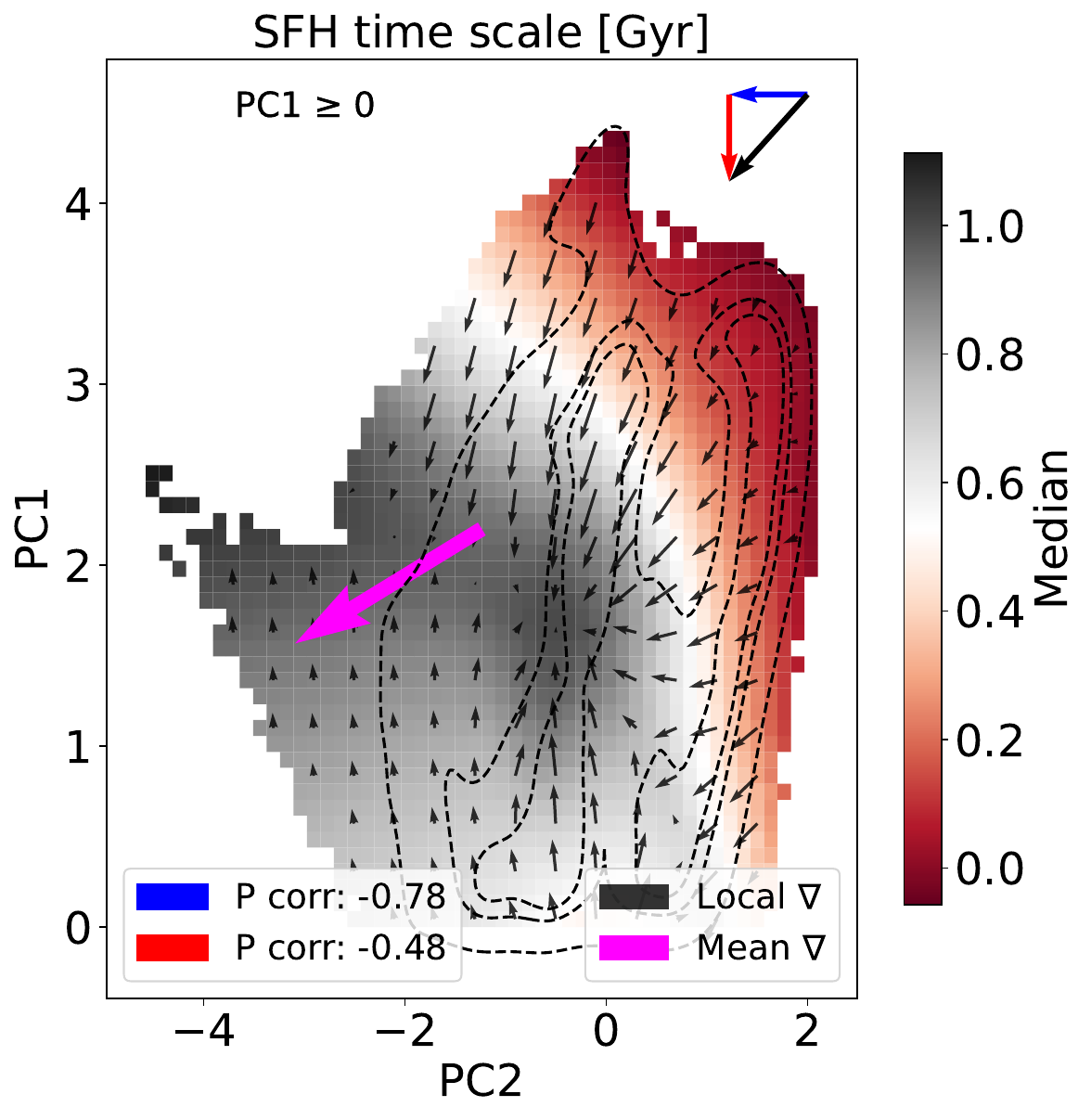}
\includegraphics[width=34mm]{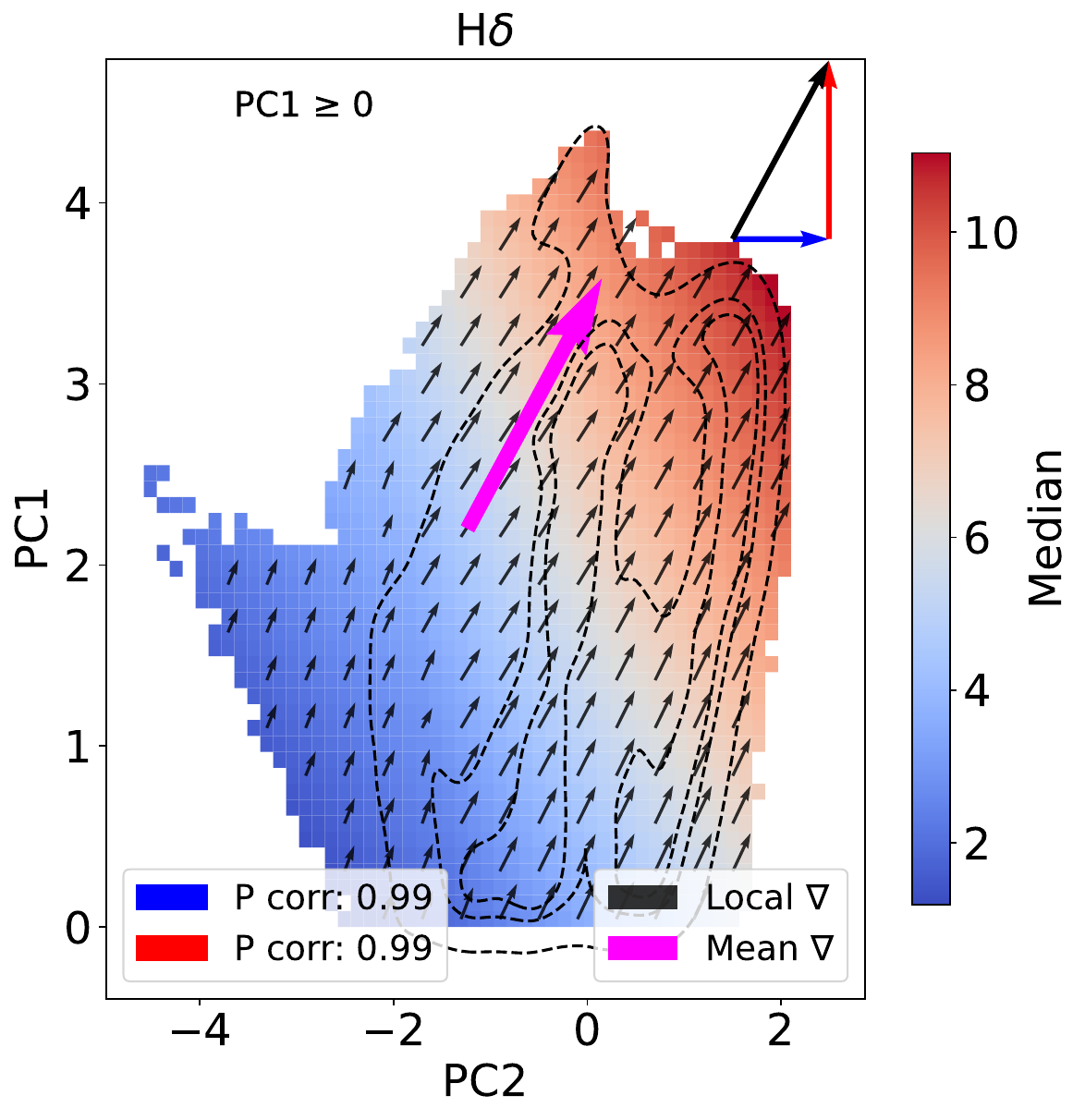}
\includegraphics[width=34mm]{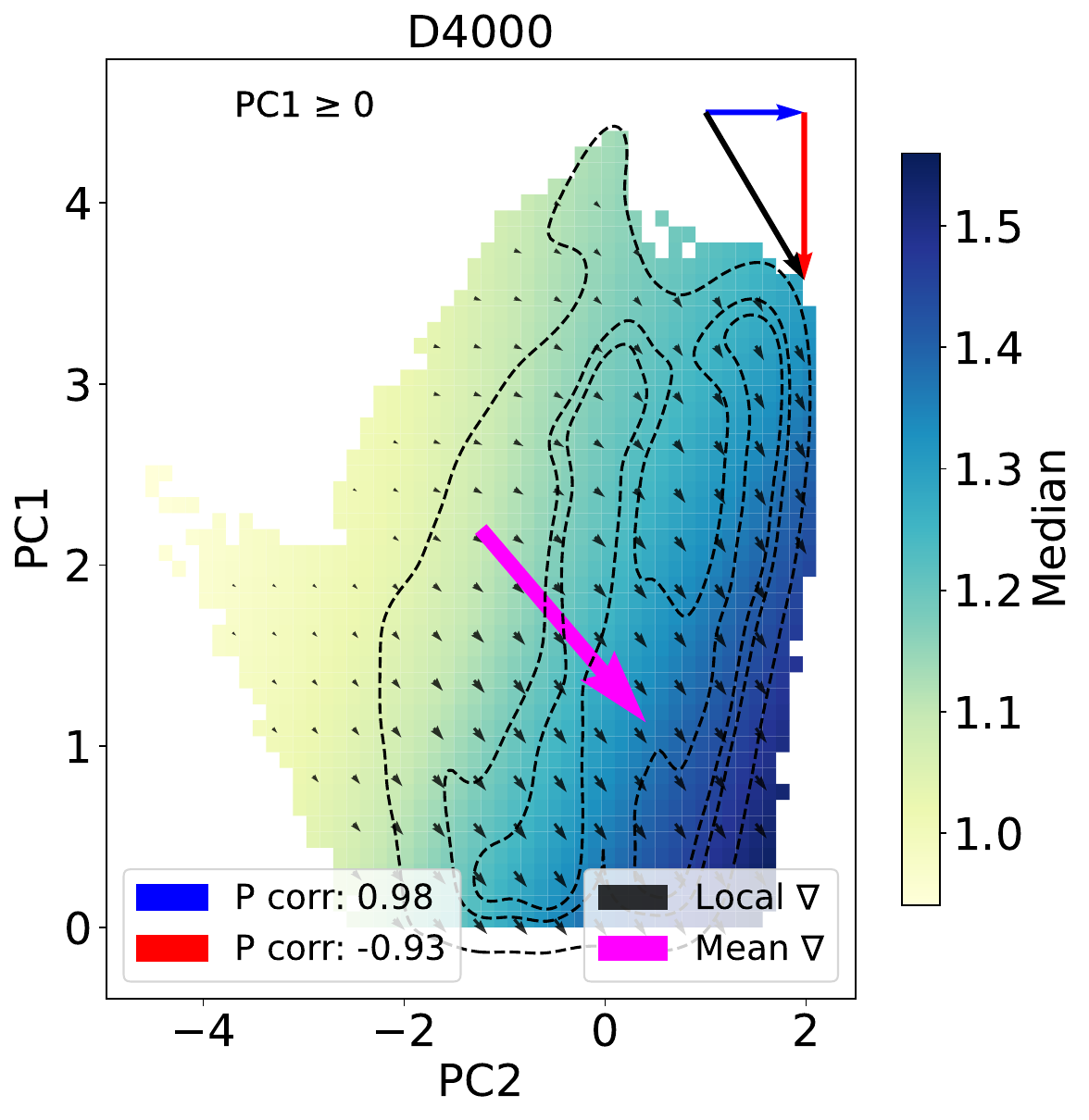}
\caption{This figure complements Fig~\ref{fig:pcmapsALL123}. Maps between PC1 and  PC2 for the regime with negative and positive PC1 values. Each panel represents the median value of the corresponding physical properties for the bin of the PC1-PC2 map.}
\label{figapp:pcmapsALL12}
\end{figure*}

\begin{figure*}
\centering
\includegraphics[width=34mm]{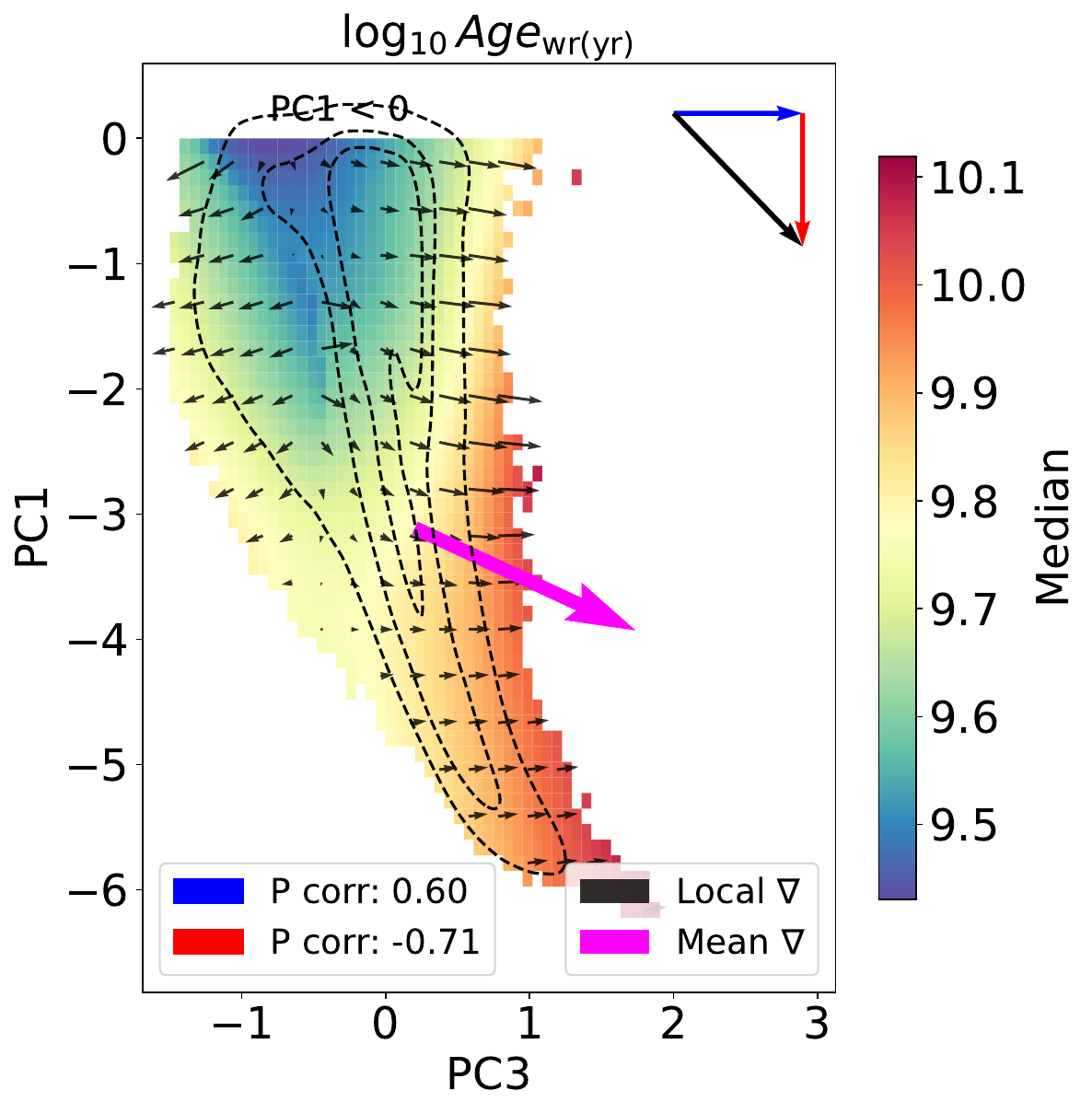}
\includegraphics[width=34mm]{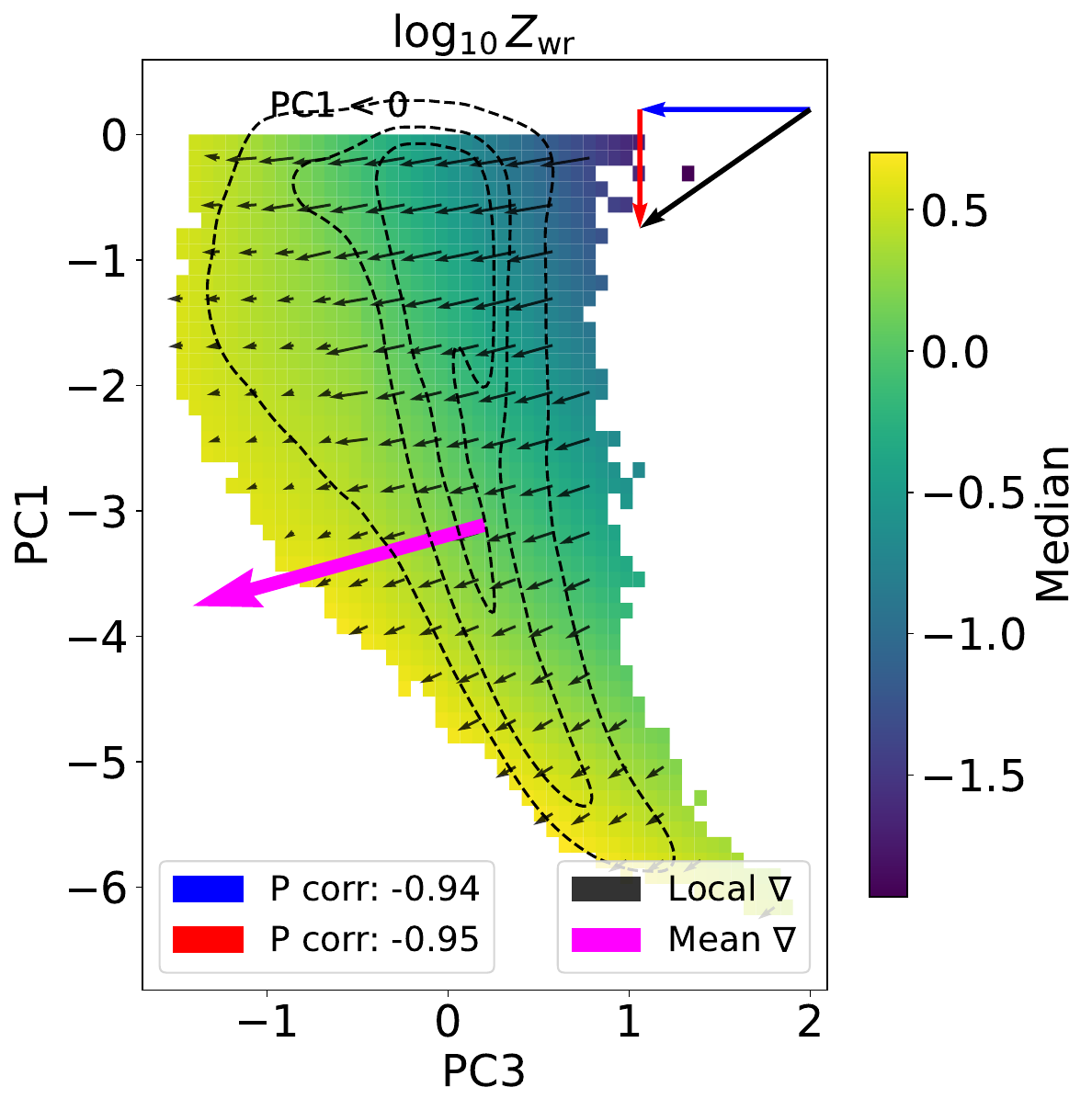}
\includegraphics[width=34mm]{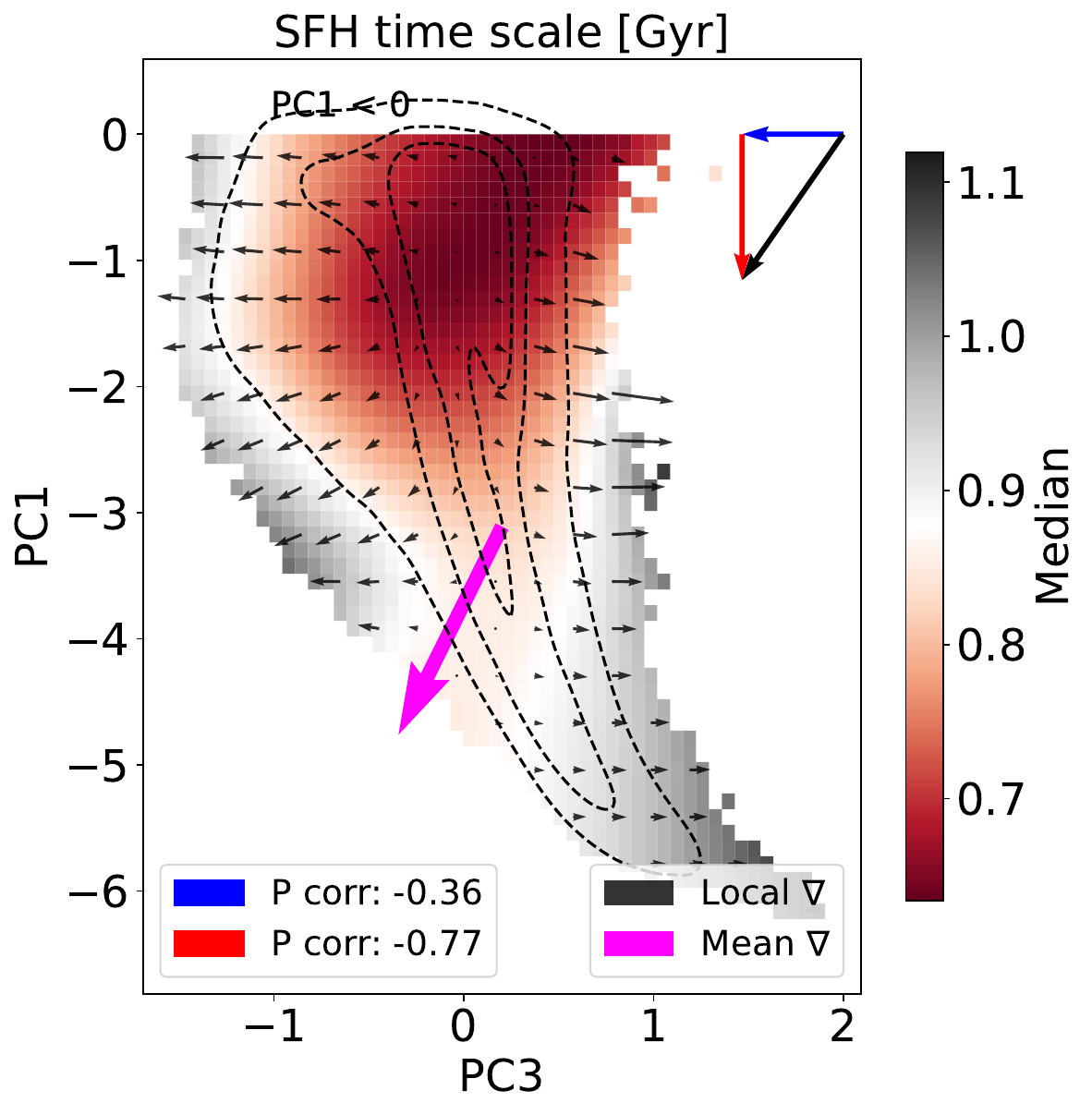}
\includegraphics[width=34mm]{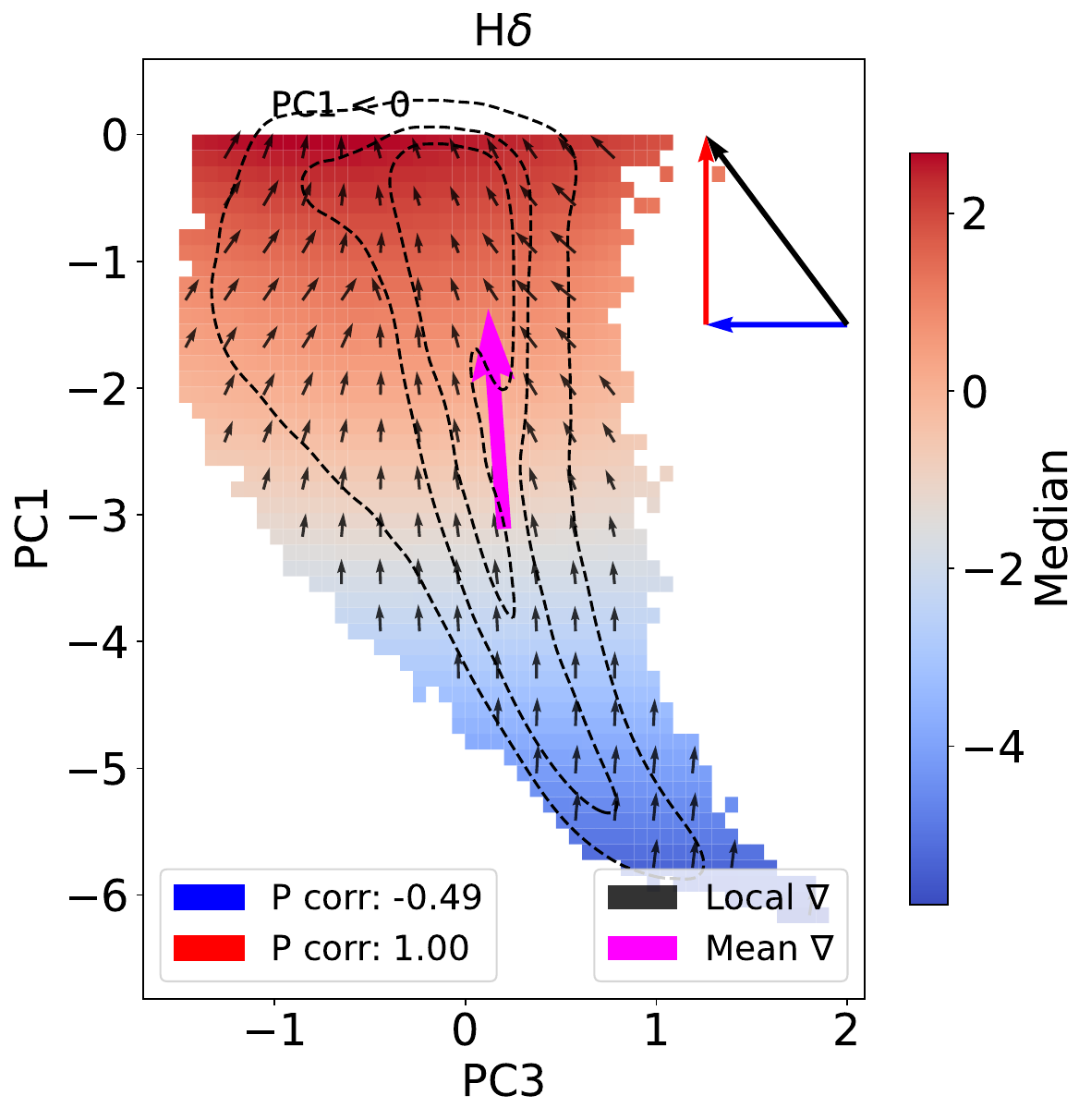}
\includegraphics[width=34mm]{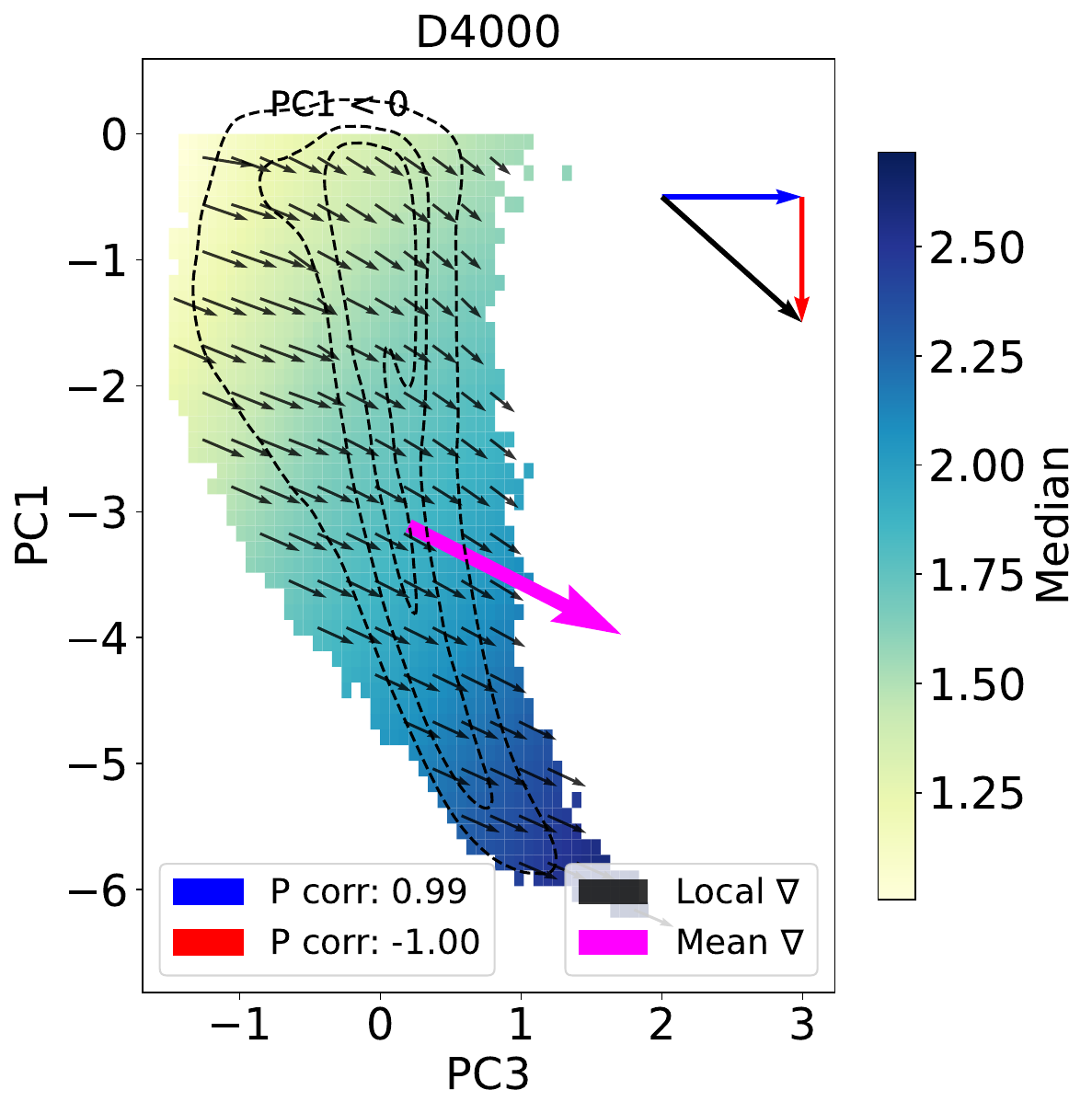}
\includegraphics[width=34mm]{figs/PCMaps/PC13p_all_Age_3.pdf}
\includegraphics[width=34mm]{figs/PCMaps/PC13p_all_Met_3.pdf}
\includegraphics[width=34mm]{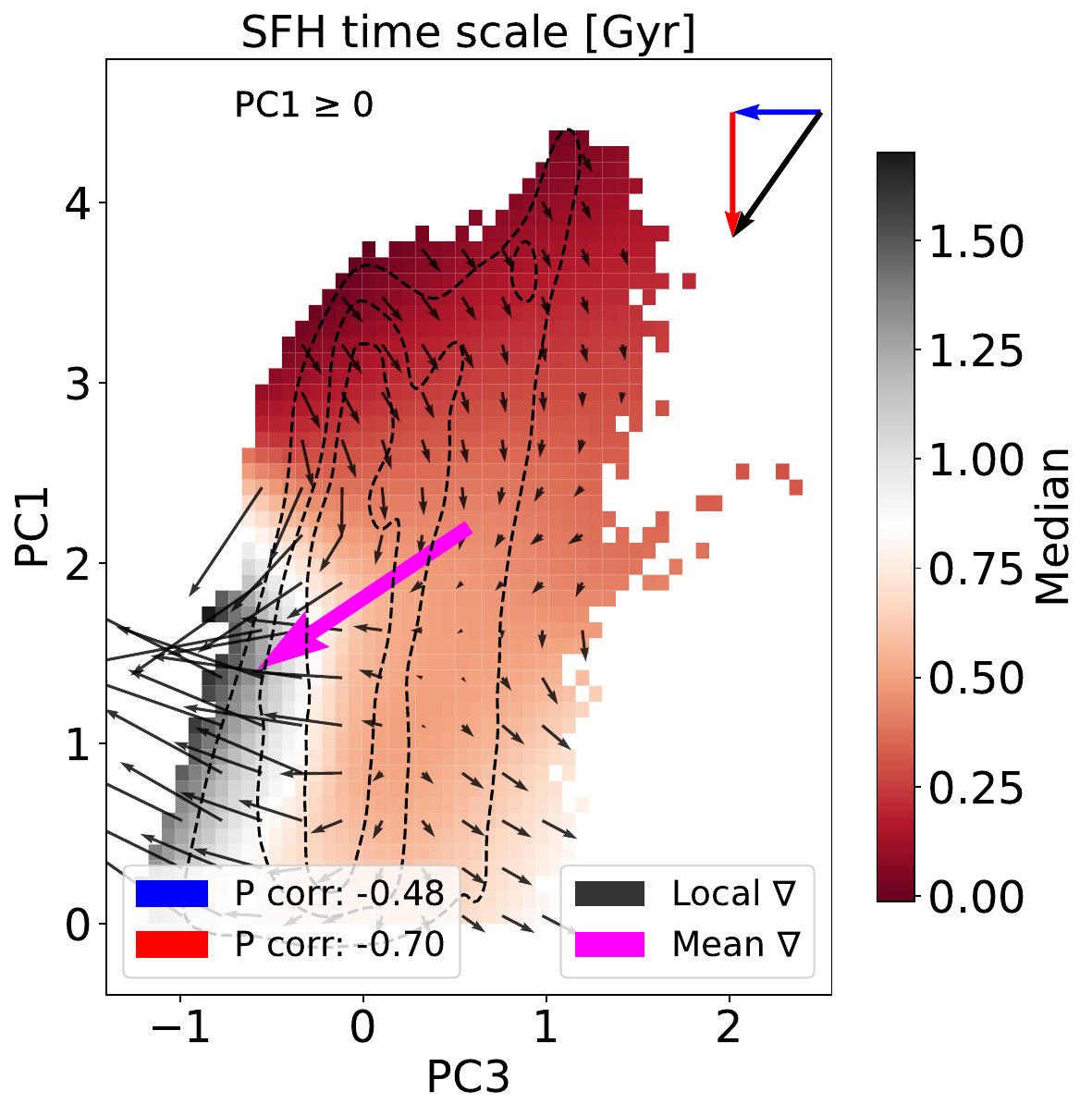}
\includegraphics[width=34mm]{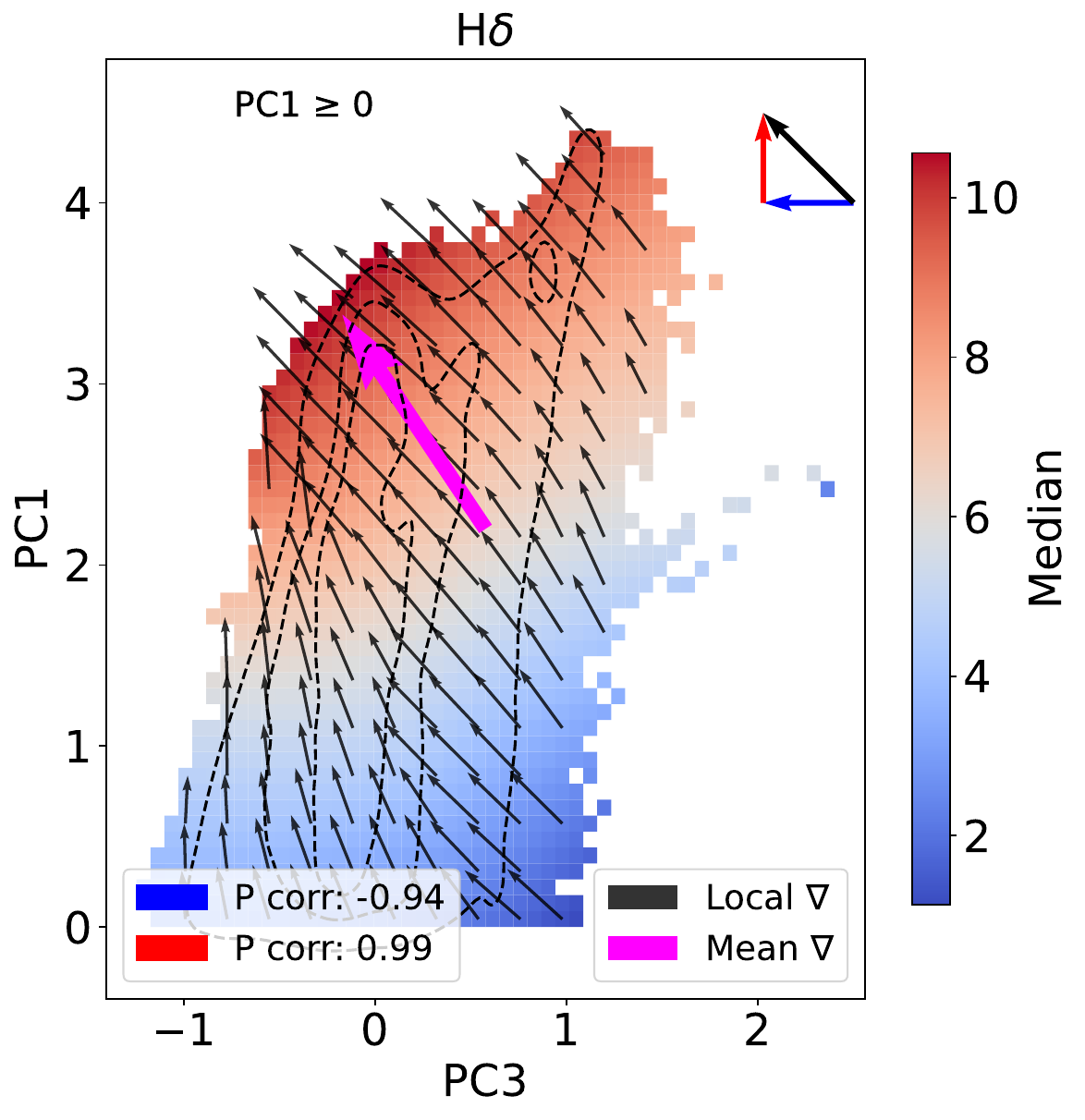}
\includegraphics[width=34mm]{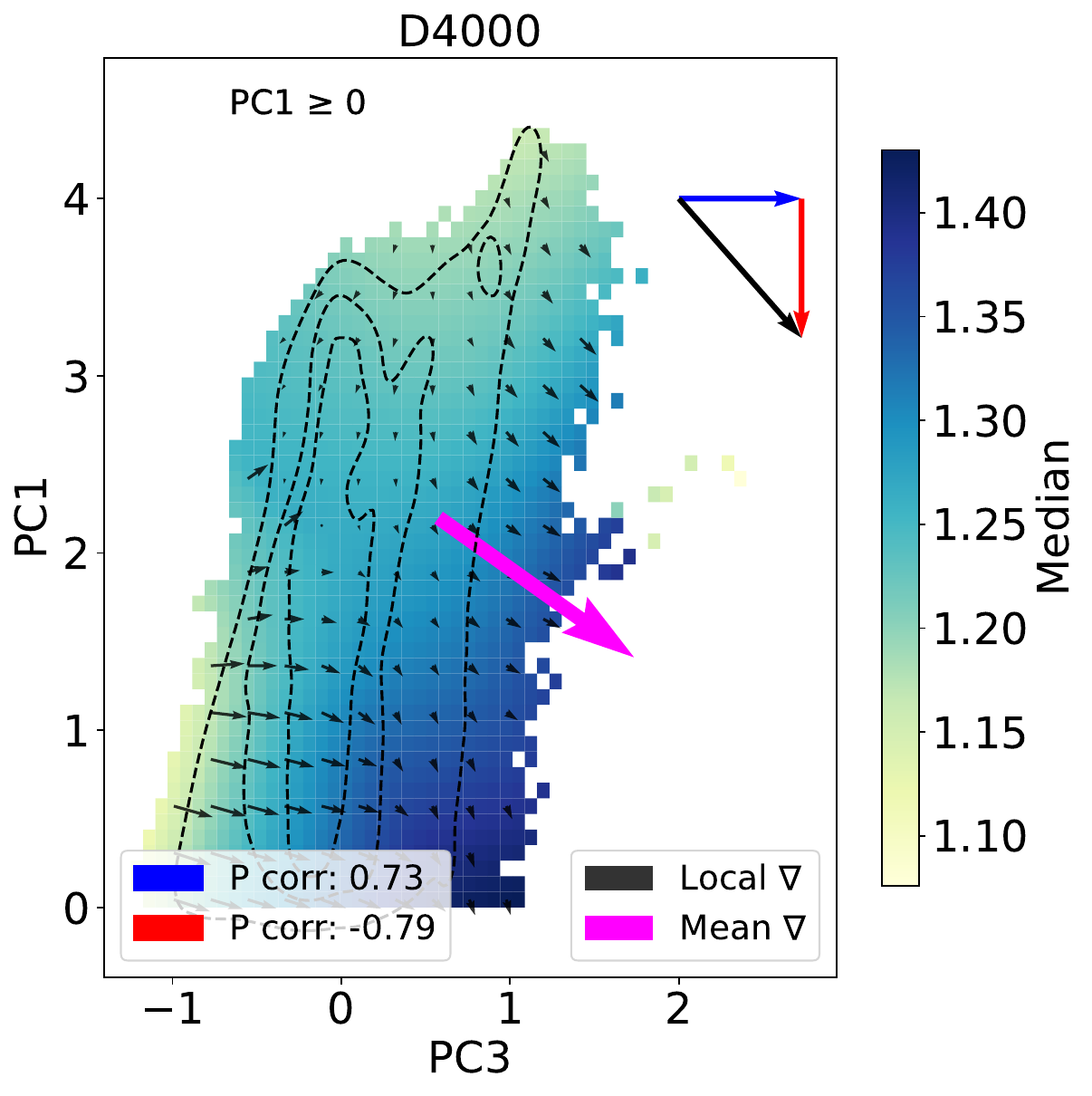}
\caption{This figure complements Fig.~\ref{fig:pcmapsALL123}. Maps between PC1 and PC3 for the regime with negative and positive PC1 values. Each panel represent the median value of the corresponding physical properties for the bin of the PC1-PC3 map.}
\label{figapp:pcmapsALL13}
\end{figure*}

\begin{figure*}
\centering
\includegraphics[width=34mm]{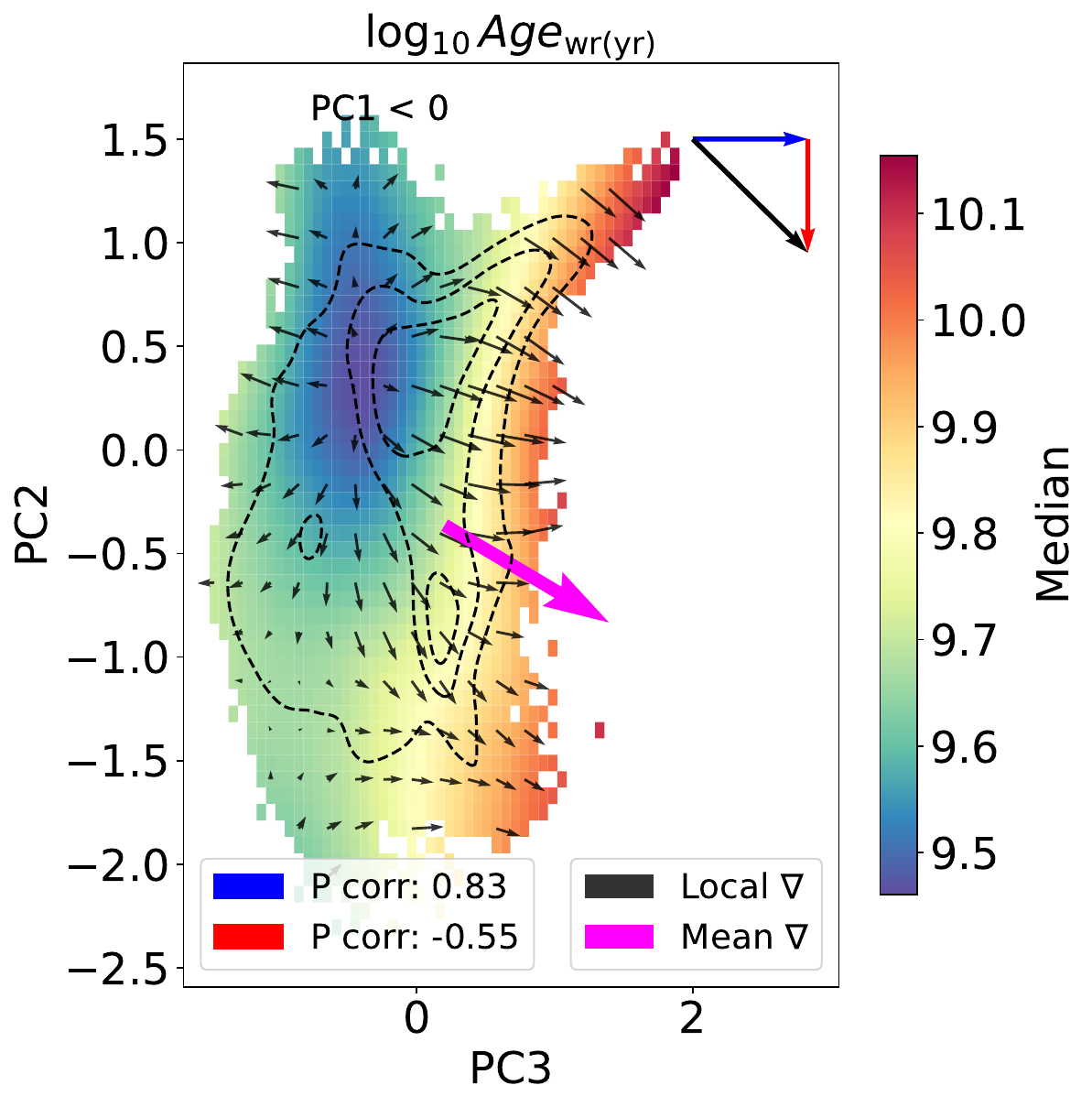}
\includegraphics[width=34mm]{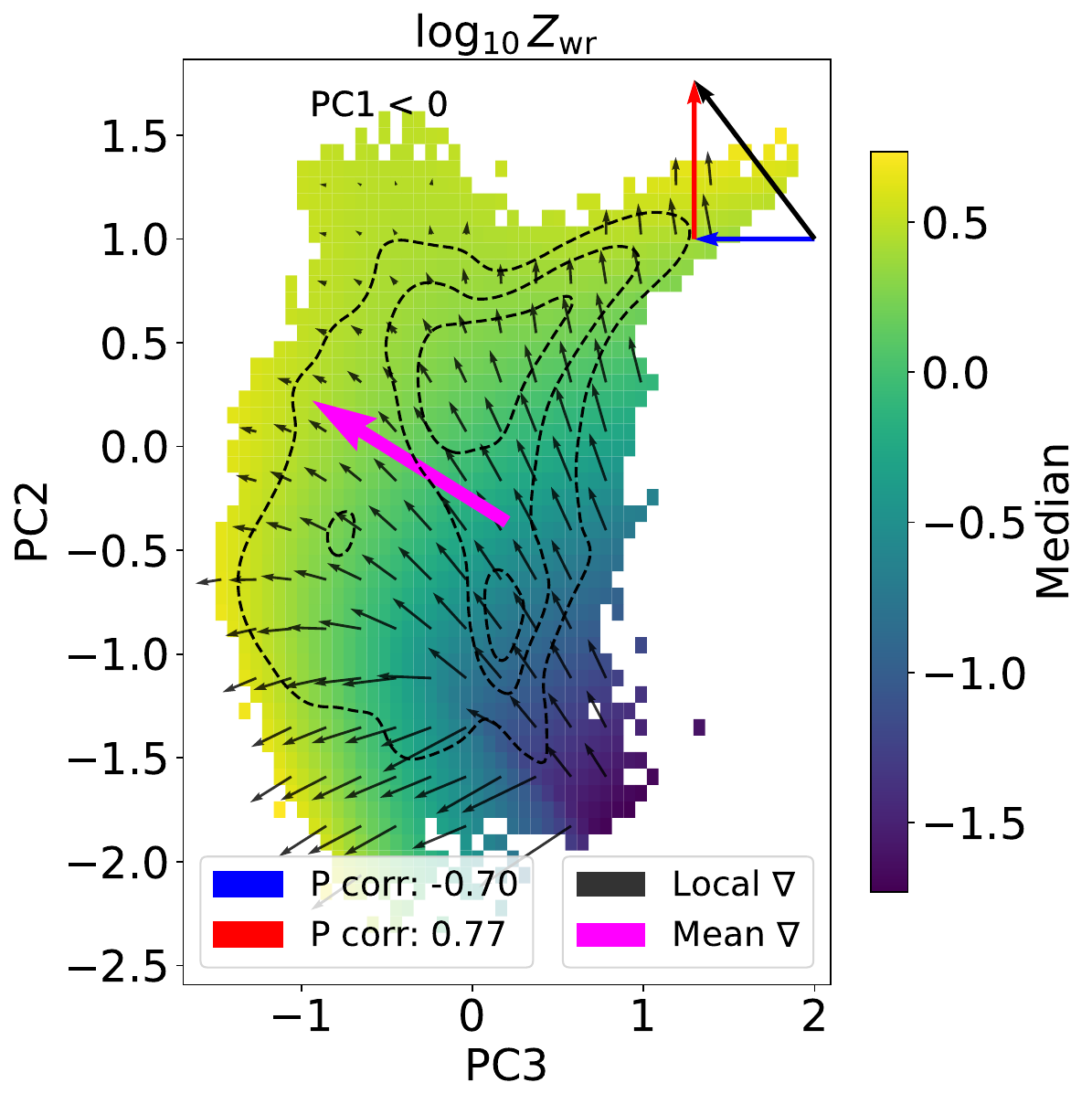}
\includegraphics[width=34mm]{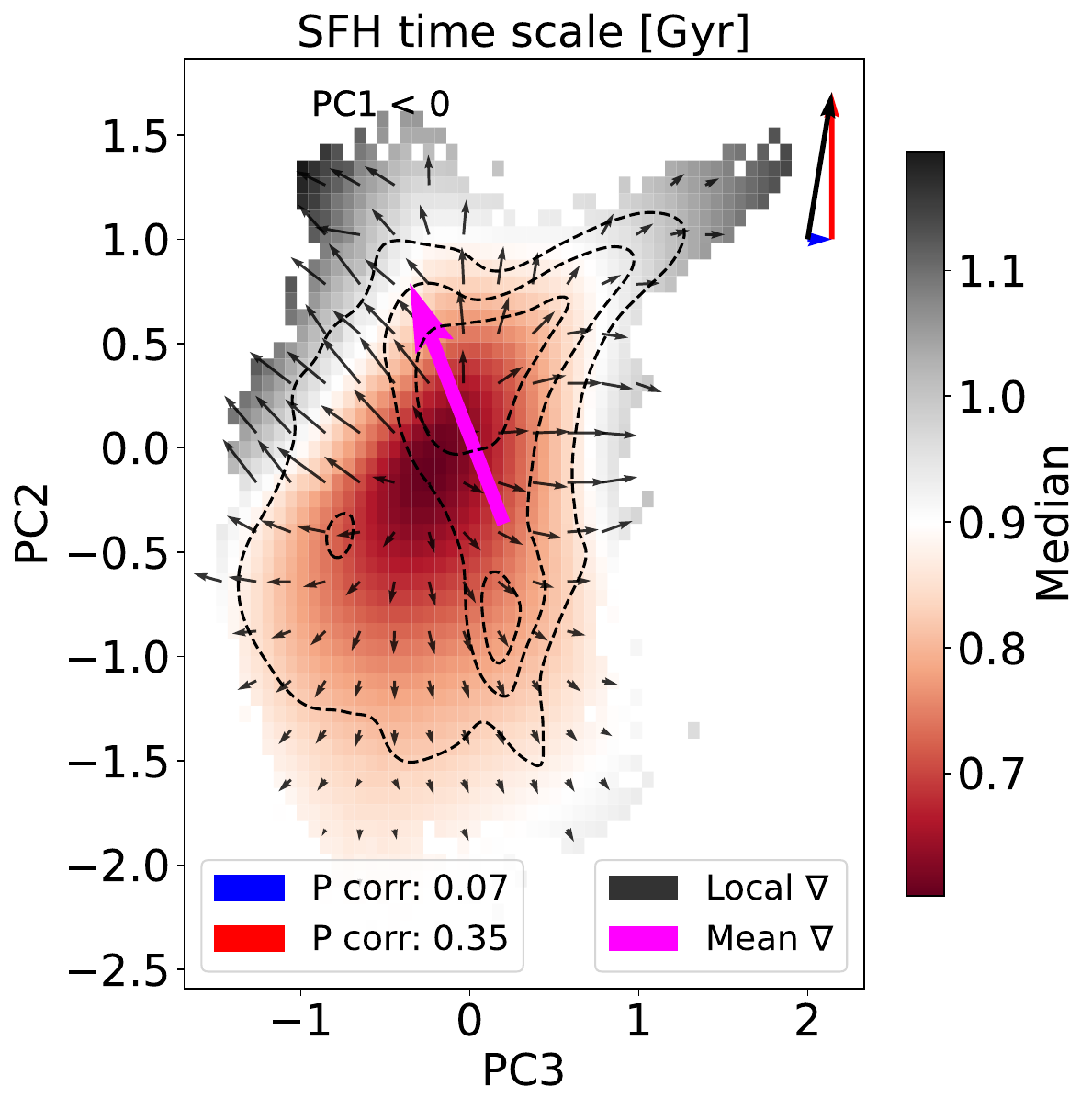}
\includegraphics[width=34mm]{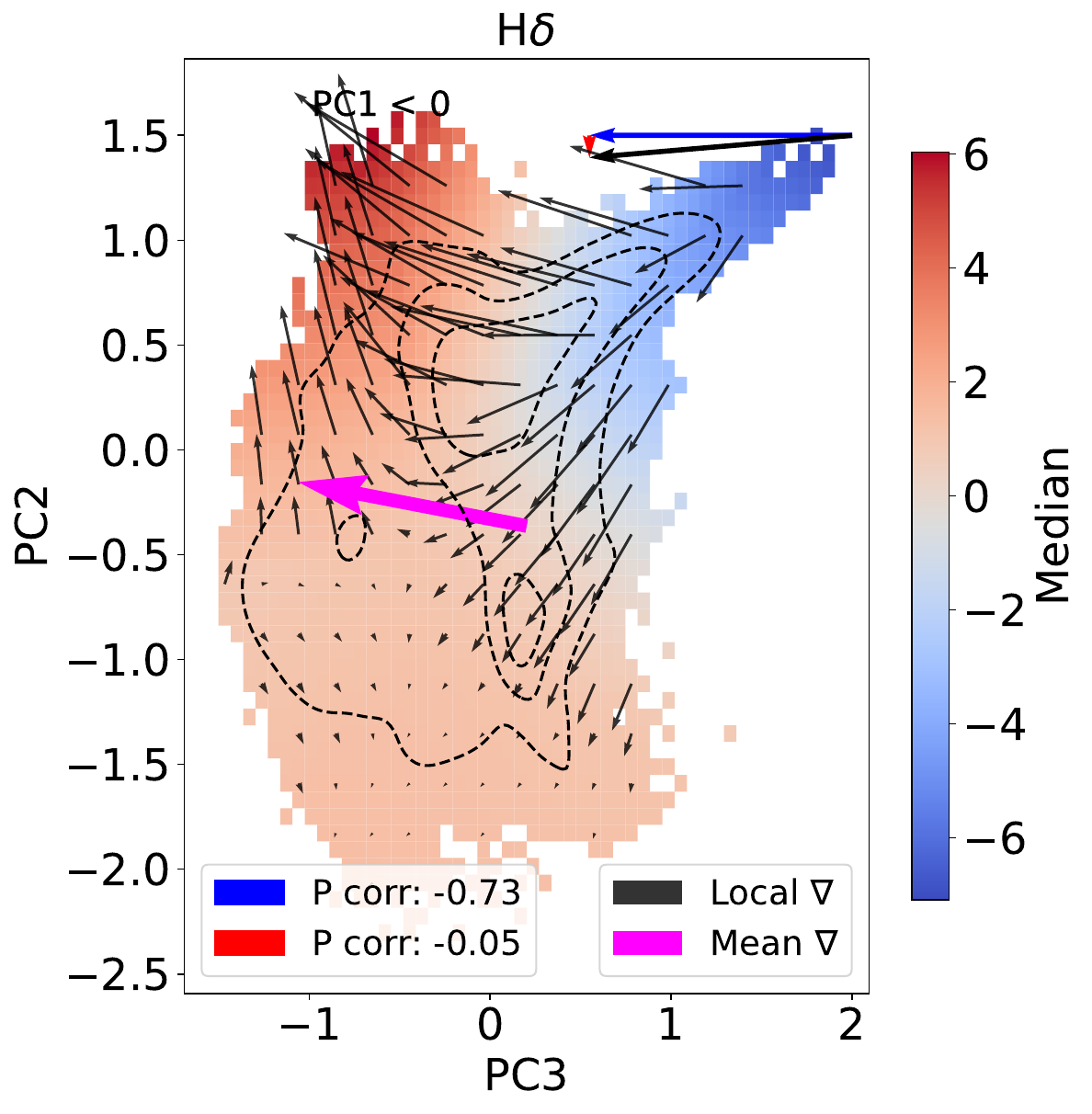}
\includegraphics[width=34mm]{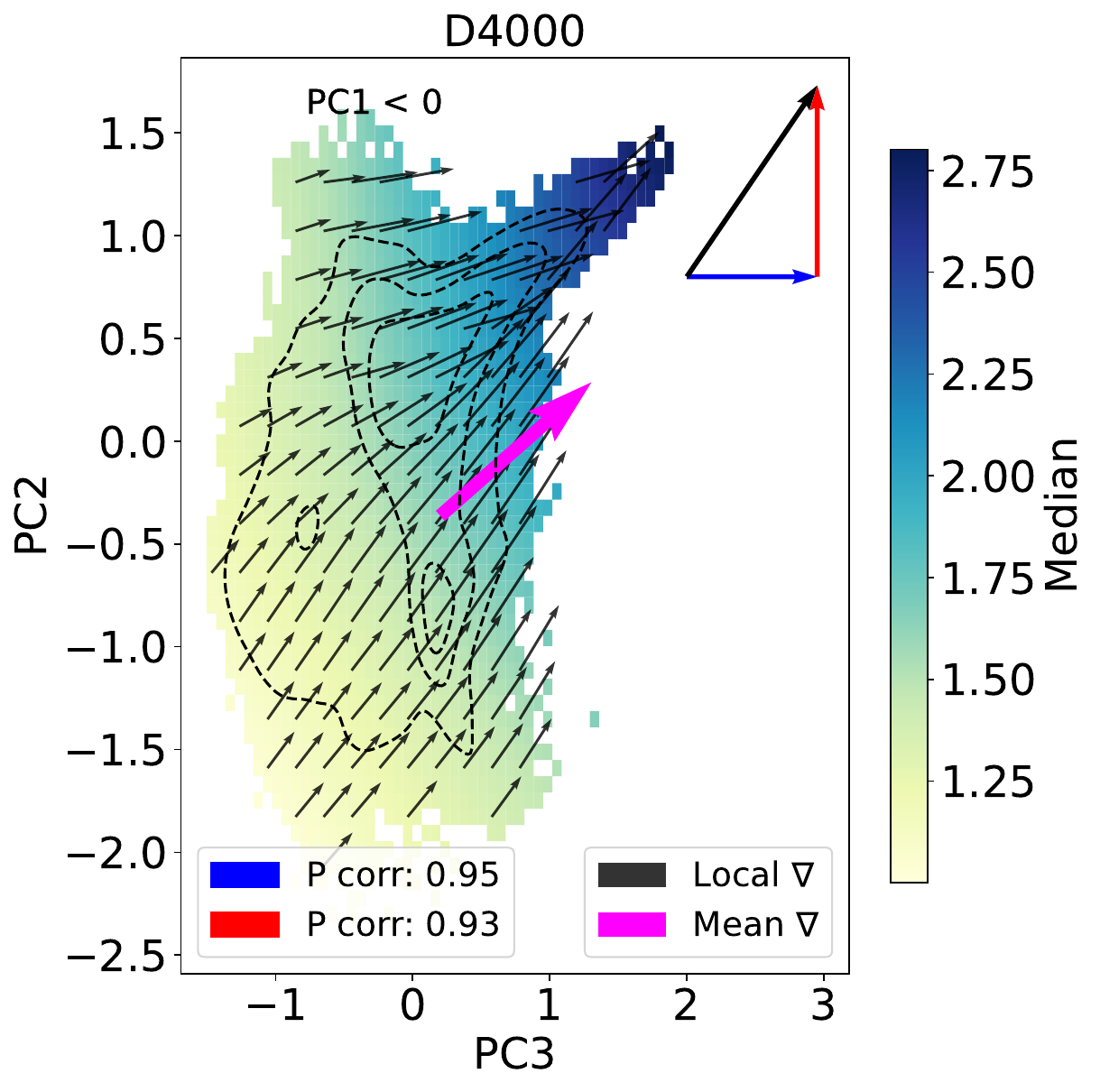}
\includegraphics[width=34mm]{figs/PCMaps/PC23p_all_Age_3.pdf}
\includegraphics[width=34mm]{figs/PCMaps/PC23p_all_Met_3.pdf}
\includegraphics[width=34mm]{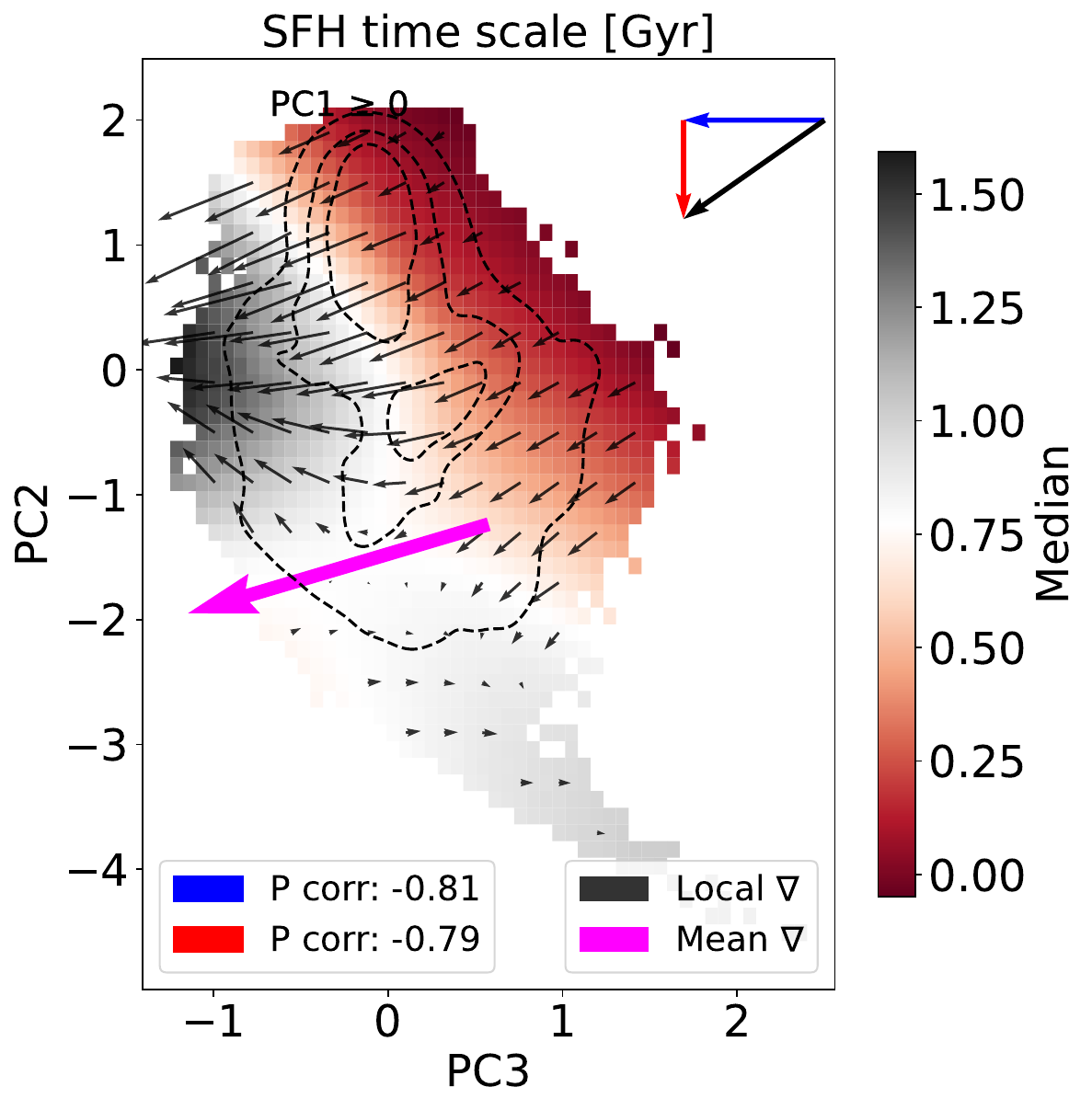}
\includegraphics[width=34mm]{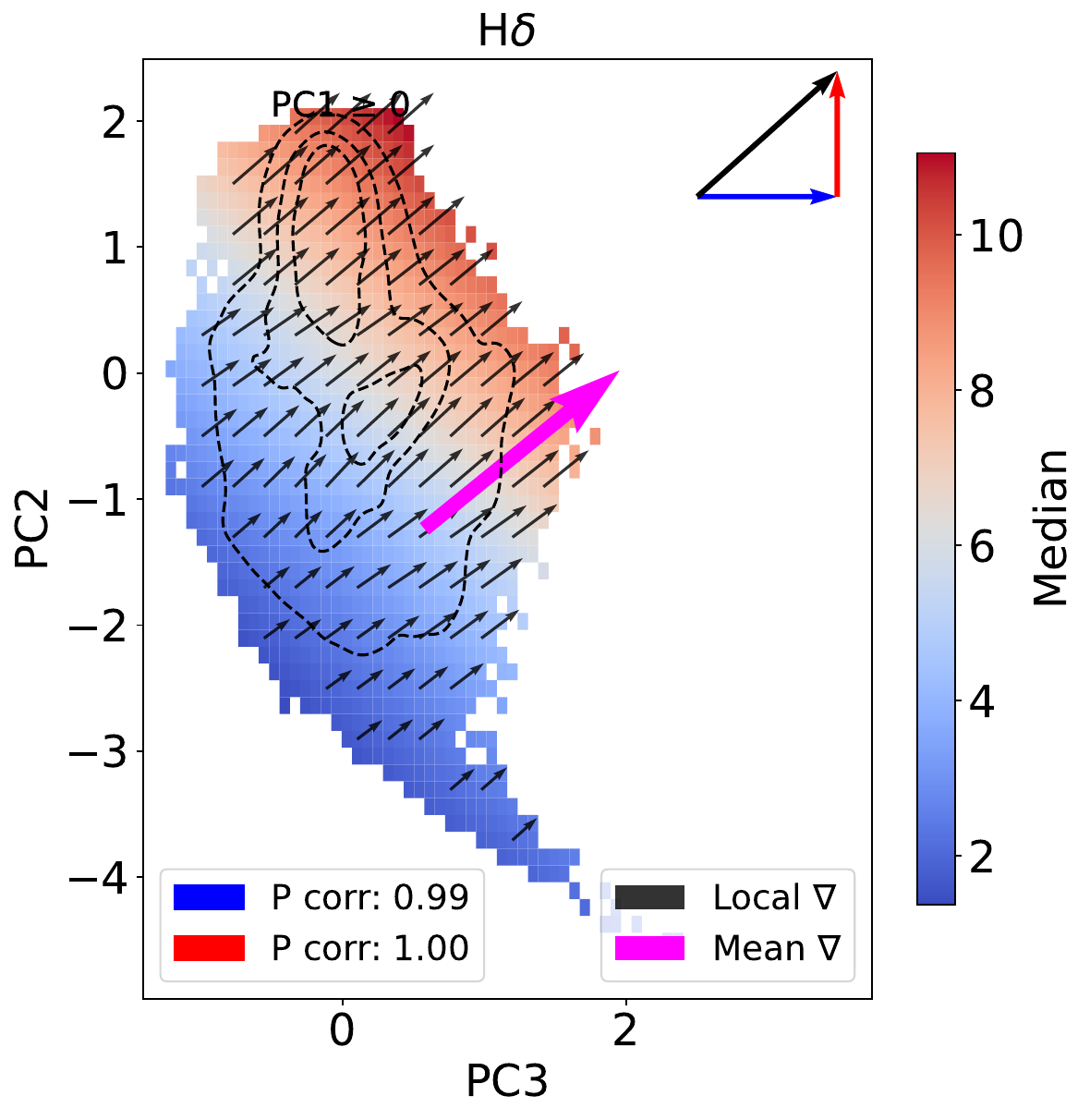}
\includegraphics[width=34mm]{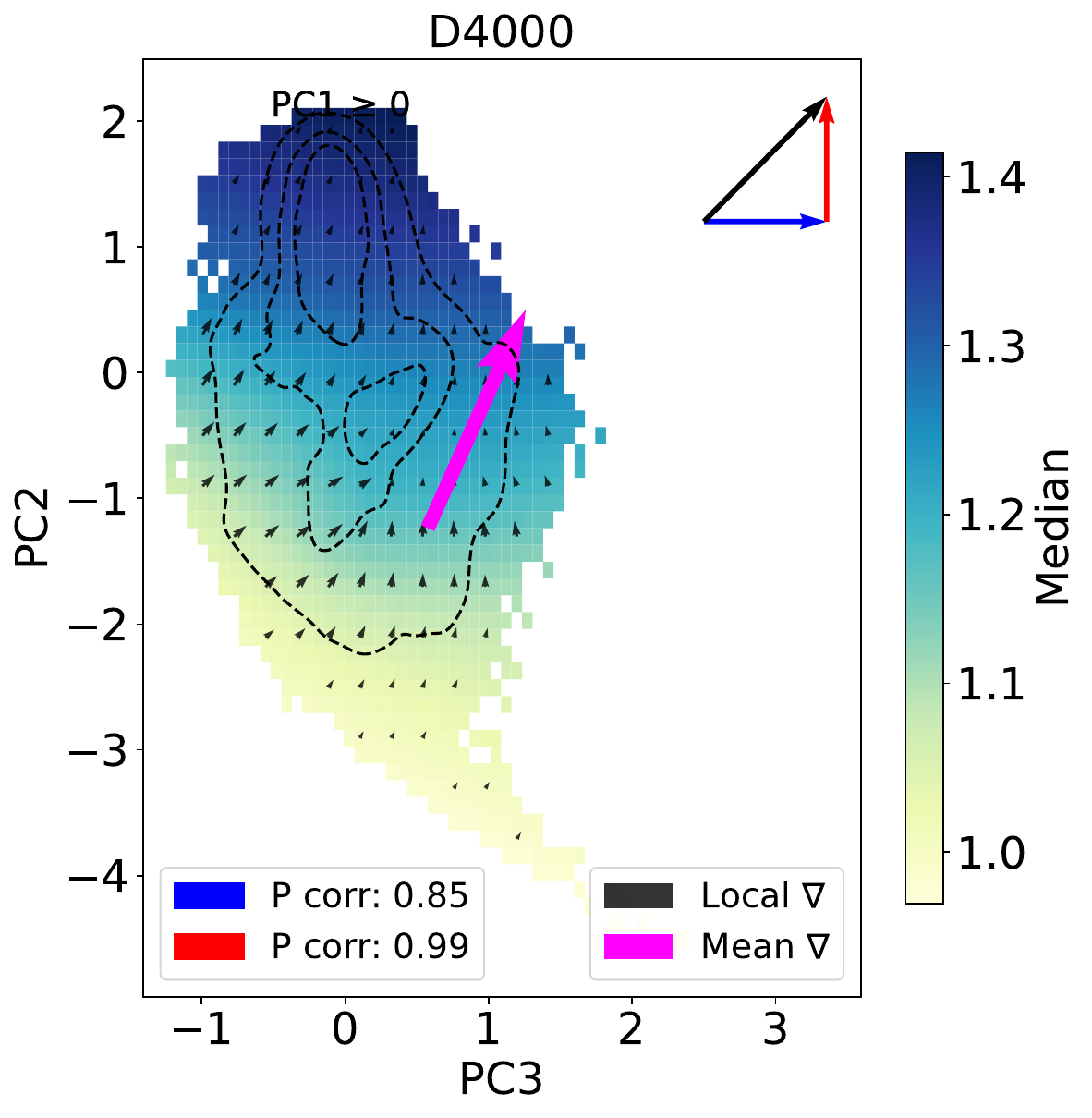}
\caption{This figure complements Fig.~\ref{fig:pcmapsALL123}. Maps between PC2 and PC3 for the regime with negative and positive PC1 values. Each panel represents the median value of the corresponding physical properties for the bin of the PC2-PC3 map.}
\label{figapp:pcmapsALL23}    
\end{figure*}

To explore these trends in more detail, Fig.~\ref{fig:pcmapsALL123} shows 2D maps in slices of latent space. We limit the analysis to projections on the first three principal components, since higher-order components encode progressively less variance. Each panel is colour-coded according to the median value of a physical property: stellar age in the top panels, and stellar metallicity in the bottom panels. The colour maps are smoothed with the locally weighted regression (LOESS) method, with code implemented by \citet{LOESS}\footnote{Available from http://purl.org/cappellari/software}. Each panel includes a pink vector representing the average colour gradient (i.e. the gradient of the physical parameters in 2D latent space). In addition, local gradients within each bin are shown as small black arrows, considering neighbouring bins, mapping the local variations. The consistency between the global and local gradients helps verify the reliability of the trends. Moreover, we also present the ($x,y$) partial correlation coefficient as vectors in the top-right corner of each panel (with the numerical values quoted in the legend). An extended version of Fig.~\ref{fig:pcmapsALL123} is presented in Figs.~\ref{figapp:pcmapsALL12}, \ref{figapp:pcmapsALL13}, \ref{figapp:pcmapsALL23} show the equivalent 2D maps for a number of model parameters, namely (from left to right) the logarithm of light-weighted age and metallicity in the SDSS $r$ band, the SFH time scale -- defined as ($T_{75} - T_{25}$), i.e. the time interval that spans the galaxy containing 25\% and 75\% of its total stellar mass \citep{Zibetti2024} -- and two key spectral indices: H$\delta_A$ and D$_n$(4000). These figures include the two subsets split with respect to the sign of PC1, as labelled.

The dense information of these plots can be fully explored in the appendix figures. In the main body of the paper, we focus on the interesting case of age (top panels) and metallicity (bottom panels) for the subsample with PC1$\geq$0 (Fig.~\ref{fig:pcmapsALL123}), that corresponds to younger populations. The figure shows the expected high entanglement of the variation (note the colour gradient in the panels) of the indices with respect to age and metallicity. However, we note that the PC1 vs PC2 graph (left panel) features a strong ``vertical'' trend for age (i.e. a dependence mostly through PC1), and the PC1 vs PC3 map (middle panel) show a ``horizontal'' variation for metallicity (in this case PC3 tracing both metallicity and age, slightly stronger for metallicity). Noting that we are dealing with the highest variance components -- i.e. more resilient with respect to noise: this shows a potentially useful methodology to obtain age and metallicity in observational data.

In contrast, the equivalent maps for the negative PC1 regime (Figs.~\ref{figapp:pcmapsALL12} and \ref{figapp:pcmapsALL13}) show a stronger age-metallicity degeneracy, as expected in older populations. However, there is a sign of degeneracy breaking even in this case. The average gradient vectors for both age (PC1 vs PC2) and metallicity (PC1 vs PC3) are more mixed. This is indicative of the fact that older populations have more prominent absorption indices that can be equally affected by later stellar types or higher metallicity. Also note that the SFH timescale (Figs.~\ref{figapp:pcmapsALL12}, \ref{figapp:pcmapsALL13}, \ref{figapp:pcmapsALL23}) timescale in these two sections of latent space follows closely the average age for PC1$<$0 models, but a substantial difference is found in the younger PC1$\geq$0 models, which also suggests a better constraining power of this timescale in younger populations. Regarding the two key indices presented in these figures, we note a stronger average trend of D$_n$(4000) with PC1. However,  H$\delta_A$) is mixed in all cases, reflecting the advantage of PCA as a method that combines all six indices to exploit the variations of specific population parameters. In other words, a naive distribution in index space would produce a more entangled multivariate distribution with respect to the PCA-based latent space of the same indices.  

Finally, Fig.~\ref{figapp:pcmapsALL23} shows the maps of PC2 versus PC3. We observe a sizeable age-metallicity degeneracy compared to the other projections. The average gradient vectors for age and metallicity in both regimes show more alignment, indicating that distinguishing between these two properties becomes rather challenging in this PC subspace. However, metallicity is also better correlated with PC3 for the PC1$\geq$0 subset. The indices also appear mixed, similarly to the previous cuts in latent space, with less consistent local gradients for H$\delta_A$ in the negative PC1 regime, reflecting an inhomogeneous distribution of populations in the PC2-PC3 bivariate. As in the previous projections, models in the negative-PC1 regime tend to be associated with shorter SFH timescales. We stress, however, that the model library does not impose any correlation between population age and SFH timescale. The observed trend, therefore, emerges from the mapping between spectral properties and the latent-space coordinates. In particular, negative PC1 is associated with weaker Balmer absorption and spectra dominated by older stellar populations, whereas extended SFHs that continue to recent epochs generally contribute younger stars and stronger Balmer absorption, shifting models toward more positive PC1 values. The relation is therefore statistical rather than deterministic, and significant overlap remains between the different SFH timescales across the PC space.

To verify that the trends observed in these 2D PC maps are not simply a consequence of projection effects, we performed a full 6D partial correlation analysis. For each principal component PC$_i$, we computed the Pearson correlation between the residuals of PC$_i$ and a given physical property ($Z$), after removing the linear dependence on the remaining five PCs. This provides the partial correlation $\rho({\rm PC}_i,Z|\mathbf{X}_{-i})$, isolating the unique contribution of each PC to the variance of the physical property. For the full sample, the resulting partial correlations are:
Age: $[-0.869, -0.547, -0.003, -0.273, 0.385, 0.211]$,
Metallicity: $[-0.631, 0.729, -0.770, 0.259, -0.352, -0.624]$,
and SFH timescale: $[0.048, 0.107, 0.115, -0.048, -0.069, 0.037]$. These results broadly confirm the trends seen in the projected PC1--PC3 maps: PC1 shows the strongest association with stellar age, while metallicity is primarily led by PC3, although its information is distributed across multiple orthogonal components. The 6D partial correlation therefore demonstrates that the relations identified in the 2D projections are not simply driven by projection effects. At the same time, the distribution of significant correlations across several PCs highlights the presence of degeneracies in the latent space, which redistribute part of the physical information among multiple components.

\subsection{The effect of bursts on the spectra}
\label{Ssec:BurstoNo}

As presented above, the large set of CSP models include a set where stochastic bursts are added, mimicking the bursty behaviour found in the SFHs of simulated galaxies. Since a burst introduces some fraction in young stars, the spectra can be significantly affected. In general, we may expect a degeneracy in the sense that a recent burst could be interpreted as an overall young (secular) component. In order to assess the ability of PCA to disentangle this degeneracy, we examine the effect of the stochastic component (random bursts) in latent space by applying PCA to the set of models without bursts, comprising 151,892 models that only feature a continuous or secular component of the star formation history. The first row of Table~\ref{tab:samples} shows the mean and standard deviation for each of the six absorption indices obtained for the CSP no-burst models. The same mean and standard deviation from these models is adopted to standardise the models with bursts, comprising  262,622 models that include  1 to 6 bursts across the entire formation time of the galaxy, in addition to the secular component. As reference, the last row of Table~\ref{tab:samples} shows the mean and standard deviation of the six spectral indices for the ensemble of model with bursts, after being standardised by the models without bursts. If the samples were statistically equivalent, we would expect zero mean and unit variance.

Table~\ref{tab:samples} thus represents how different the models with bursts are with respect to the smooth, continuous SFH models. In the bursty models, D$_n$(4000) shows a strong negative mean, suggesting that these models have systematically lower D$_n$(4000), i.e. younger populations. Similarly, the Balmer indices exhibit positive shifts from the non-bursty models, also a signature of more recent star formation. Note that the sensitivity of the different Balmer indices is not identical, a point that will be discussed further below. [MgFe]$^\prime$ and [Mg$_2$Fe] both show strong negative shifts from the non-bursty models. We note that these indices are affected by the age–metallicity degeneracy, and part of the observed shift may be driven by the presence of younger stellar populations in bursty models. In addition, the implementation of bursts introduces stars at specific metallicities, which can lead to an excess of lower-metallicity stars relative to the final enrichment level.

\begin{figure}
  \includegraphics[width=\columnwidth]{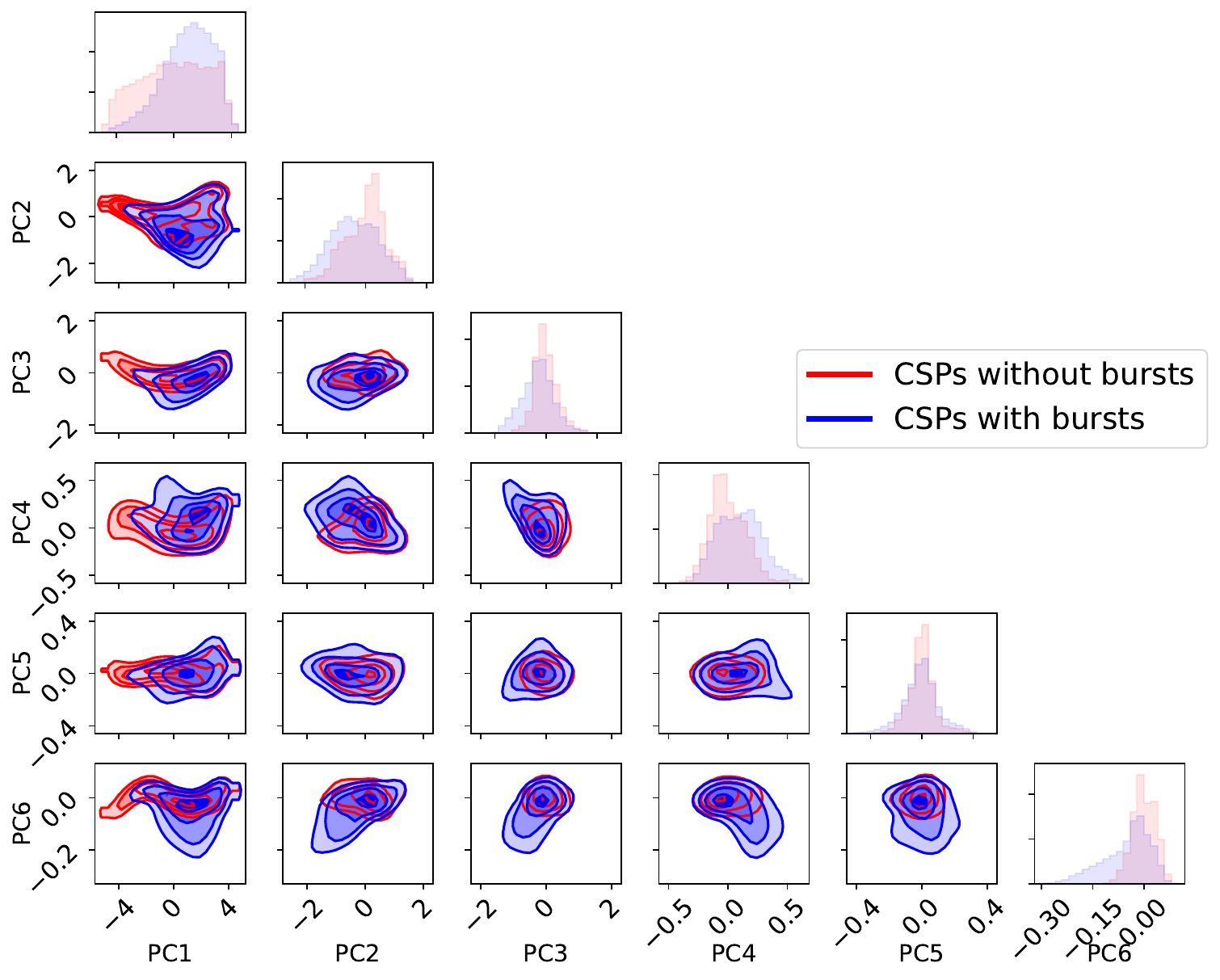}
  \caption{Distribution of the projections of the index measurement for the CSPs onto the principal components, colour coded as models with burst (blue), and without bursts(red). The contours encompass 25\%, 50\%, 75\%, and 90\% of each subsample.}
  \label{fig:latent}
\end{figure}

By projecting the absorption indices onto the eigenvectors from the covariance of the secular, non-bursty models, we visualise the latent space of the models without bursts, to be contrasted with the models with bursts. The latent space is illustrated in Fig.~\ref{fig:latent} for both sets of models, as labelled. The fraction of the explained variance (scree plot) is as follows: 90.42\%, 7.24\%, 1.89\%, 0.34\%, 0.09\%, and 0.02\%. Note that the models are only limited by the noise of the stellar  libraries, therefore, even at the lowest levels of variance, the information of the eigenvectors is expected to be meaningful and not just caused by noise. Fig.~\ref{fig:latent} shows a strong overlap between the models with and without bursts, with subtle variations in the distributions, an illustration of the rather large entanglement found in spectra with respect to the underlying star formation histories. In this case, the degree of entanglement between the two sets of models is strong enough that separating different bursty models with different numbers of bursts and secular models is  difficult. Note that PC1 shows a more extended distribution for non-bursty models, whereas the opposite is true for PC2, PC3 and PC6. The other components are harder to separate. We will refer below to the prominent trend towards negative values of PC6 in the subset of bursty models.

\begin{figure}
  \includegraphics[width=85mm]{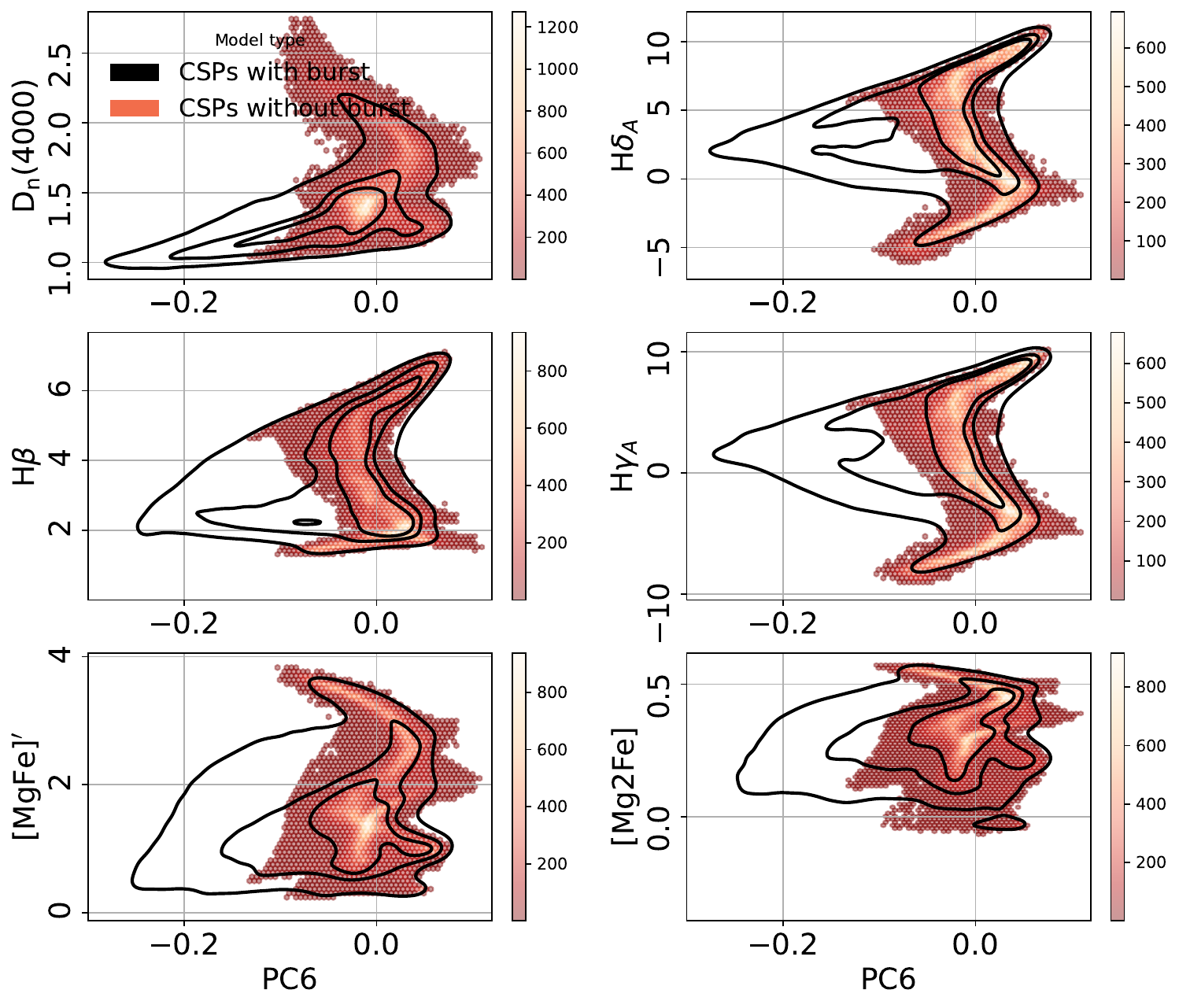}
  \caption{Correlation of the index measurements with principal component number 6. The black contours with four levels illustrate the distribution for models with bursts, while the underlying density distribution, represented by the red points, shows the models without bursts. The colour bar indicates the number of galaxies in each bin, while there are 50 horizontal bins for each panel. Here we show PC6, which is most sensitive to bursting behaviour. The full version with all six principal components can be found in the appendix.}
  \label{fig:Cmodels}
\end{figure}

Fig.~\ref{fig:Cmodels} shows the relation between the principal component number 6 projection and the spectral absorption indices (with the full set of PCs shown in the appendix: Fig.~\ref{figapp:Cmodels}). In all panels, the black contours represent the distribution of bursty models, while the coloured regions represent the density of models without bursts. A strong overlap is found between the two types of models. The effect of star formation bursts on the absorption index measurements is not strong enough to clearly separate the models into distinct regions of parameter space. This behaviour is expected, since bursts can occur at any time and their impact on the spectrum depends both on their age and their strength. However, in the regions where the models do not overlap, these are likely associated with the youngest and/or most intense bursts, where the presence of a young stellar population produces a stronger imprint on the spectral indices. PC6 clearly identifies a range where only bursty models are found, also evident in figure \ref{fig:latent}. Specifically, Fig.~\ref{fig:Cmodels} highlights a clear tail formed by the bursty models, characterised by intermediate Balmer feature strengths, low D$_n$(4000), and lower [MgFe]$^\prime$ and [Mg$_2$Fe]. These properties are indicative of intermediate-age stellar populations that have undergone recent bursts of star formation.

\begin{table*}
\centering
\begin{tabular}{|l|r|r|r|r|r|r|}
 Dataset & D$_n$(4000) & H$\delta_A$ & H$\gamma_A$ & H$\beta$ & [MgFe]$^\prime$ & [Mg$_2$Fe] \\
\hline
No burst models & 1.58/0.31 & 2.67/4.26 & 0.52/5.17 & 3.38/1.56 & 1.90/0.86 & 0.34/0.13\\
All models      & 1.44/0.29 & 3.36/3.64 & 1.57/4.44 & 3.50/1.43 & 1.64/0.79 & 0.31/0.13\\
\hline
 & \multicolumn{6}{c}{Standardised distributions}\\
SDSS            & 1.03/0.58 & $-$0.98/0.54 & $-$1.15/0.52 & $-$0.91/0.47 & 1.63/0.60 & 1.32/0.41\\
LEGA-C          & 0.89/0.79 & $-$0.56/0.61 & $-$0.80/0.66 & $-$0.54/0.64 & 1.32/0.84 & 1.10/0.70\\
Bursts          & $-$0.74/0.77 & 0.26/0.74 & 0.32/0.74 & 0.12/0.86 & $-$0.48/0.82 & $-$0.46/0.94\\
\end{tabular}
\caption{Mean and standard deviation of the indices from the models, as labelled. The mean and standard deviation corresponding to the observations in SDSS and LEGA-C are also shown, after standardisation using the mean and standard deviation of the set including all models. The last row represents the equivalent mean and standard distribution of the set of bursty models when standardised according to the ``no burst'' models.}
\label{tab:samples}
\end{table*}

\section{Confronting real galaxies in latent space}
\label{Sec:Real}

\begin{figure}
\centering
\includegraphics[width=85mm]{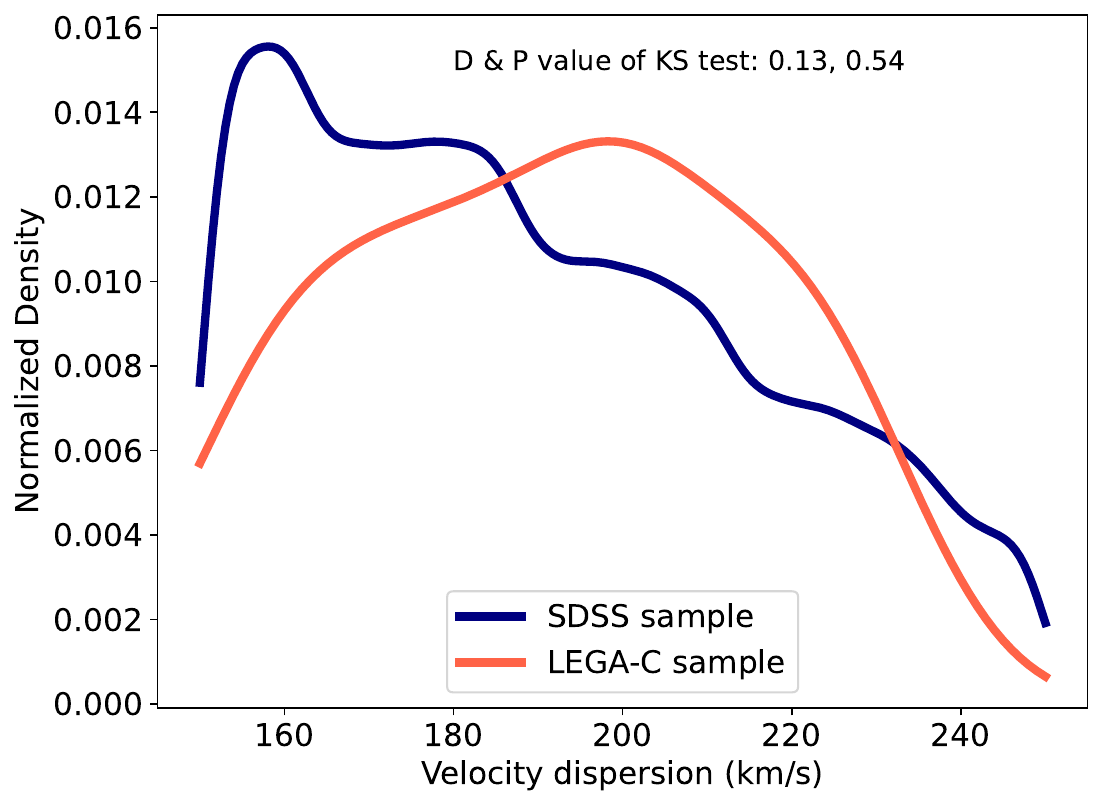}
\caption{Histogram of stellar velocity dispersion of the selected samples from SDSS (blue) and LEGA-C (orange). In both sets, we impose the requirement of having all six indices, adopted to define the latent space in this study.}
\label{Vdisp}
\end{figure}

Thus far, we have explored the properties in latent space of the synthetic models. This section focuses on projecting data from real galaxies in the same latent space. We focus on two well established data sets, to assess the validity of this approach to compare galaxies at two  different cosmic epochs. We consider a subset of galaxies in the legacy SDSS  sample and in LEGA-C, as presented in \S\S\ref{Ssec:SDSS} and \S\S\ref{Ssec:LEGAC}. We first examine the velocity dispersion distribution, as shown in Fig.~\ref{Vdisp}. The SDSS sample is represented in blue, and the distribution of LEGA-C galaxies is shown in orange. We note that the LEGA-C sample is substantially smaller, especially as we impose that all six absorption indices must be available. Consequently, because of the limited statistics, the selected SDSS sample and LEGA-C sample do not appear as strictly drawn from the same parent population. This is reflected in the KS test, which yields a $p$-value of 0.54. However, the samples are still sufficiently comparable for our purposes, as shown by the histograms. We project the SDSS and LEGA-C data onto the eigenvectors from the analysis of the whole set of the CSP models.

To accurately project SDSS and LEGA-C galaxies into the latent space produced by the models, the input indices are transformed into the same statistical space -- i.e., standardised using the same mean and standard deviation as the models, otherwise the projections will be biased due to the different scale and offset mismatch between the synthetic models and the observations. The mean and standard deviation of each index in the SDSS and LEGA-C samples, after this model-based standardisation, are reported in Table~\ref{tab:samples}. The D$_n$(4000) index shows a small increase in mean values in both the SDSS and LEGA-C samples, indicating higher values compared to the model mean. While the increase is stronger in the SDSS sample, the LEGA-C data  show a larger standard deviation. This indicates older stellar populations for SDSS galaxies, as shown by higher values of D$_n$(4000). The Balmer absorption features (H$\delta_A$, H$\gamma_A$ and H$\beta$) exhibit negative mean values in both samples, with a stronger deviation from the models in SDSS. Both samples have comparable standard deviations for these indices. This is consistent with the SDSS sample being dominated by older/quiescent galaxies with weaker Balmer absorption than the younger populations and CSP models. The mean values for both metallicity-sensitive indices, [MgFe]$^\prime$ and [Mg$_2$Fe], are positive in both samples, but notably higher in SDSS, indicating an older, more metal-rich population. The LEGA-C sample, on the other hand, exhibits larger standard deviations for these indices, indicating a wider spread in metallicity-sensitive features. Overall, these results suggest that the LEGA-C index measurements are somewhat closer to the general distribution of the CSP models. This may reflect the fact that $\sim\sfrac{2}{3}$ of the CSP models represent younger, bursty populations that are more consistent with higher-redshift galaxies, such as those in LEGA-C \citep{Gallazzi:25}. In contrast, SDSS galaxies, that map the lower redshift Universe, are more likely to be quiescent \citep{Zibetti2026}.

\begin{figure*}
    \centering
    \includegraphics[width=15cm]{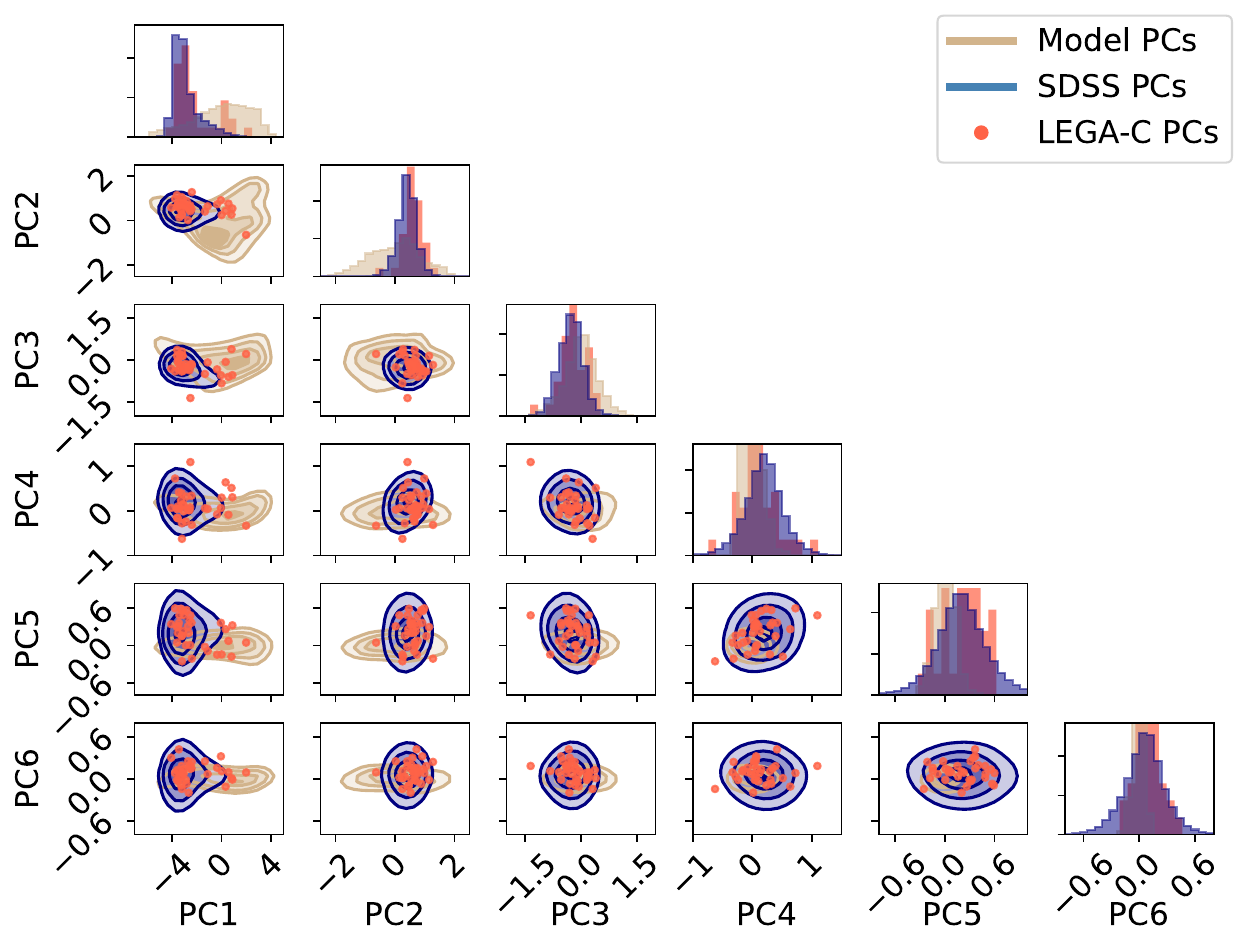}
    \caption{Distribution of the projections of the index measurement for the CSPs, SDSS and LEGA-C onto the information vectors obtained by applying PCA on the CSPs, colour coded as
models PCs (tan), SDSS PCs(blue) and LEGA-C PCs(red). The contours encompass 25\%, 50\%, 75\%, and 90\% of each subsample.}
    \label{Corner_sdss_legac}
\end{figure*}

After standardising the index sets of the SDSS and LEGA-C samples using the mean and standard deviation of the CSP models, we project these standardised observations onto the eigenvectors defined by the CSP models. The corner plot of the projected components (PCs) is shown in Fig.~\ref{Corner_sdss_legac}. PC1, which captures most of the variance ($\sim 80\%$) in the input model space, reveals that the SDSS and LEGA-C samples partially overlap. However, there is a subset of LEGA-C galaxies clearly separated from the SDSS distribution. SDSS galaxies tend to align with the left side of the CSP model distribution -- an area that, based on Fig.~\ref{fig:latent}, corresponds to lower model density and to CSPs without recent bursts. In contrast, the distinct LEGA-C subset from SDSS populates a region closer to the high-density area of the model space, associated with bursty and younger populations. Considering the distribution of the first three PCs and the maps created between them, we observe that both observational samples (SDSS and LEGA-C) partially populate the same general region as the CSP models. However, in higher-order components (PCs beyond PC3), the observational samples begin to populate a somewhat separate subspace, almost orthogonal to the main distribution of the models. This suggests that while the latent space created by the CSP models captures the variance present in the observations, there are aspects of the SDSS and LEGA-C data that may not be fully explained by the model grid. We also explored the effect of noise—absent in the models (see Appendix.\ref{noise})—and found that the mismatch between the models and the observational samples cannot be fully explained by noise alone. It is important to note that CSP models are constructed such that there is no physical coupling between the age of the stellar population and the duration or shape of the star formation history (SFH), unlike in the real Universe. This lack of physical correlation may lead to a broader distribution in the model space, from which the observed data can be drawn, but which does not necessarily reproduce the detailed structure of the observational distributions.

\begin{figure*}
\centering
\includegraphics[width=34mm]{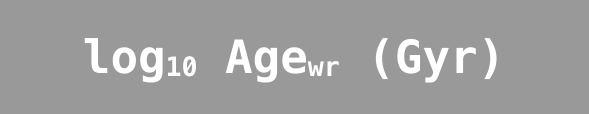}
\includegraphics[width=34mm]{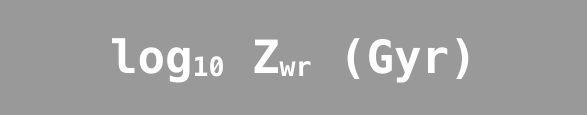}
\includegraphics[width=34mm]{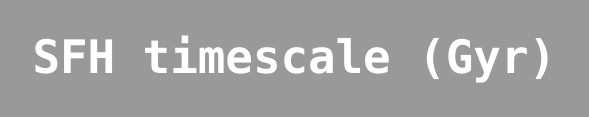}
\includegraphics[width=34mm]{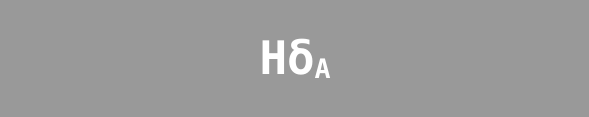}
\includegraphics[width=34mm]{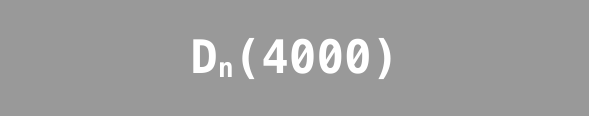}
\includegraphics[width=34mm]{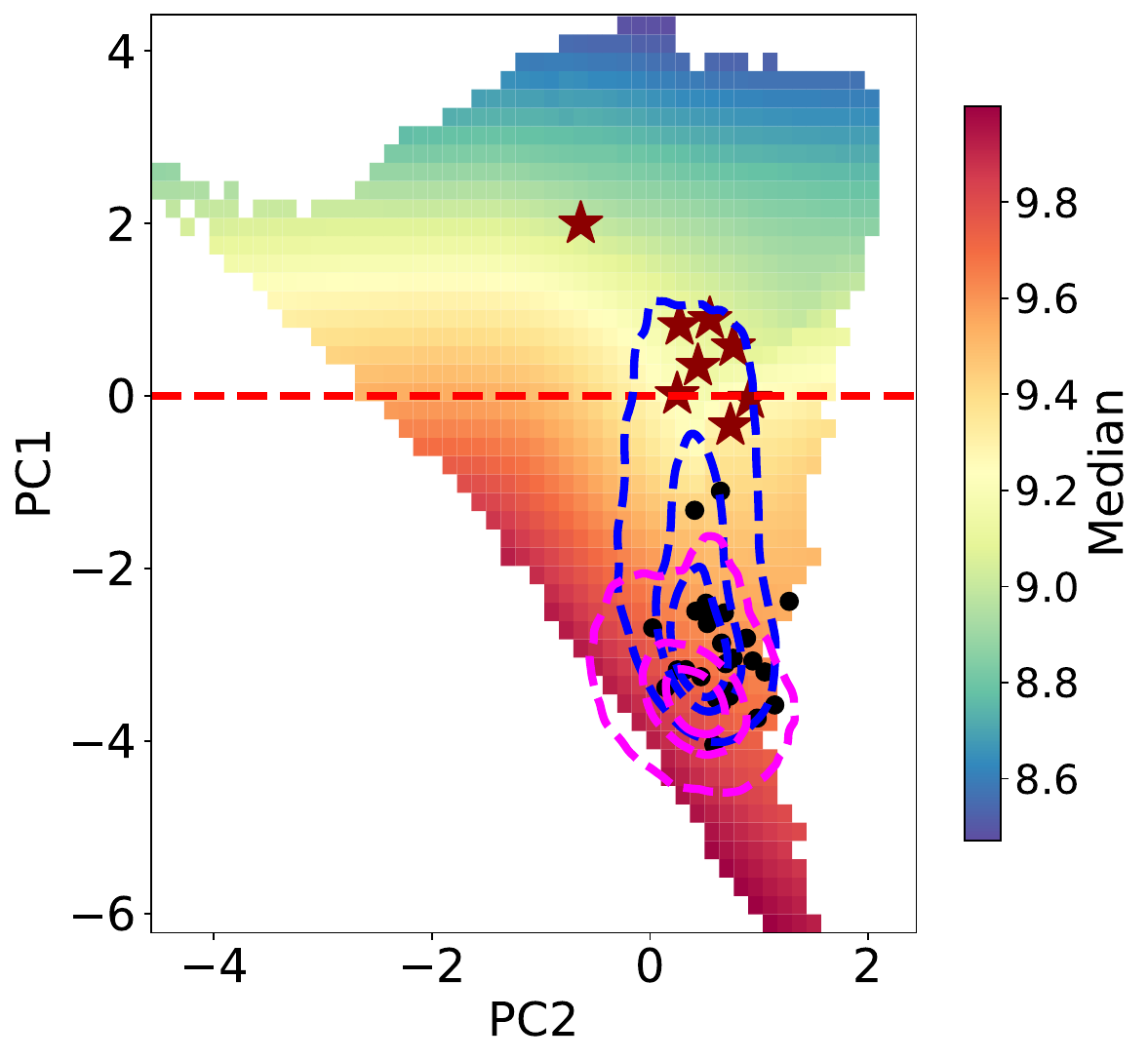}
\includegraphics[width=34mm]{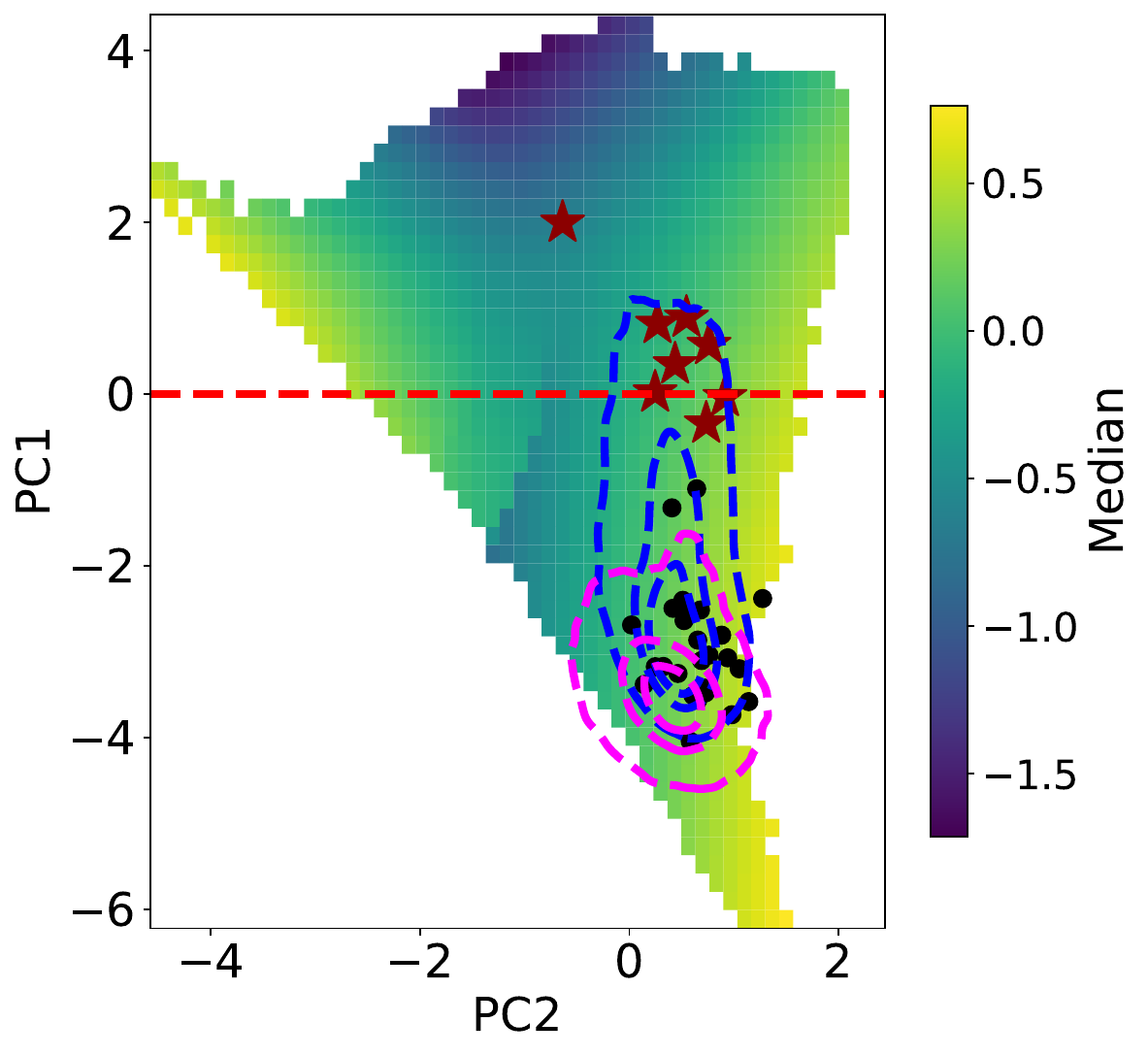}
\includegraphics[width=34mm]{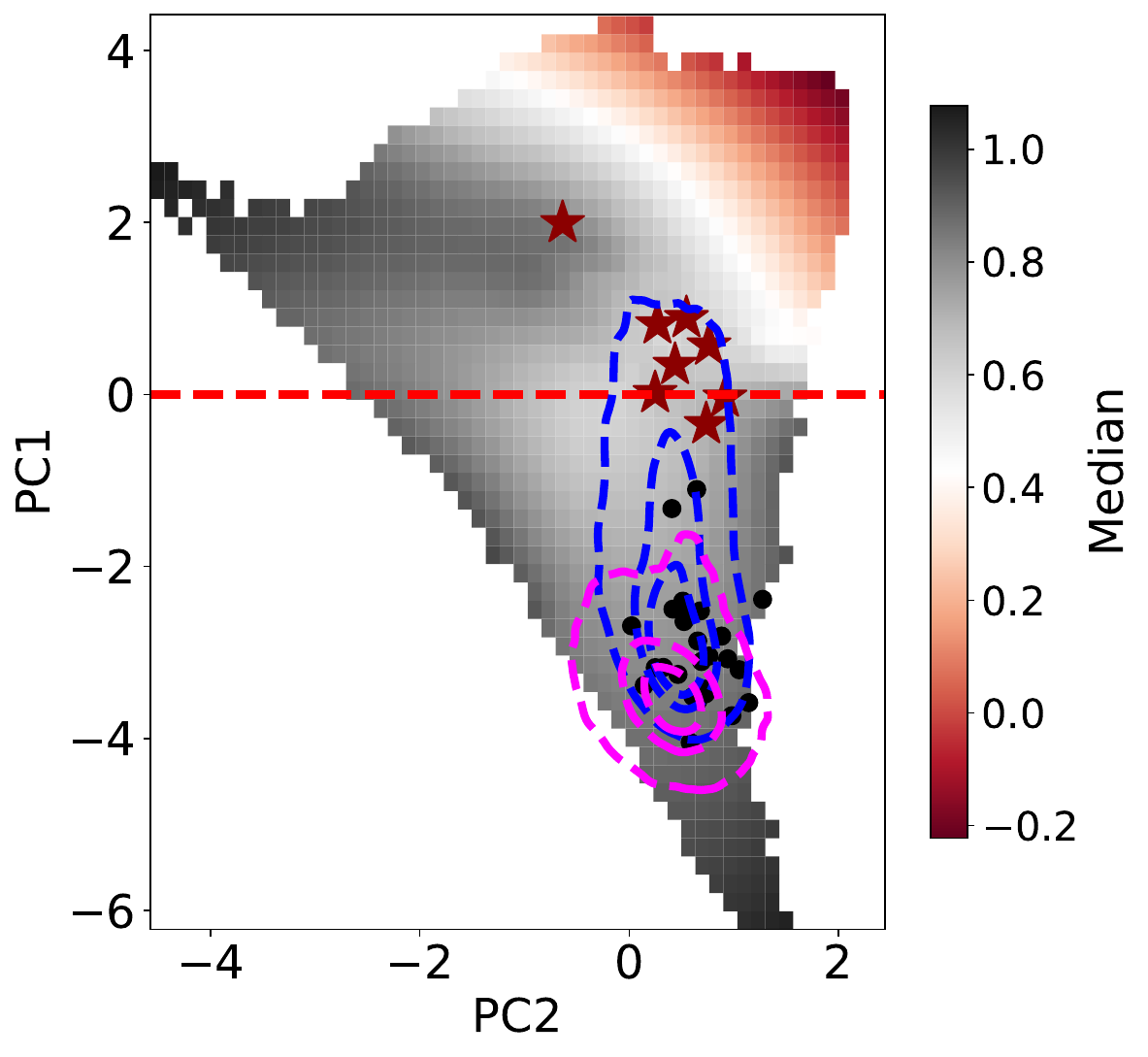}
\includegraphics[width=34mm]{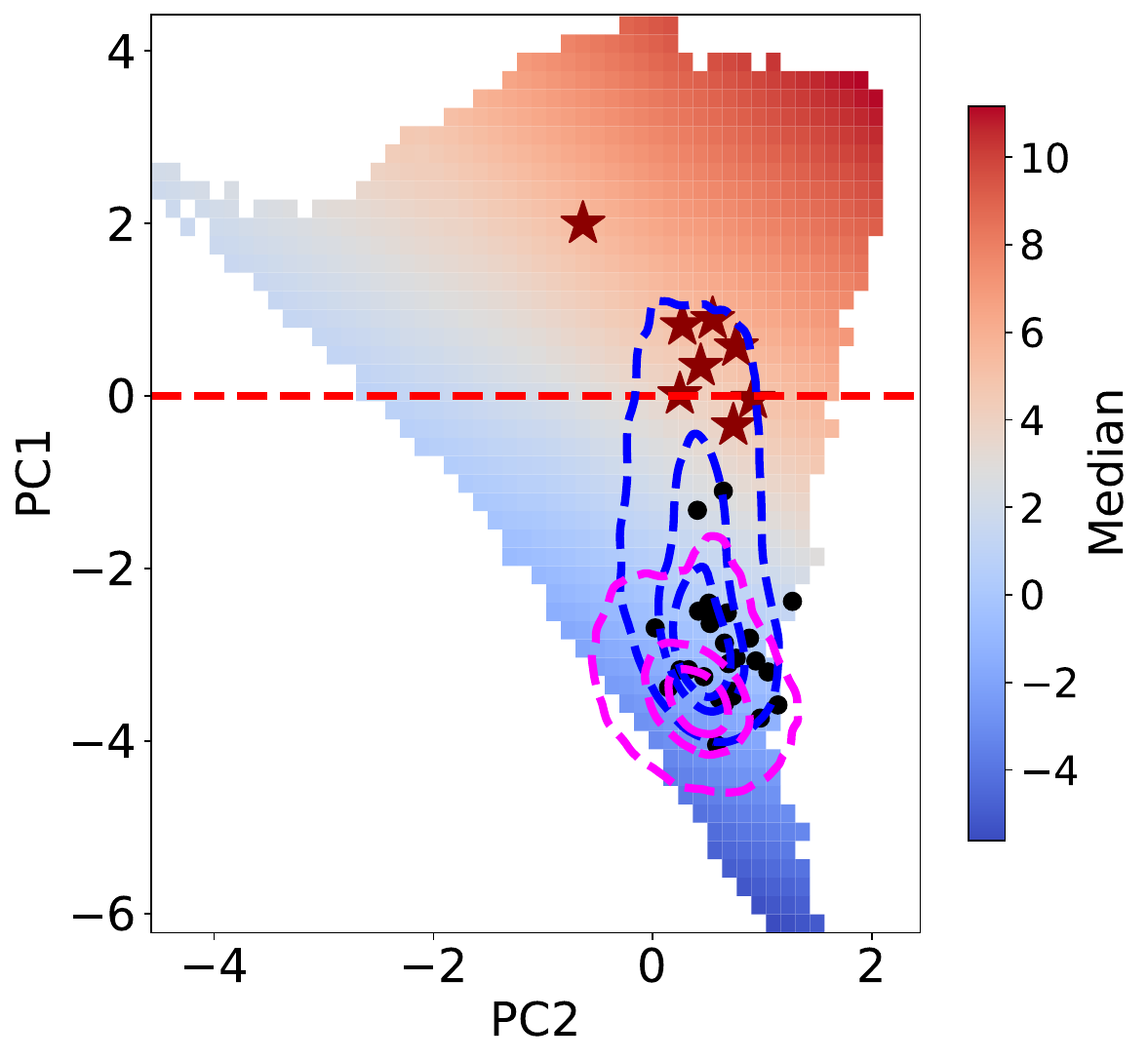}
\includegraphics[width=34mm]{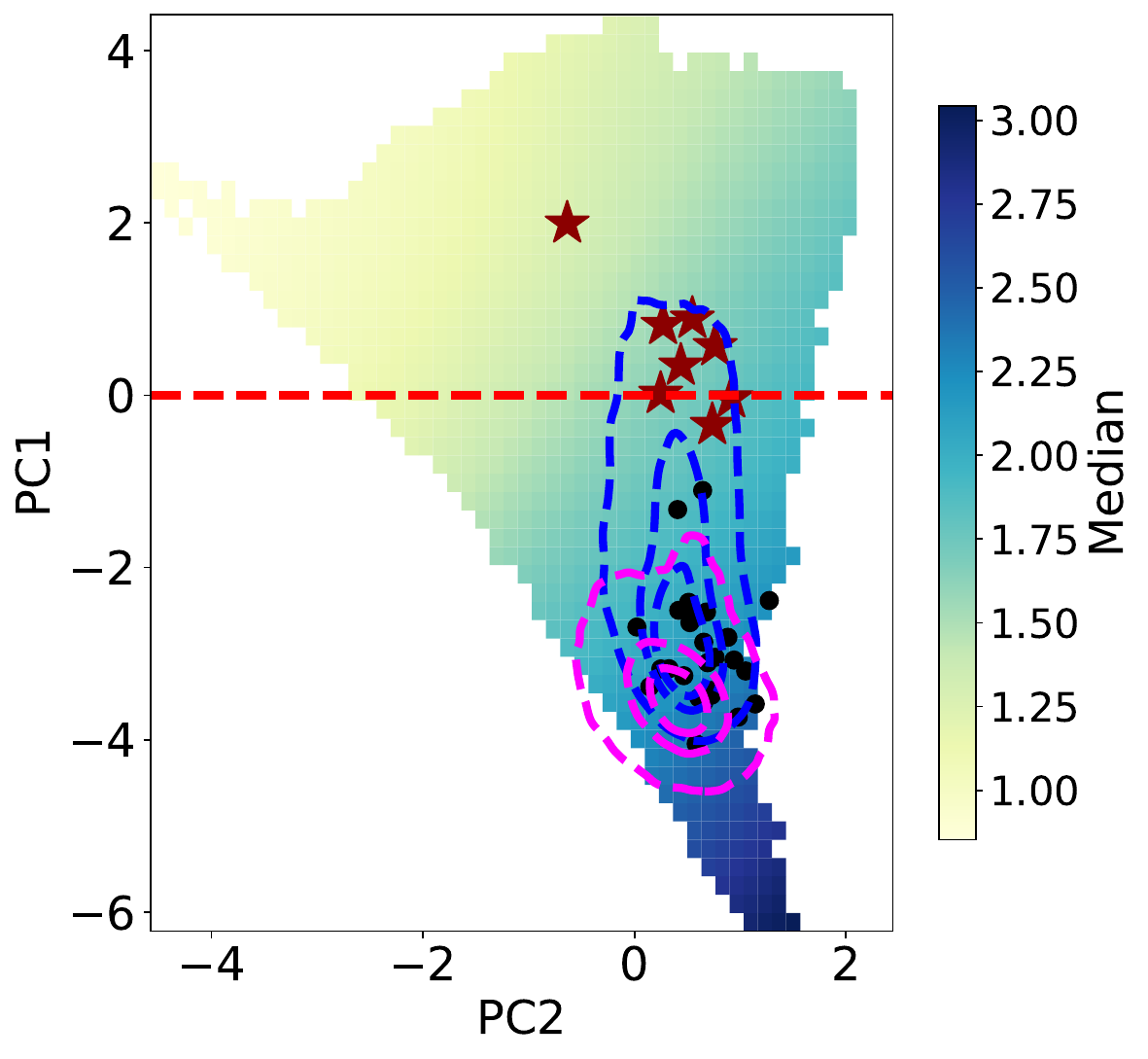}
\includegraphics[width=34mm]{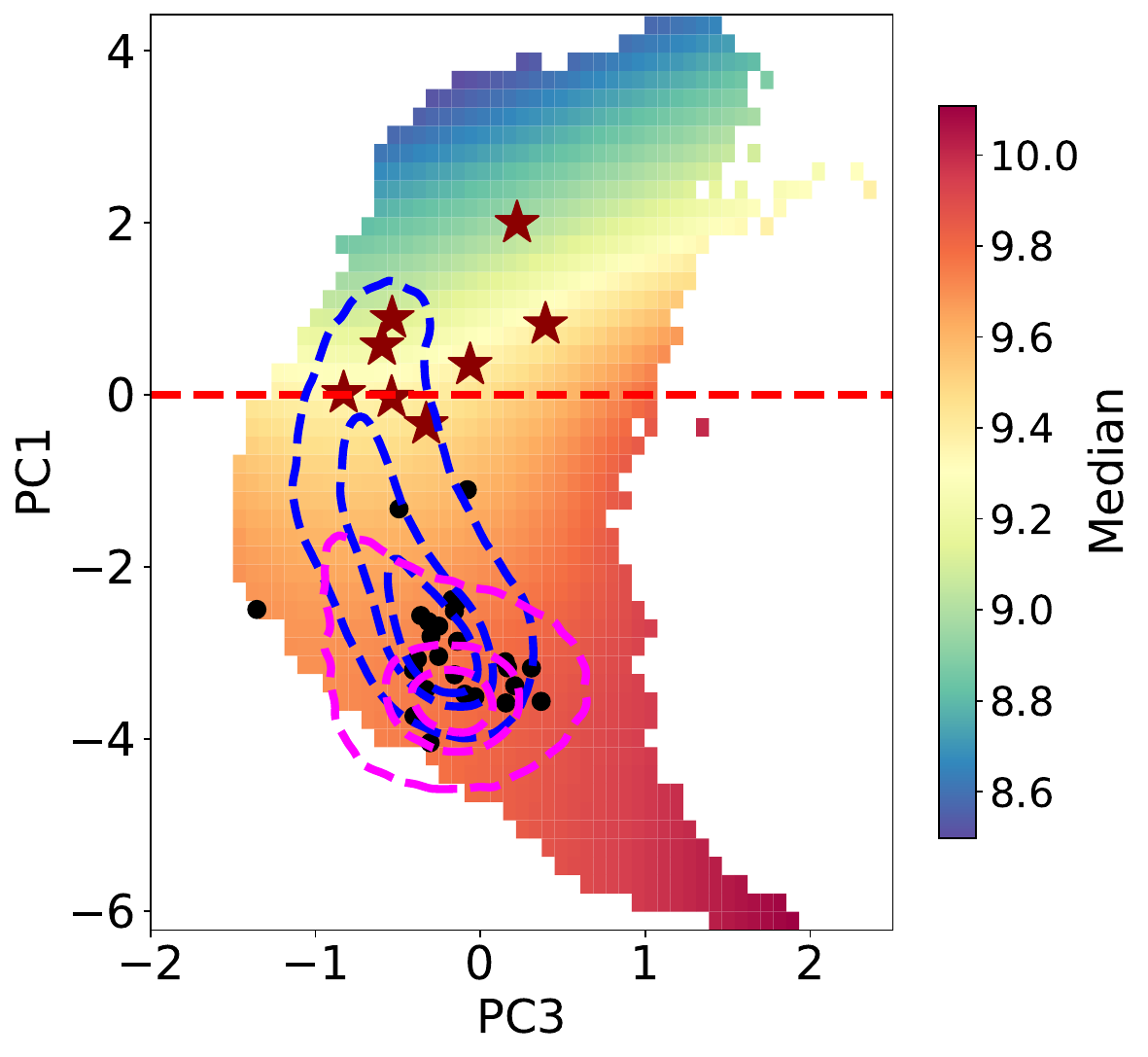}
\includegraphics[width=34mm]{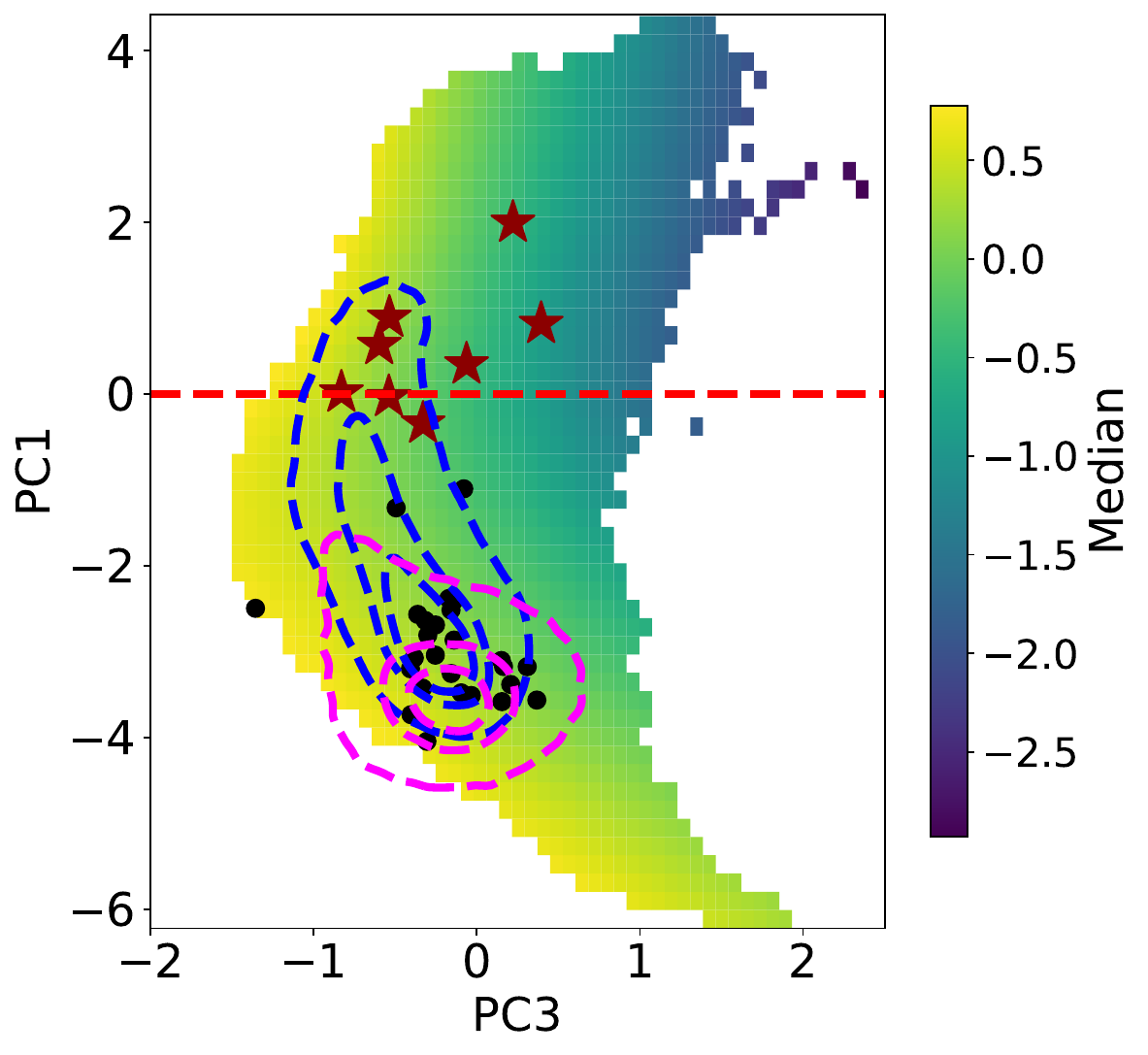}
\includegraphics[width=34mm]{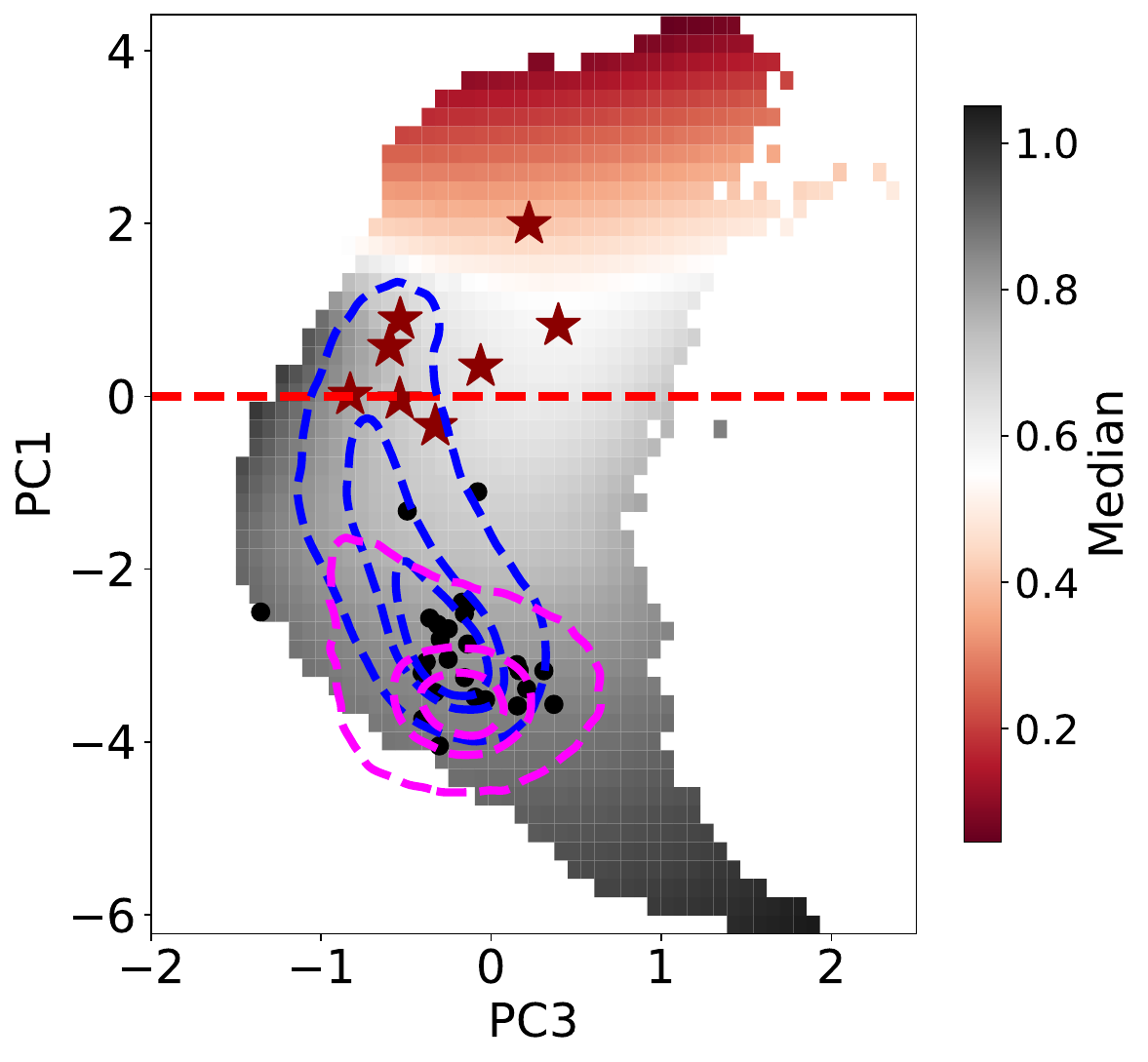}
\includegraphics[width=34mm]{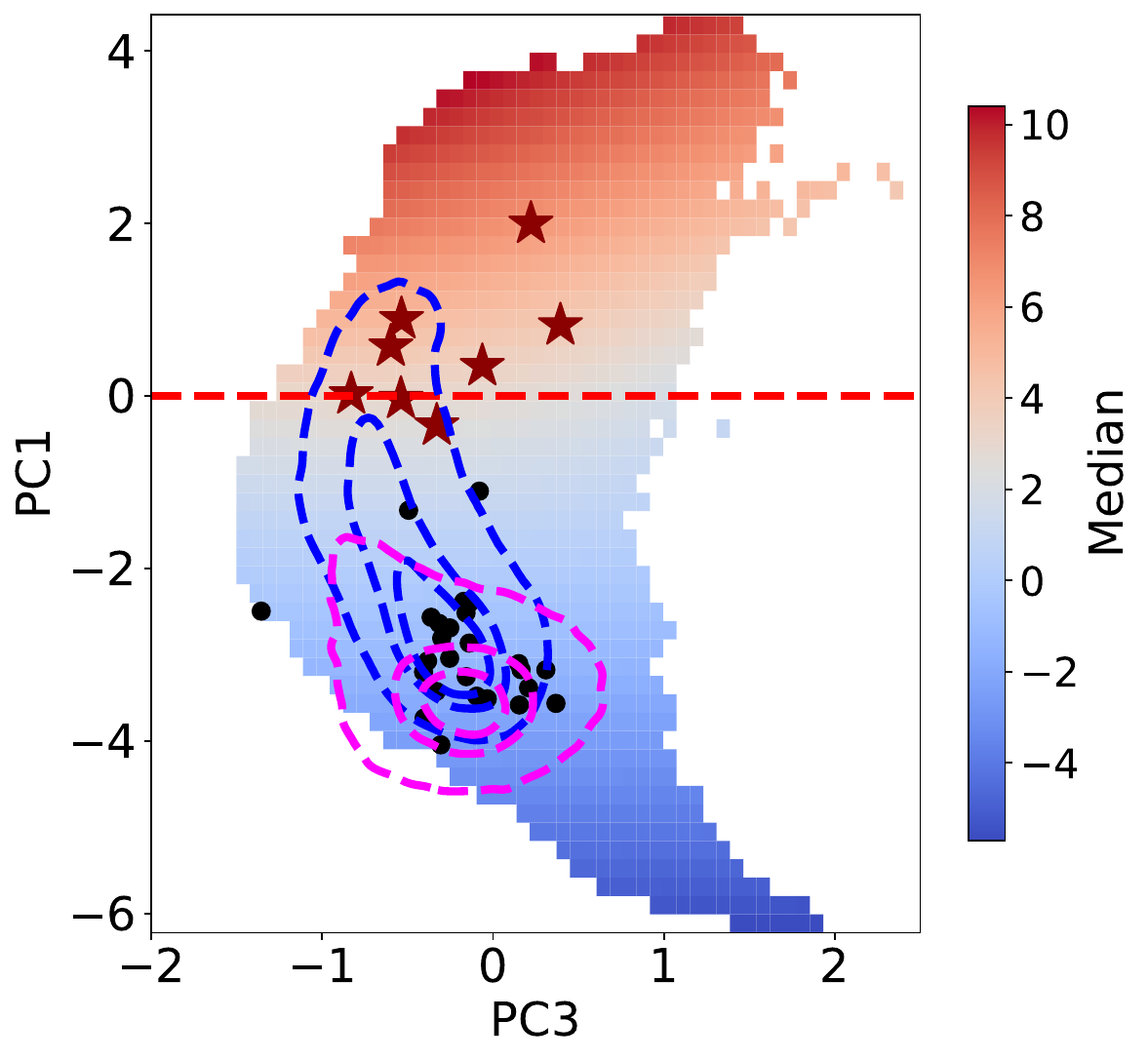}
\includegraphics[width=34mm]{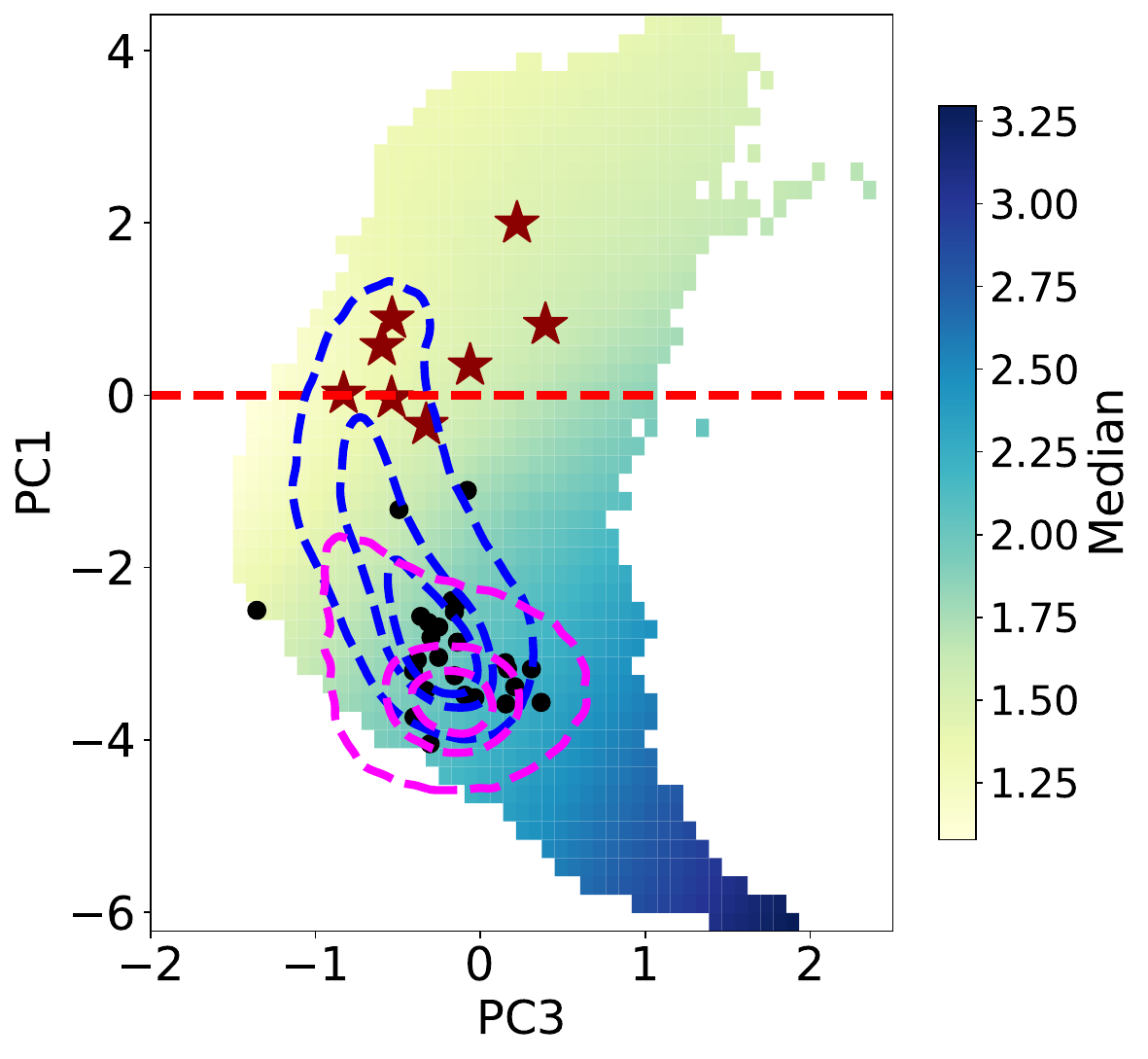}
\includegraphics[width=34mm]{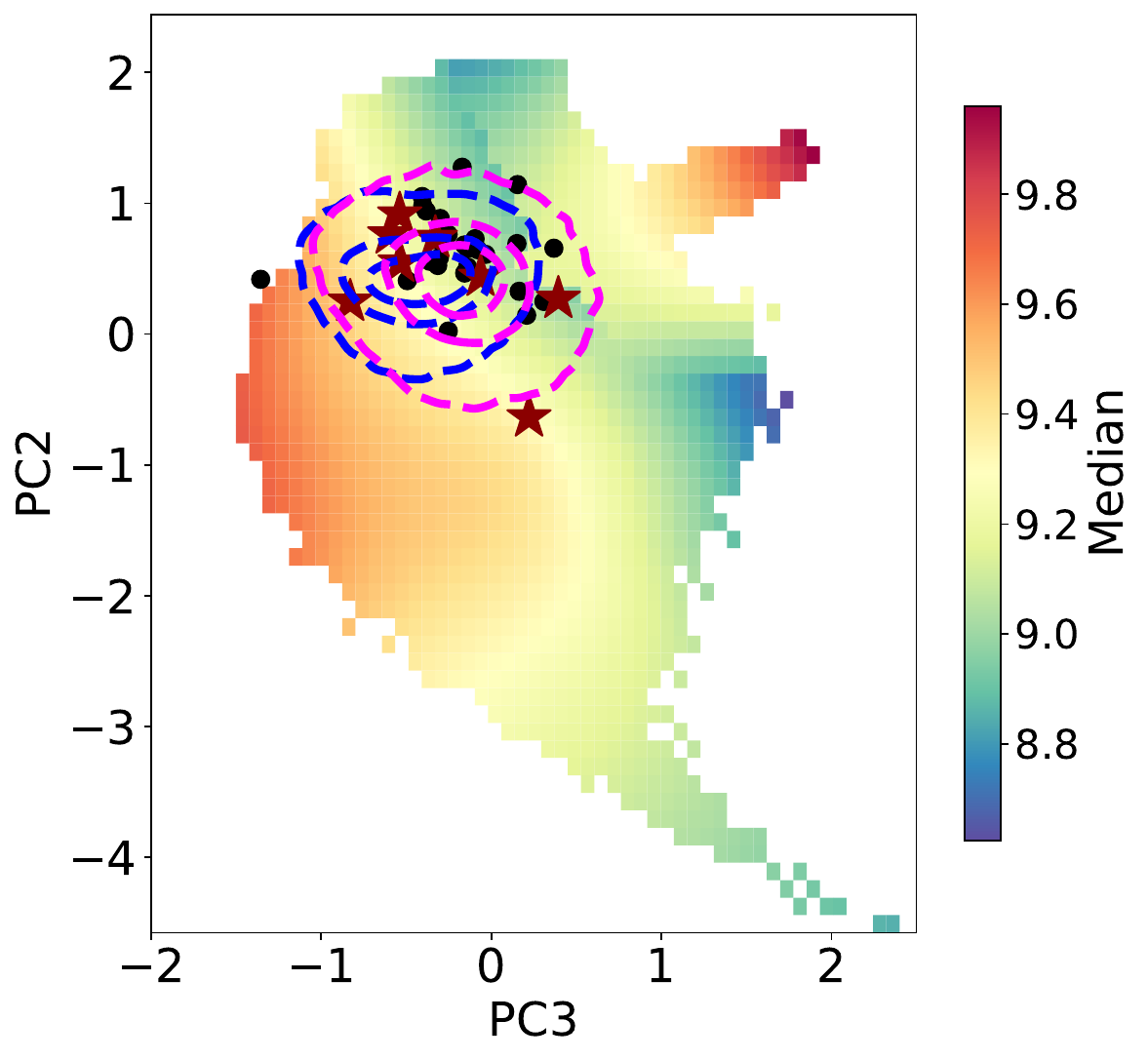}
\includegraphics[width=34mm]{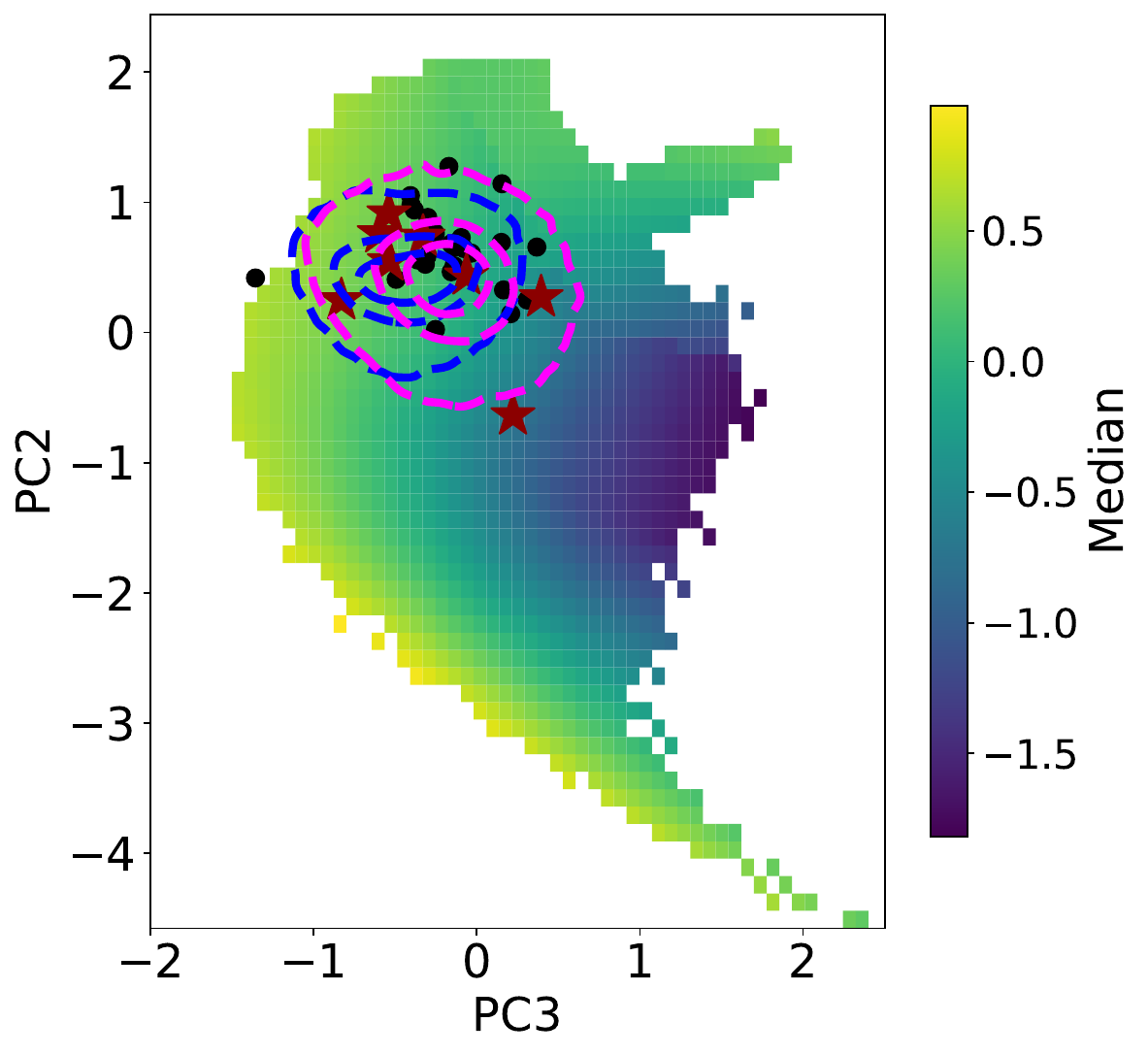}
\includegraphics[width=34mm]{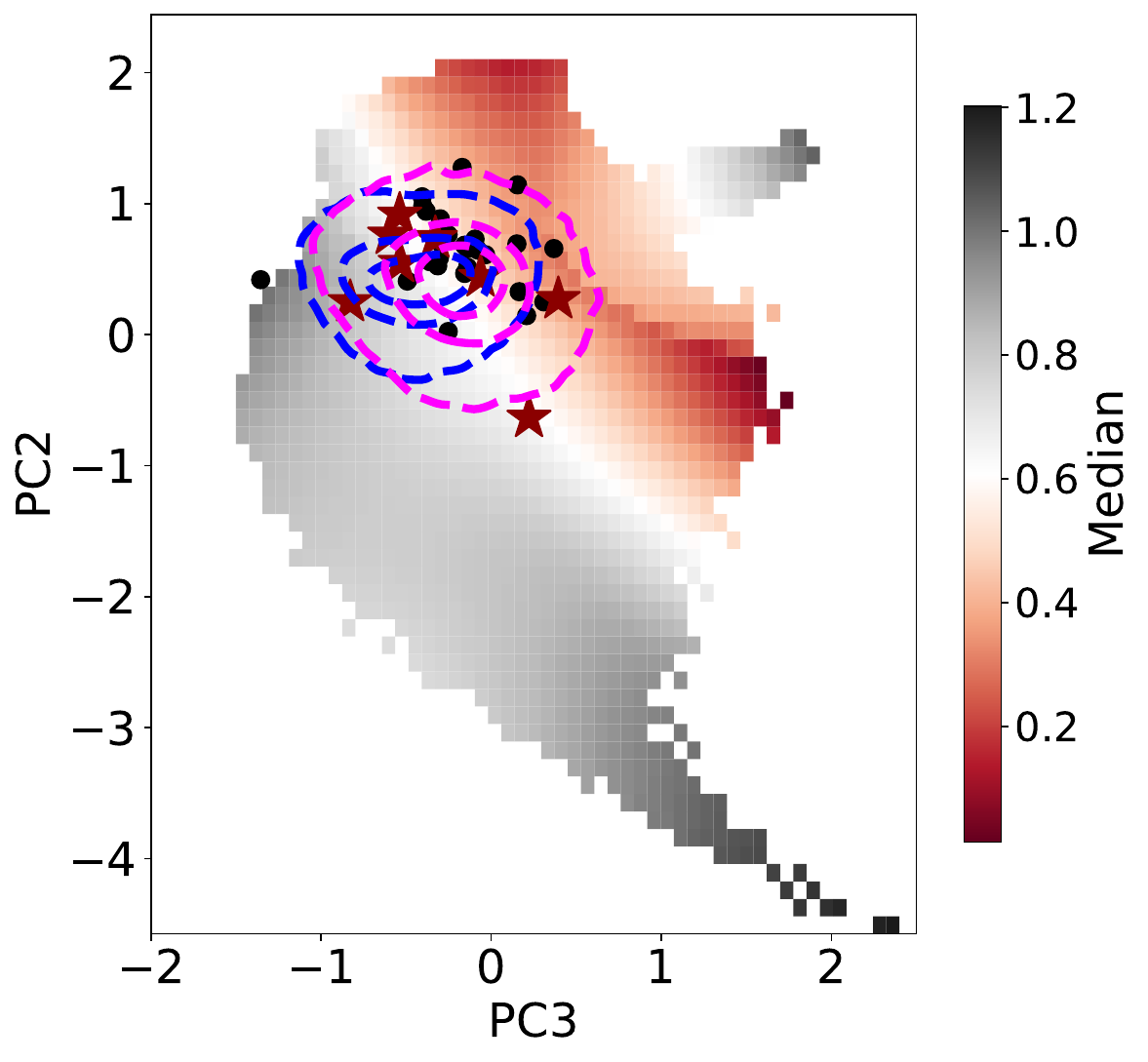}
\includegraphics[width=34mm]{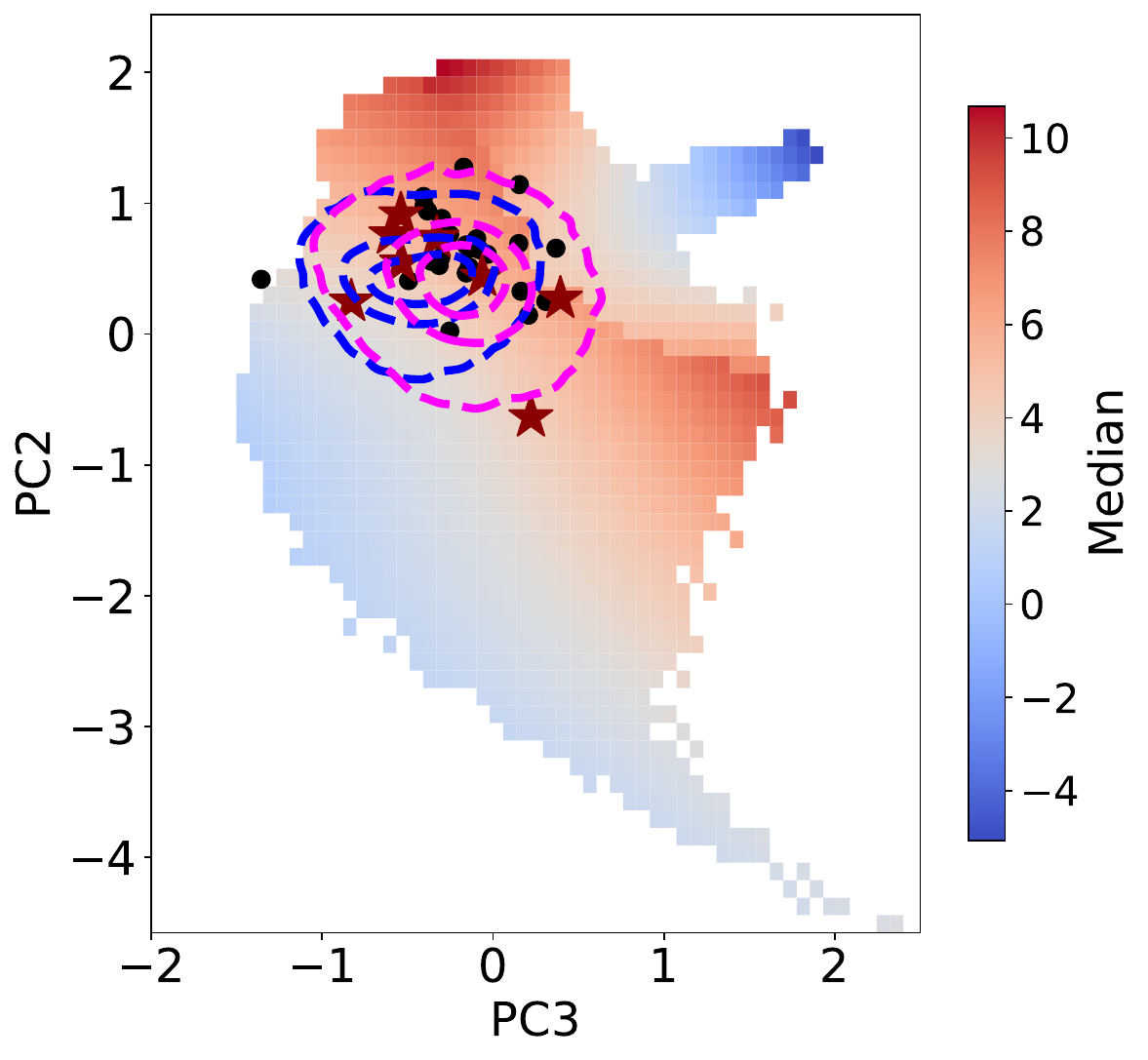}
\includegraphics[width=34mm]{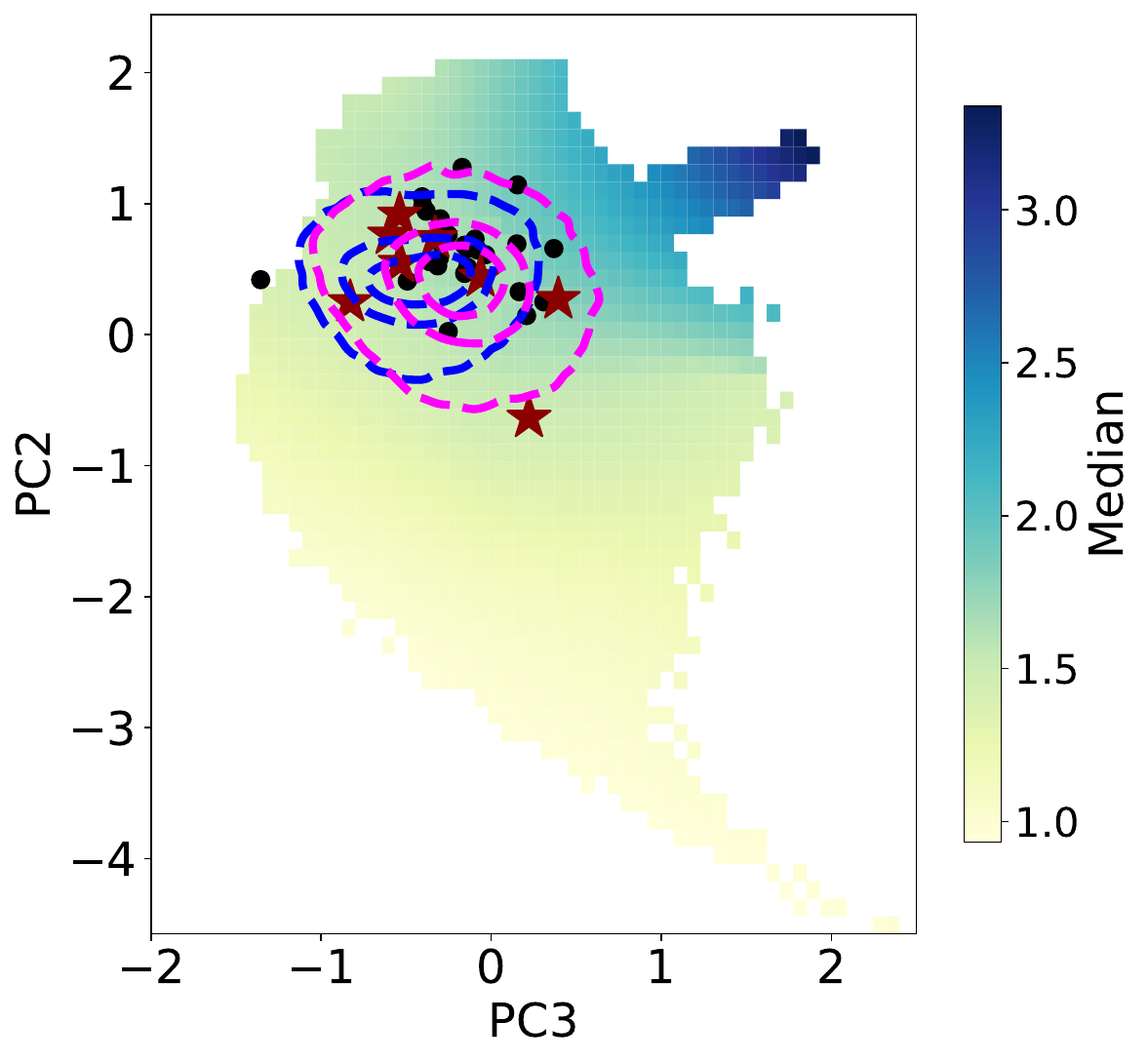}
\caption{Maps between PC1-PC2, PC1-PC3 and PC2-PC3 from top to bottom, which are color-coded based on the median values of the synthetic models for (from left to right) the log of light weighted age and metallicity in the SDSS $r$ band, star formation history time scale -- defined as T75-T25 -- H$\delta_A$ and D$_n$(4000). Blue and magenta contours locate the projection of the SDSS data (blue: young, magenta: old), and LEGA-C galaxies are shown as points. The black points correspond to LEGA-C galaxies with PC1$<-0.9$, and share a similar location in latent space with the SDSS sample. In contrast, the red stars, with PC1$>-0.9$, do not have an equivalent SDSS sample in this space. The red dashed horizontal lines in the first two rows indicate the positive and negative regimes based on PC1, which represents a dynamical range separation of the models (see text for details).}
\label{SDSS-LEGAC}
\end{figure*}

The structure of the CSP model grid in PCA space is further explored through maps between the first three principal components, that capture most of the variance in the index input space (see Fig.~\ref{SDSS-LEGAC}). Each row in Fig.~\ref{SDSS-LEGAC}, from top to bottom, shows maps between PC1–PC2, PC1–PC3, and PC2–PC3, respectively. The columns represent different physical properties: from left to right, the maps colour-code the CSP models with respect to age, metallicity, star formation history (SFH) timescale, H$\delta_A$, and D$_n$(4000). Overlaid on these maps, the KDE-smoothed blue/magenta contours represent the SDSS sample, while black and red points mark the LEGA-C galaxies. The SDSS sample is split with respect to the bimodal distribution found in the stellar ages (blue: young, magenta: old) with respect to stellar mass, as presented in \citet{Mattolini:25}\footnote{The actual threshold to separate SDSS galaxies into young and old is age=0.082$\log($M$_s$/M$_\odot$)+8.72, where $M_s$ is the stellar mass of the galaxy.}, that follow a Bayesian index fitting technique. The red points correspond to those LEGA-C galaxies that are separated from the SDSS distribution in latent space, as shown in Fig.~\ref{Corner_sdss_legac}, based on a threshold in PC1. Note that the histograms of PC1 projections feature a tail of LEGA-C galaxies that do not overlap with the distribution of PC1 values for SDSS galaxies. The black points represent the remaining LEGA-C galaxies that overlap with SDSS. As discussed earlier in \S\ref{CSPs_all}, PC1  captures the dynamical range of the indices in the CSP models. Splitting the models by the sign of PC1 reveals two distinct branches in the higher-order components, with each branch corresponding to either positive or negative PC1. In Fig.~\ref{SDSS-LEGAC}, the red horizontal dashed lines in the PC1–PC2 and PC1–PC3 panels mark this separation. By moving from the shared SDSS–LEGA-C region toward the isolated LEGA-C red points (especially in the PC1–PC2 and PC1–PC3 panels), and focusing on the first column (stellar age), we observe a clear trend: the stellar populations shift from older ages ($\sim$10\,Gyr) to younger ones ($\lesssim$1\,Gyr). Note that, while the majority of LEGA-C data overlap the SDSS contours, their barycenter aligns with the young SDSS sub-population rather than the old one. In the second row, the PC1–PC3 map (which better reflects metallicity structure, as PC3 correlates more strongly with metallicity) shows that while the SDSS–LEGA-C common region corresponds to solar ($\log Z = 0$) and supersolar metallicities, the distinct LEGA-C subpopulation has a slight skew toward subsolar metallicities. In the third column (SFH timescale), the SDSS and overlapping LEGA-C galaxies are associated with longer star formation timescales, suggesting more extended star formation histories. In the isolated LEGA-C group (red points), the timescales become shorter. The common SDSS/LEGA-C region is characterised by high D$_n$(4000) and low H$\delta_A$, consistent with older, more quiescent populations. In contrast, the separated LEGA-C subpopulation exhibits lower D$_n$(4000) and higher H$\delta_A$, consistent with younger, post-starburst or star-forming populations. By comparison, the PC2–PC3 maps do not separate  SDSS and LEGA-C galaxies. These projections show a more complex and less physically interpretable structure.

\section{Discussion and Conclusions}
\label{Sec:Disc}

In this paper, we explore an alternative methodology to study the stellar population content of galaxies by producing a PCA-based latent space from the absorption indices of a comprehensive set of composite stellar population (CSP) models typically used for the analysis of survey data \citep[e.g.][]{Gallazzi:05,Gallazzi2014}. This approach extracts the fundamental units for comparison -- the principal components -- in an entirely data-driven manner. Moreover, in comparison with deep learning algorithms, PCA has the advantage that no parameters have to be fitted, and only the statistical properties of the sample are used, avoiding any  bias derived from, for instance, exposing the algorithm to training sets.

The CSP models are constructed by combining Simple Stellar Populations (SSPs) based on parametric Star Formation Histories (SFHs) and Chemical Enrichment Histories (CEHs), and by incorporating dust attenuation \citep[see][for details]{Zibetti2017, Mattolini:25}. Only six targeted absorption indices are adopted, instead of applying PCA to the full spectrum, focusing on regions where prior physical knowledge suggests a high information content \citep{Ferreras2023}. The selected indices include Balmer absorption features (H$\beta$, H$\gamma_A$, H$\delta_A$), the prominent 4000\AA\ break, D$_n$(4000), and two indices involving the Mg-Fe complex around 5000-5500\AA: [MgFe]$^\prime$, and [Mg$_2$Fe]. These indices are not only well-established spectral diagnostics but also provide strong constraints on stellar ages and metallicities. Our goal is to understand how, at the deepest level, the variations in the absorption indices map into stellar population parameters. Details of the generation of this latent space are presented in \S\S\ref{Ssec:PCA}.

\subsection{PC1: natural discriminator between different evolutionary stages of stellar populations}
In \ref{CSPs_all}, we delve into the physical interpretation of the principal components and the resulting information vectors in greater detail. We run PCA on the whole set of CSP models. PC1, as the component with the largest share of variance, does not isolate specific spectral features. Instead, it reflects the overall trend of the combined indices, capturing the coherent variation across them (see Fig.~\ref{figapp:heatmap_allmodels}). This is particularly noteworthy given that the indices were standardised prior to applying PCA; the persistence of such a collective structure in PC1 suggests the presence of an underlying physical driver, such as an age-metallicity continuum or the general evolutionary trend of stellar populations. When PC1 values are separated into two groups: positive and negative, the branching structures observed in the correlation plots between various indices and higher order principal components (see Fig.~\ref{fig:allmodels}) can be explained: each branch corresponds to one of the two PC1-based groups. This structural consistency across principal component space confirms that PC1 not only encapsulates the dominant mode of variance, but also serves as a natural discriminator between different evolutionary stages of stellar populations.

\subsection{Partial breaking of the age–metallicity degeneracy}
As illustrated in Figs.~\ref{figapp:pcmapsALL12} and \ref{figapp:pcmapsALL23}, PC2 and PC3, as the second and third higher-order components, capture subtler but physically important secondary effects. As discussed in \ref{CSPs_all}, the projection maps in PC1-PC2 and PC1–PC3 space (Fig.~\ref{fig:pcmapsALL123}), colour-coded by age and metallicity, suggest that a partial breaking of the age–metallicity degeneracy can be achieved, with the former mapping nearly PC2 independent tracks with respect to age, and the latter mapping PC1 independent tracks regarding metallicity. This interpretation is supported by two key observations. First, the negative PC1 regime corresponds to older populations with short star formation timescales, whereas the positive PC1 regime is associated with younger populations and more extended star formation histories. Second, PC1 shows a consistent correlation across all six indices, reinforcing the idea that it reflects a broad, underlying driver, most likely stellar age. This finding is consistent with previous studies of absorption features in galaxy spectra, including the analysis of underlying variance in SDSS galaxy spectra \citep{variance}, as well as follow-up investigations using synthetic spectra from the EAGLE and IllustrisTNG simulations \citep{Sharbaf2025}. Our study supports the conclusion from these works that the highest-variance principal component can be interpreted as an evolutionary sequence among galaxies, with a strong correlation to stellar age, and to a second order with metallicity. This reflects a bimodality: spectra dominated by cool stars (due to older ages, but also higher metallicity) versus spectra dominated by hot stars (mainly young populations). In agreement with our results, \citet{Chen2012} also found that the first principal component from a sample of massive galaxy spectra is relatively featureless and acts as a first-order proxy for the age of the stellar populations, being strongly correlated with both the 4000\AA\ break and Balmer absorption indices. Furthermore, they showed that the third principal component carries information about velocity dispersion and metallicity, which aligns with our interpretation of the PC3 as a component that captures the metallicity (see also \citealt{Rogers2007}, where the third principal component is also shown to correlate with velocity dispersion).

\begin{figure}
    \centering
    \includegraphics[width=\columnwidth]{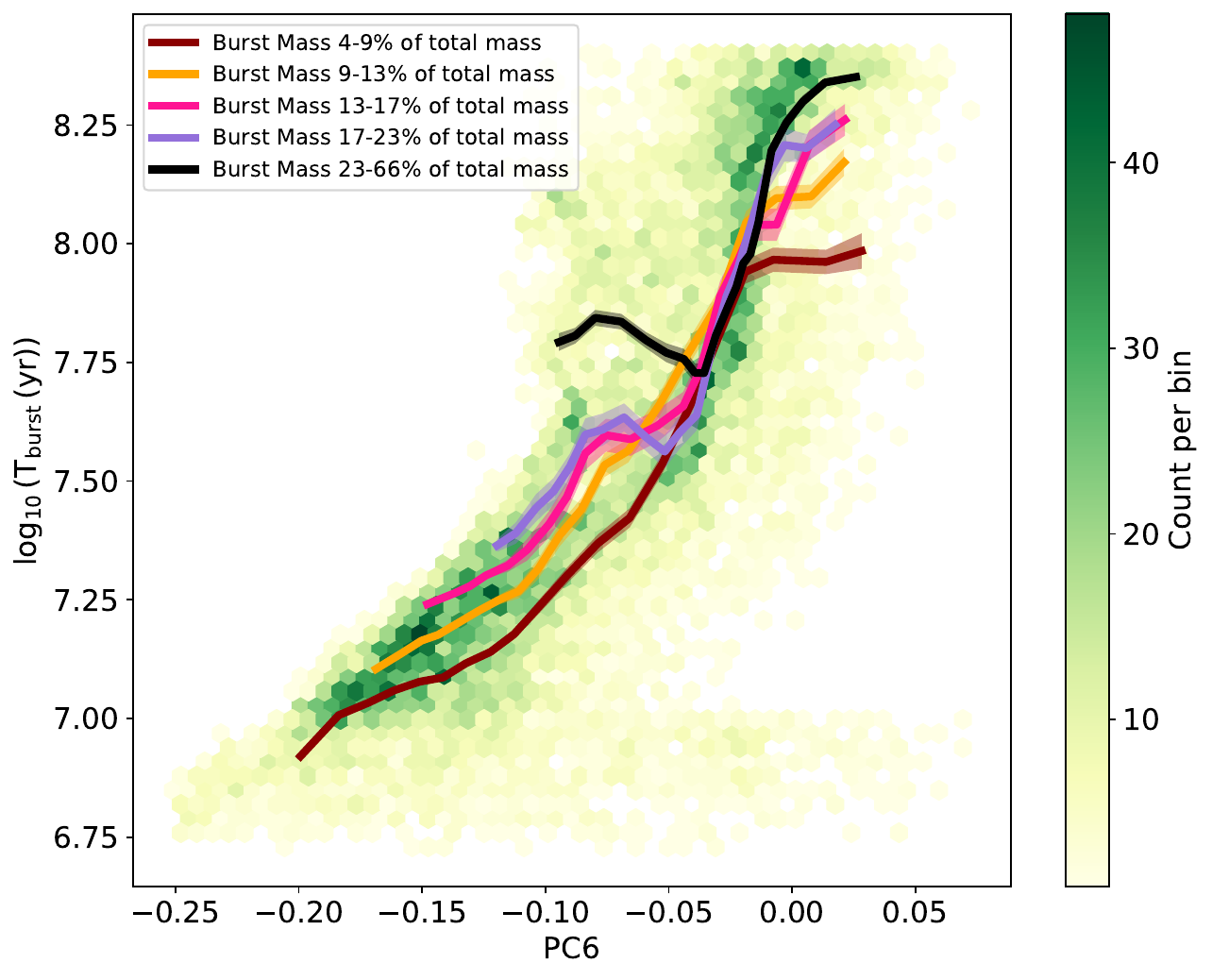}
    \caption{Correlation of PC6 with recent bursting activity, shown as a function of the age of the most recent burst ($log_{10}(T_{\rm burst})<8.4$ yr). A density plot is shown, with tracks tracing a running median for cases selected with respect to the burst mass, as labelled.}
    \label{fig:PC6burst}
\end{figure}

\subsection{Effect of bursts on CSPs models}
In \S\S\ref{Ssec:BurstoNo} we explore the properties of CSP models, splitting them into continuous SFH and those with added random bursts. We apply PCA only to the former, in order to assess the potential for disentangling bursty and non-bursty models. It is worth contrasting in Table~\ref{tab:samples}  the mean and standard deviation of the standarisation of the models without bursts (row 1), and the resulting mean and standard deviation of the bursty models after standardisation by the former (row 5). These deviations highlight differences in the indices due to the presence of one or more bursts. As expected, we find that bursts make the Balmer absorption features more pronounced. The standard deviation of H$\beta$ is higher, which may result from its sensitivity to a wide range of early post-starburst ages, making it more variable. In contrast, H$\gamma_A$ and H$\delta_A$, while also sensitive to A-type stars, are relatively less influenced by young populations. Their absorption features tend to peak slightly later and are less variable in post-burst scenarios \citep{Blake2004, Goto2005, Goto2007, Wilkinson2022}. The D$_n$(4000) break and metallicity-sensitive indices are generally weaker in models with bursts, as younger and lower-metallicity populations contribute to the integrated light \citep[see, e.g.,][]{Gallazzi2014}. As shown in Fig.~\ref{fig:latent}, although the bursts are randomly distributed over the secular SFHs, and we project the bursty models on eigenvectors derived from burst-free models, the resulting variance structure remains entangled. Even after reorganisation via PCA, the overlap between the two populations is substantial, making it difficult to fully disentangle the effects of bursting events from the underlying secular evolution.

\subsection{PC6: discriminator of recent bursts}
It is worth noting, however, that the highest order eigenvector (corresponding to the lowest variance) shows an intriguing tail in the distribution of bursty models, as shown in Fig.~\ref{fig:latent}, at PC6$<-0.1$. Therefore, PC6 can potentially serve as a useful diagnostic of recent starburst activity and may be employed to derive a lower bound on the number of galaxies with recent bursts. Note that even though the amount of information contained in the higher-order principal components (PC4 to PC6) is lower than in the first three, this cannot be attributed to noise, as PCA is applied to synthetic spectra that are virtually free from observational uncertainties. The trend picked up by the sixth eigenvector consists of a meaningful trend, although represented by a minute variation in the absorption indices. Alas, this is the bane of spectral fitting, as variations of important population parameters are mapped into small fractional deviations \citep[as in the percent-level changes of very specific line strengths to constrain the low mass end of the IMF, see, e.g.][]{IF:13,LaBarbera2013}. The density plot in Fig.~\ref{fig:PC6burst} shows in more detail the dependence of PC6 on the presence of bursts, with a strong trend with the age of the most recent burst ($\log_{10}(T_{\rm burst})<8.4$ yr). There is an expected degeneracy between T$_{\rm burst}$ and its fractional mass contribution, as shown by the coloured lines that represent running median values for subsets within an interval of the burst mass fraction, as labelled. Regardless of this degeneracy, we can conclude that models with PC6$<-0.1$ unequivocally correspond to the presence of recent bursts. Since the weights of PC6 mainly ``oppose'' the H$\gamma_A$ and H$\delta_A$, with all other indices playing a minimal role in this projection (Fig.~\ref{figapp:heatmap_allmodels}), we decided to compare in Fig.~\ref{fig:PC6HgHd} the variation of these two Balmer indices with respect to stellar age for an SSP from the \citet{Bruzual2003} models at solar metallicity. The figure traces the time evolution of both indices, normalized to their maximum values, and the inset shows the respective time derivatives. The figure zooms in on the time interval where the variations of these two indices differ, and suggests that this variation is what causes the sensitivity of PC6 to the presence of recent bursts. At a more fundamental level, we note that similar trends between Balmer indices of different order can be caused by phases beyond the main sequence turnoff, such as blue HB stars \citep{Schiavon:04,Schiavon:07}. We note that the correlation between PC6 and metallicity requires confirmation using more detailed modelling of the full galaxy spectra and therefore warrants further investigation.

\subsection{Comparison between SDSS and LEGA-C sample}
To study the statistical difference between samples of galaxies at different redshift in this framework, we chose two samples from SDSS and LEGA-C, as described in \S\S\ref{Ssec:SDSS} and \S\S\ref{Ssec:LEGAC}, respectively. Galaxies were selected within a velocity dispersion range of 150 to 250\,km/s to ensure the differences are not caused by the well known mass-age or mass-metallicity relation. This range was chosen because, at higher redshift, the availability of high-quality spectra for low-mass galaxies decreases. Restricting the sample in velocity dispersion range ensures a sufficient number of high-quality spectra across both datasets. The spectral index measurements of these galaxies were standardised using the mean and standard deviation from the CSP models, ensuring consistent scaling between the models and the observational samples. The offsets in mean and standard deviation for each index—comparing the observational samples to the CSP model distributions—are reported in Table~\ref{tab:samples}. This analysis reveals that the LEGA-C sample, which probes higher redshift, displays spectral index properties more closely aligned with the wide range of CSP model space. This is particularly evident in its broader index distributions (see Fig.~\ref{Corner_sdss_legac}) and closer agreement with the Balmer absorption features, suggesting a stellar population with more active or recent star formation. These characteristics are consistent with the majority of CSP models, which include bursty and younger components. This is in agreement with previous studies based on PCA of spectra that also suggest a higher fraction of galaxies with broader index distributions at higher redshift \citep{Wild2009, Rowlands:2018}. In contrast, the SDSS sample exhibits stronger and more systematic deviations from the CSP model mean, especially D$_n$(4000) and the metallicity-sensitive indices, indicating more quiescent, metal-rich, and evolved stellar populations \citep{Gallazzi:05}. Together, these findings show an evolutionary sequence from active star formation at earlier cosmic times to more quenched systems in the local Universe \citep{wu2018, Wu2021}. Although not explicitly built into the models, the alignment of observed galaxies along this sequence suggests that it captures a physically meaningful axis of variation that is consistent with galaxy evolution, even if it does not represent a direct evolutionary pathway within the model framework. Fig.~\ref{SDSS-LEGAC} reinforces this finding, showing a bimodal distribution within the LEGA-C galaxy sample. One population of LEGA-C galaxies overlaps with the young SDSS sample in latent space. Based on the underlying structure of the CSP model distribution, this overlapping region corresponds to older, metal-rich populations with extended star formation histories. In contrast, a second group of LEGA-C galaxies occupies a distinct region in latent space, associated with younger, metal-poor populations and shorter star formation histories. The rapid change in the SFH time scale of the LEGA-C galaxies compared to that of lower-redshift SDSS galaxies is supported by previous studies \citep[e.g.,][]{Kauffmann2003,Kauffmann2014,Guo2016}. This bimodality in the LEGA-C sample supports the interpretation of ongoing galaxy evolution across cosmic time, as well as the validity of the model-derived latent structure in separating populations by physical properties. A previous study of the LEGA-C galaxies reveals a clear bimodal distribution in the D$_n$(4000)-EW(H$\delta_A$) diagnostic, indicative of a light-weighted age bimodality similar to that seen locally, but with stronger H$\delta_A$ absorption in the younger group \citep{wu2018}. The analysis of the LEGA-C DR3 data further indicates that these two populations occupy distinct locations in age-metallicity space: one is younger, metal-poor and star-forming, while the other is older, metal-rich and quiescent \citep{Nersesian2025, Gallazzi2026}. Comparisons with SDSS galaxies \citep{Gallazzi:05,Gallazzi:25} confirm that the intermediate-redshift galaxy population in LEGA-C is more actively forming stars and has broader index distributions, consistent with cosmic evolution toward more quiescent systems at later cosmic time \citep[e.g.,][]{MD:14}.

\begin{figure}
    \centering
    \includegraphics[width=\columnwidth]{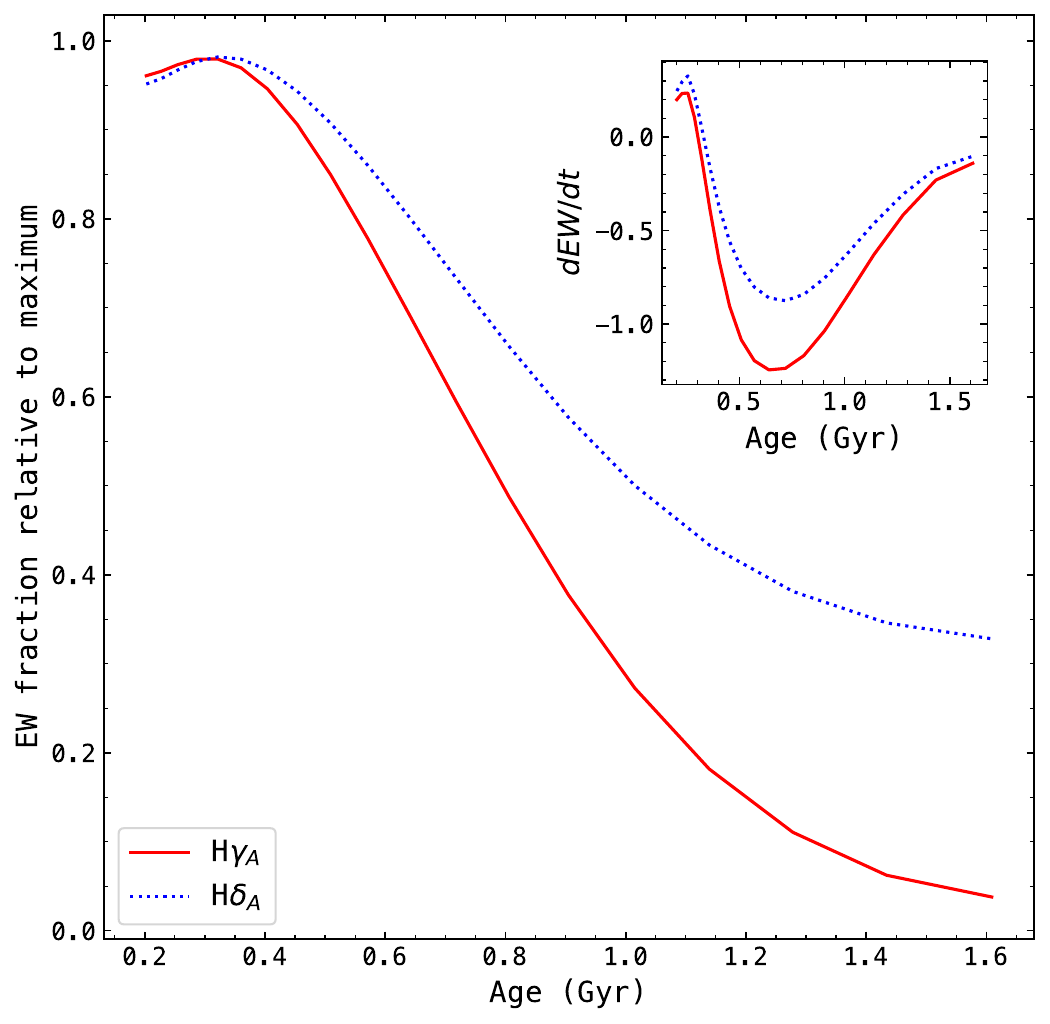}
    \caption{Variation of H$\gamma_A$ and H$\delta_A$ with respect to stellar age for a solar metallicity SSP from the models of \citet{Bruzual2003}. The main panel shows the absorption index measured as a ratio with respect to the maximum in each case. The inset shows the time derivative of this ratio in the region of interest.}
    \label{fig:PC6HgHd}
\end{figure}

To validate these data-driven results, we examined the physical properties of the two galaxy populations identified within the LEGA-C sample. We confirm that the observed bimodality is not driven by signal-to-noise ratio, redshift, or stellar mass, but instead arises from differences in stellar age. Specifically, the  light-weighted average age of the galaxies overlapping with the SDSS sample (black points in Fig.~\ref{SDSS-LEGAC}) is 4.6\,Gyr, while the distinct group (dark red points in Fig.~\ref{SDSS-LEGAC}) has a younger average age of 1.4\,Gyr. These age estimates come from the spectral fitting analysis of \citet{Gallazzi:25}, and they show strong agreement with the age distributions inferred from the principal component (PC) maps. Although both the PCA-based analysis and the fitting method use the same underlying set of CSP models, it is important to note that the PCA method does not involve direct spectral fitting. Therefore, this agreement underscores the power of PCA as a statistical tool for identifying population differences, especially when comparing observational samples or assessing simulations \citep[e.g.][]{Sharbaf2025}. This work highlights the potential of variance-based approaches like PCA in capturing meaningful physical differences, even when applied in a model-driven but non-parametric framework. Finally, concerning the link between burstiness and PC6, we note that the distribution of observational data for this component (Fig.~\ref{Corner_sdss_legac}) is rather wide, possibly due to noise -- not present in the models -- so that a careful analysis of high quality data is needed to explore this further, beyond the scope of this paper.

The results of this study demonstrate that galaxy evolution across redshift can be effectively described using a single principal component, with a strong correlation with stellar age. Although the standard model of galaxy formation within a cold dark matter framework suggests that galaxy properties are shaped by a wide range of physical parameters \citep[e.g.,][]{Dalcanton1997,Mo1998,Baugh2006}, our findings align with the conclusions of \citet{Disney2008}, who argued that a single underlying parameter could explain most of the fundamental scaling relations observed in galaxies. Similarly, our analysis reveals that the dominant component of variance, closely tied to stellar age, captures a significant portion of the structural and evolutionary information encoded in galaxy spectra across the entire redshift range.

\section*{Acknowledgements}
ZS acknowledges support from grant CNS2024-154592 financed by MICIU/AEI/10.13039/501100011033. 
ZS and IF acknowledge the support of the Spanish Research Agency of the Ministry of Science and Innovation (AEI-MICINN) under grant PID2019-104788GB-I00. ARG acknowledges support from the INAF-Minigrant-2022 ``LEGAC'' 1.05.12.04.01. SZ acknowledges support from the INAF-Minigrant-2023 ``Enabling the study of galaxy evolution through unresolved stellar population analysis'' 1.05.23.04.01. LSD acknowledges support from PRIN-MUR project ``PROMETEUS'' financed by the European Union – Next Generation EU, Mission 4 Component 1 CUP B53D23004750006. ZS thanks, Pedro. H Cezar, Eirini Angeloudi and Francesco Sinigaglia for their valuable comments and suggestions. Funding for SDSS-III has been provided by the Alfred P. Sloan Foundation, the Participating Institutions, the National Science Foundation, and the U.S. Department of Energy Office of Science. The SDSS-III web site is \href{http://www.sdss3.org/}{http://www.sdss3.org/}.

\section*{Data availability}
The original SDSS galaxy spectra were retrieved from the publicly available SDSS DR7 archive. The spectra of LEGA-C galaxies used in this analysis, as well as the estimates of the spectral indices and relative errors, are taken from the DR3 of the LEGA-C survey, and are publicly available. SDSS and LEGA-C catalogues used here are available through the website: \href{https://www.basta.inaf.it/}{https://www.basta.inaf.it/}. Other supporting material related to this article is available on request to the corresponding author.

%
\bibliographystyle{mnras}
\bibliography{PCA_CSP_Final}

\appendix

\section{Principal component projections}
Fig.~\ref{figapp:Cmodels} is a full version of Fig.~\ref{fig:Cmodels} in the main
body of the paper. Here the results are shown for the projections on all six
eigenvectors (PC1$\cdots$6). 

\begin{figure*}
    \centering
    \begin{subfigure}[t]{0.41\textwidth}
        \centering
        \captionsetup{justification=centering}
        \caption*{PC1 vs Spectral Indices}
        \includegraphics[width=\textwidth]{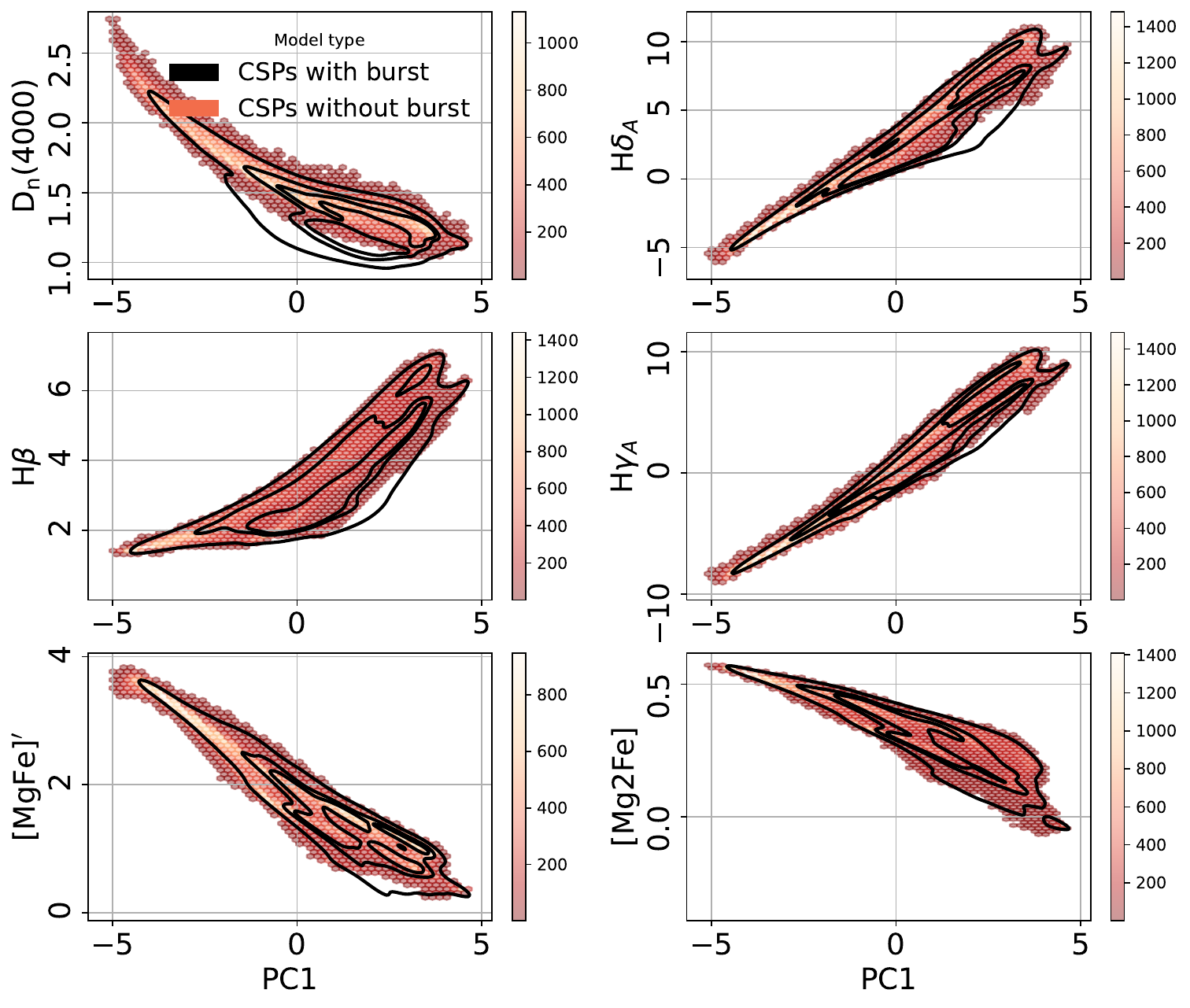}
        \label{fig:panel-a}
    \end{subfigure}\hfill
    \begin{subfigure}[t]{0.41\textwidth}
        \centering
        \captionsetup{justification=centering}
        \caption*{PC2 vs Spectral Indices}
        \includegraphics[width=\textwidth]{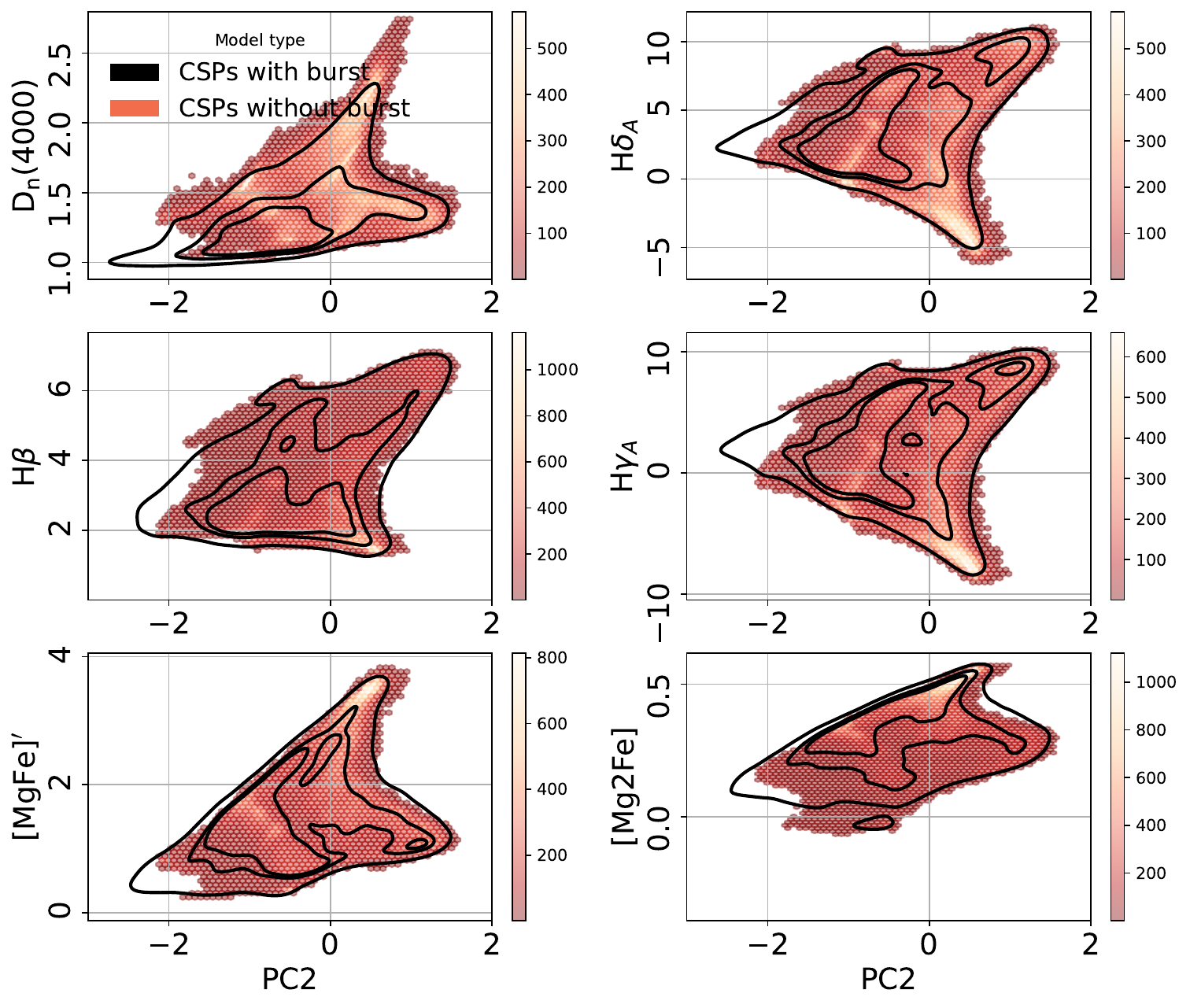}
        \label{fig:panel-b}
    \end{subfigure}
    \begin{subfigure}[t]{0.41\textwidth}
        \centering
        \captionsetup{justification=centering}
        \caption*{PC3 vs Spectral Indices}
        \includegraphics[width=\textwidth]{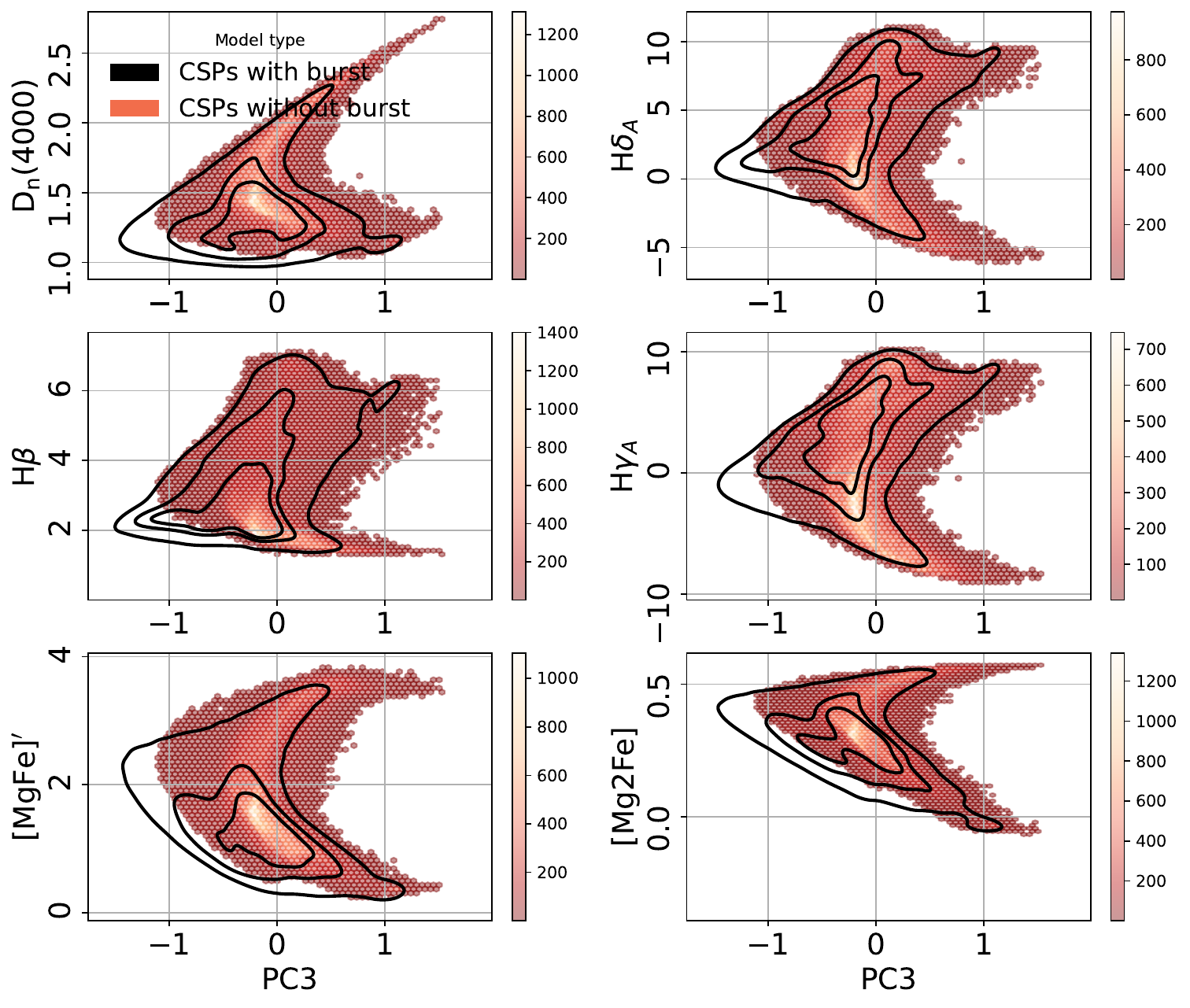}
        \label{fig:panel-c}
    \end{subfigure}\hfill
    \begin{subfigure}[t]{0.41\textwidth}
        \centering
        \captionsetup{justification=centering}
        \caption*{PC4 vs Spectral Indices}
        \includegraphics[width=\textwidth]{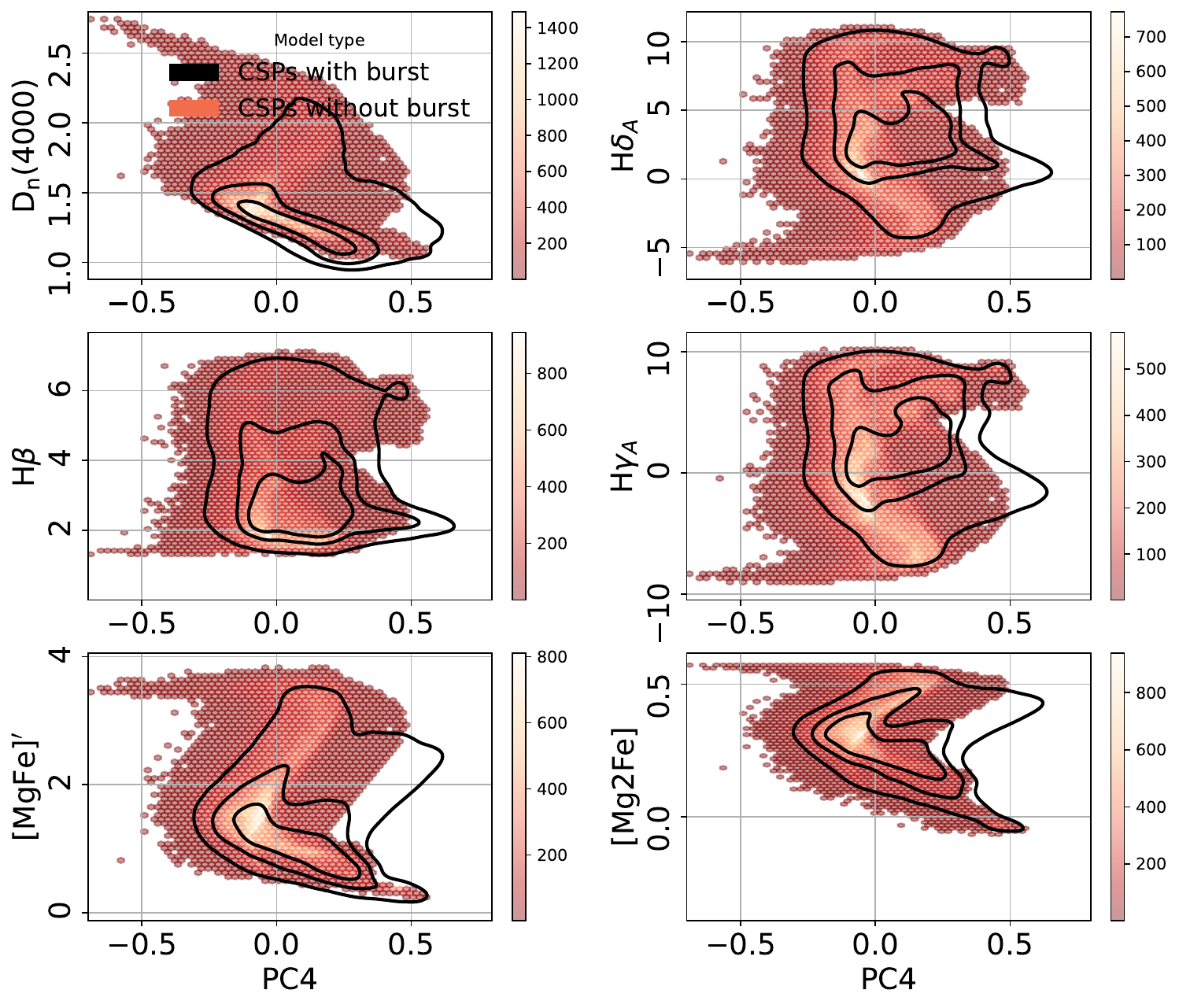}
        \label{fig:panel-d}
    \end{subfigure}

    \begin{subfigure}[t]{0.41\textwidth}
        \centering
        \captionsetup{justification=centering}
        \caption*{PC5 vs Spectral Indices}
        \includegraphics[width=\textwidth]{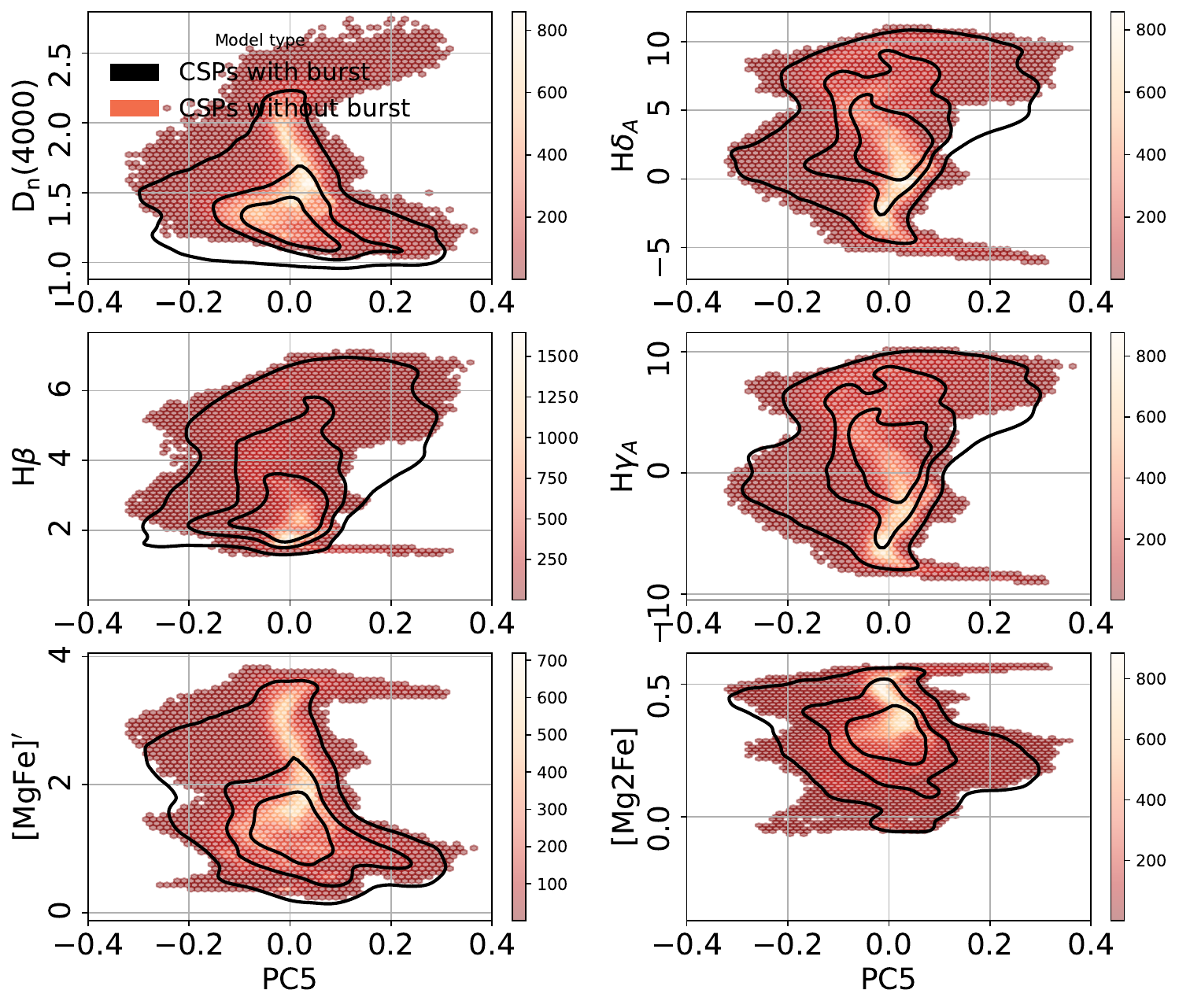}
        \label{fig:panel-e}
    \end{subfigure}\hfill
    \begin{subfigure}[t]{0.41\textwidth}
        \centering
        \captionsetup{justification=centering}
        \caption*{PC6 vs Spectral Indices}
        \includegraphics[width=\textwidth]{figs/PC6vLS_nobrst.pdf}
        \label{fig:panel-f}
    \end{subfigure}
    \caption{Full version of Fig.~\ref{fig:Cmodels}. Correlation of the index measurements with principal components when PCA is only applied to models without bursts, but projection to the derived eigenvectors is for models with and without bursts. The black contours with four levels illustrate the distribution for models with bursts, while the underlying density distribution, represented by the red points, shows the models without bursts. The color bar indicates the number of galaxies in each bin, while there are 50 horizontal bins for each panel.}
    \label{figapp:Cmodels}
\end{figure*}

\section{Effect of noise on the latent space comparison}\label{noise}

To assess the impact of observational uncertainties on the latent space, we propagated the measurement errors of the spectral indices into the principal components (PCs). For each galaxy, we generated 100 Monte Carlo realisations of its six index vector by perturbing each index with Gaussian noise scaled to its corresponding measurement error. These perturbed realisations were then standardised using the same mean and STD derived from the CSP models. This ensures that all noisy realisations are projected into the same standardised space as the PCs without noise realisations. 
Each of the 100 perturbed standardised index vectors is projected onto the PCA eigenvectors derived from the whole set of the CSP models, giving 100 PC scores per galaxy.
For each galaxy, the uncertainty in each PC score was estimated as the standard deviation of its 100 Monte Carlo realisations. As expected, galaxies with larger index measurement errors exhibit a wider spread in their realisations, resulting in larger uncertainties in PC space.

The effect of these uncertainties is illustrated in the corner plots shown in Figures \ref{fig:corner_SDSS_noisy} and \ref{fig:corner_LEGA_noisy}. In these figures, a random subset of 50 SDSS galaxies, along with the full LEGA-C sample, are compared against the CSP model distribution in PC space. Error bars corresponding to the propagated uncertainties in each PC have been included for the observational data. Overall, while measurement noise introduces scatter in the PC projections, it does not fully account for the systematic differences observed between the model and observational latent spaces.

\begin{figure} 
    \centering
    \includegraphics[width=\columnwidth]{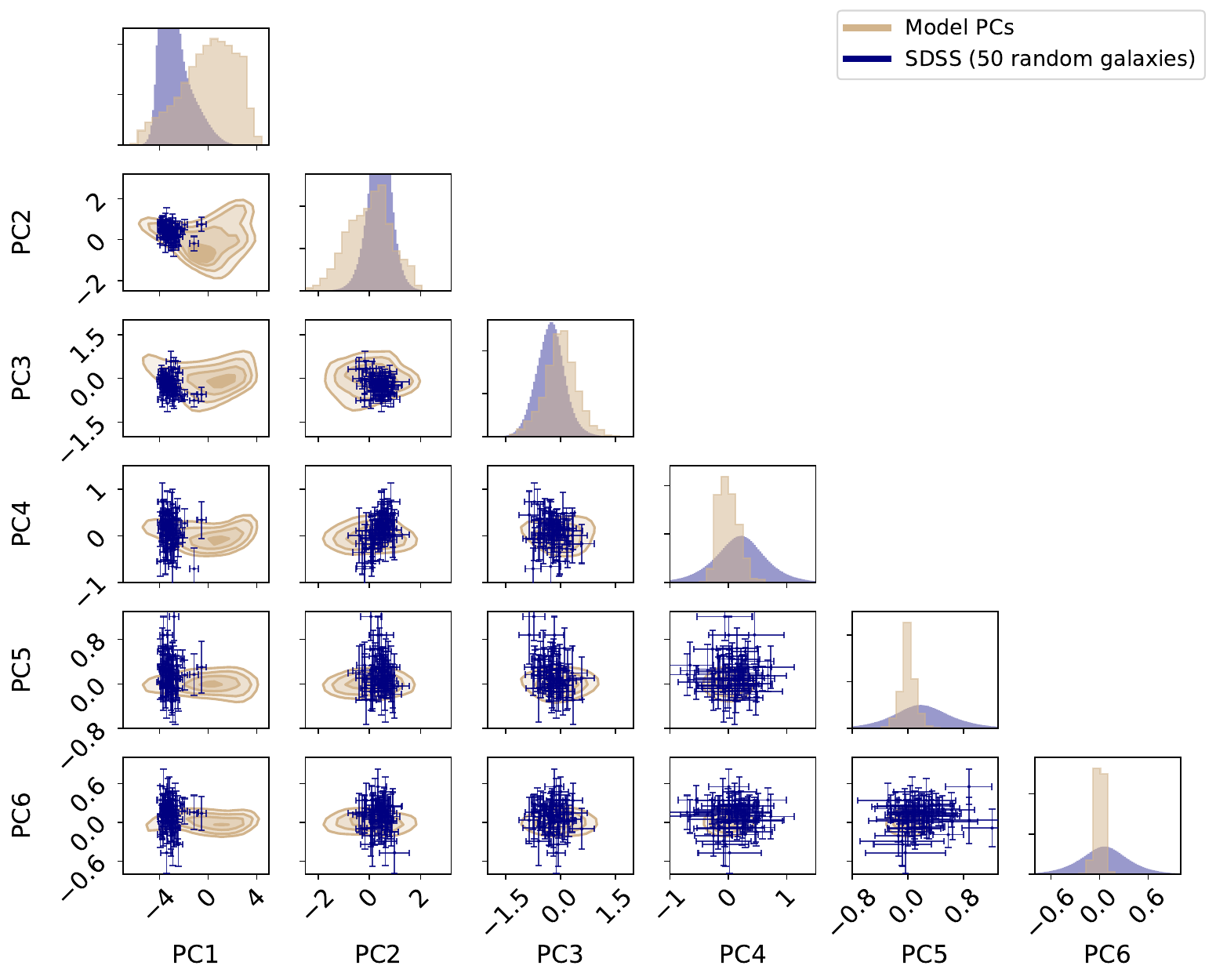}
    \caption{Comparison of the distribution of PCA-based latent space between the synthetic models (tan) and a randomly selected sample of SDSS galaxies, including error bars (blue).}
    \label{fig:corner_SDSS_noisy}
\end{figure}

\begin{figure} 
    \centering
    \includegraphics[width=\columnwidth]{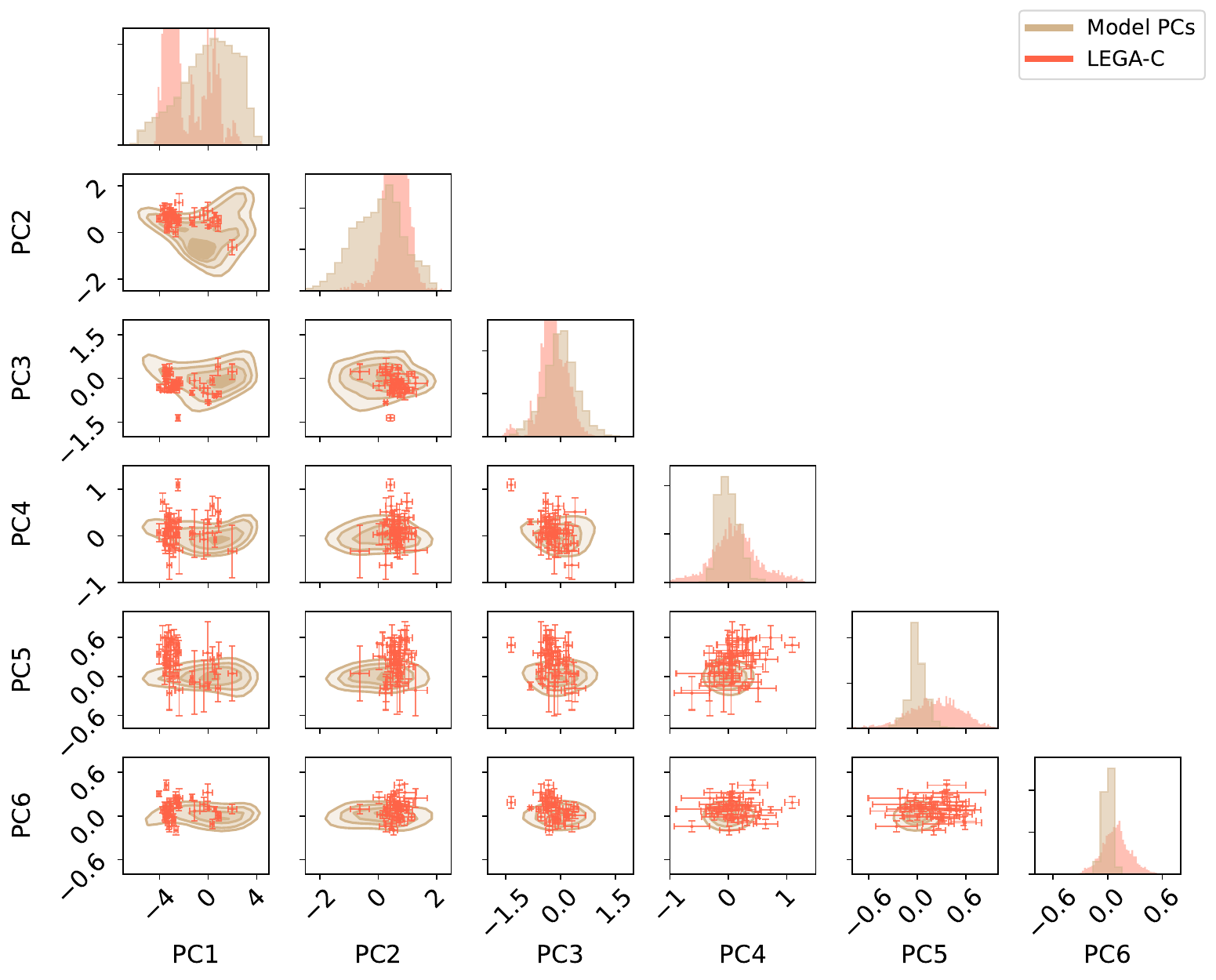}
    \caption{Equivalent of Fig.~\ref{fig:corner_SDSS_noisy} for LEGA-C galaxies (red).}
    \label{fig:corner_LEGA_noisy}
\end{figure}

\bsp	
\label{lastpage}

\end{document}